\documentclass[reprint, superscriptaddress, twocolumn, pra, nofootinbib]{revtex4-2}
\usepackage{amsmath, amssymb}
\usepackage{graphicx}
\usepackage[usenames, dvipsnames]{color}
\usepackage{mathtools}
\usepackage{hyperref}
\hypersetup{colorlinks=true, linkcolor = blue, citecolor = blue, urlcolor = blue}
\usepackage{orcidlink}
\usepackage{xcolor}
\usepackage{bbold}
\newcommand{\opH}{\hat{\mathcal{H}}}

\newcommand{\opD}{\hat{\mathcal{D}}}

\newcommand{\opR}{\hat{R}}

\newcommand{\opS}{\hat{\mathcal{S}}}

\newcommand{\opN}{\hat{N}_\mathrm{ex}}

\newcommand{\Tr}{\mathrm{Tr}} 
\newcommand{\dd}{\mathrm{d}}
\newcommand{\ee}{\mathrm{e}}
\newcommand{\ii}{\mathrm{i}}

\newcommand{\CC}{\mathbb{C}}

\newcommand{\omegar}{\omega_\mathrm{r}}

\newcommand{\Lgr}{L} 						

\newcommand{\res}{\mathrm{r}}  				
\newcommand{\pd}{\mathrm{p}}				
\newcommand{\cpl}{\mathrm{c}}				

\newcommand{\zpf}{\mathrm{zpf}}				

\newcommand{\cpc}{C}						
\newcommand{\ind}{L}						
\newcommand{\JJ}{\mathrm{J}}				
\newcommand{\Cs}{C_\mathrm{s}}				
\newcommand{\LG}{L_\mathrm{G}}				
\newcommand{\EJ}{E_\JJ}

\newcommand{\Phib}{\widetilde{\Phi}_\mathrm{b}}

\newcommand{\Phir}{\Phi_\res}

\newcommand{\phib}{\widetilde{\phi}_\mathrm{b}}
\newcommand{\mnm}{\mathrm{m}}
\newcommand{\phim}{\phi^\mnm}

\newcommand{\PP}{\mathcal{P}} 				
\everymath{\displaystyle}

\begin{document}
\title{Photon-number resolution with microwave Josephson photomultipliers}
\author{E.~V.~Stolyarov \orcidlink{0000-0001-7531-5953}}
\affiliation{Bogolyubov Institute for Theoretical Physics, National Academy of Sciences of Ukraine, Vulytsya Metrolohichna 14-b, 03143 Kyiv, Ukraine}
\affiliation{Institute of Physics, National Academy of Sciences of Ukraine, Prospect Nauky 46, 03028 Kyiv, Ukraine}
\author{O.~V.~Kliushnichenko \orcidlink{0000-0003-4806-8971}}
\affiliation{Institute of Physics, National Academy of Sciences of Ukraine, Prospect Nauky 46, 03028 Kyiv, Ukraine}
\author{V.~S.~Kovtoniuk \orcidlink{0009-0008-8363-0898}}
\affiliation{Bogolyubov Institute for Theoretical Physics, National Academy of Sciences of Ukraine, Vulytsya Metrolohichna 14-b, 03143 Kyiv, Ukraine}
\author{A.~A.~Semenov \orcidlink{0000-0001-5104-6445}}
\affiliation{Bogolyubov Institute for Theoretical Physics, National Academy of Sciences of Ukraine, Vulytsya Metrolohichna 14-b, 03143 Kyiv, Ukraine}
\affiliation{Institute of Physics, National Academy of Sciences of Ukraine, Prospect Nauky 46, 03028 Kyiv, Ukraine}
\affiliation{Department of Theoretical and Mathematical Physics, Kyiv Academic University, Boulevard Vernadskogo 36, 03142 Kyiv, Ukraine}
\begin{abstract}
    We study counting photons confined in a mode of a microwave resonator via repeated measurements by a Josephson photomultiplier (JPM).
    The considered JPM is essentially a flux-biased phase qubit operating as a single-photon detector.
    We identify optimal operational regimes that maximize photon-number resolution within a predetermined range.
    Two counting techniques are studied.
    The first is to count the total number of clicks in the measurement sequence.
    The second involves counting the number of clicks until the occurrence of either the first no-click event or the end of the measurement sequence.
    Our theoretical methods employ the derived positive operator-valued measures for the considered photocounting techniques and the introduced measure of the photon-number resolution.
    The results reveal that the resolution decrease in both cases is mainly caused by the JPM relaxation. 
    As an example, we show how the obtained results can be used for practical testing nonclassical properties of electromagnetic radiation in a microwave resonator.
\end{abstract}

\setcounter{page}{1}\maketitle

\section{Introduction} \label{sec:intro}


    The measurement theory \cite{busch2016} is a vital part of quantum physics, linking quantum states to experimentally accessible measurement outcomes and their probabilities.
    Photocounting measurements, the theoretical description of which was introduced in Refs.~\cite{mandel1964, kelley1964, lamb1969, mandelwolf1995book}, are an example of measurements that play a key role in quantum optics.
    These measurements are essential for both quantum technologies \cite{knill2001, kok2007, gisin2007, hadfield2009, natarajan2012, you2020, giovannetti2011, degen2017, pirandola2020, polino2020, xu2020} and experiments testing quantum theory \cite{shalm2015, bohmann2018}.

    In an ideal scenario, the presence of $n$ photons in the electromagnetic radiation must always be converted into $n$ clicks of the photocounter.
    However, practical situations often differ significantly from this idealized situation. 
    Even the simplest imperfection---detection losses---leads to the possibility of registering fewer clicks than the number of photons present \cite{herzog1996}.
    On the other hand, dark counts may increase the number of clicks \cite{karp1970, lee2004,semenov2008,*semenov2008err}.
	
    Photocounters commonly used in quantum optics, including photomultiplier tubes, avalanche photodiodes, and superconducting nanowire single-photon detectors, typically operate as single-photon on-off detectors \cite{hadfield2009} with no photon-number resolution at all.
    Nevertheless, the usage of various experimental approaches enables approximating a challenging experimental issue of resolving between adjacent numbers of photons.
    One such method entails dividing the light beam into multiple spatial or temporal modes and detecting each of them with on-off detectors \cite{paul1996, achilles2003, fitch2003, rehacek2003, castellano2007, schettini2007, blanchet2008, hlousek2019}.
    Another approach involves counting the number of photocurrent pulses in measurement time windows \cite{saleh1978book}.
    However, none of these techniques provides reasonable photon-number resolution \cite{sperling2012, kovalenko2018, uzunova2022, semenov2023}.
    This means that the number of detected clicks may differ considerably from the actual number of photons.


    Alternatively to the optical domain, one can use microwave radiation, which has been extensively studied within circuit quantum electrodynamics (QED); see, e.g., Refs.~\cite{wendin2007ltp, *clarke2008, gu2017, *blais2021rmp} for a review.
    In this context, a number of theoretical and experimental advances have been demonstrated in quantum state engineering \cite{makhlin2001rmp, hofheinz2008, hofheinz2009, premaratne2017, pfaff2017, leghtas2013, eickbusch2022}, quantum computation and simulation \cite{devoret2013, *wendin2017, *blais2021rmp, lamata2018, kreula2016, potocnik2018, mezzacapo2015, brennen2016}, etc.
    A promising approach involves utilizing Josephson photomultipliers (JPM) \cite{yfchen2011, oelsner2013, oelsner2017, golubev2021, guarcello2021, opremcak2018, opremcak2021} as single-photon detectors for electromagnetic radiation confined in a microwave resonator \cite{poudel2012, govia2012, govia2014, oelsner2017, schondorf2018a, golubev2021} or propagating in a microwave waveguide \cite{romero2009, romero2009ps, peropadre2011, schondorf2018b}.
    For example, the use of JPMs can be advantageous for dispersive qubit readout \cite{govia2014, opremcak2018,opremcak2021, schondorf2018a, sok2020}.
    
    The JPM is an on-chip device compatible with other circuit QED components.
    Technically, it is a phase qubit \cite{martinis2009} that acts as a narrowband absorbing photodetector.
    In general, there are two types of JPMs. 
    The first type \cite{yfchen2011, poudel2012, govia2012, oelsner2013, govia2014, schondorf2018b, oelsner2017, golubev2021, guarcello2021} is based on a single current-biased Josephson junction (JJ) \cite{martinis1985}.
    While compact and simple, this type of JPM has an essential disadvantage.
    When a photon is absorbed by such a JPM, it switches to the voltage state, resulting in an outburst of quasiparticles which strongly affect the next measurements.
    The quasiparticles have relaxation times of milliseconds \cite{lenander2011, wang2014} which decreases the measurement repetition rate for this JPM design to $\textstyle \sim 1\,\mathrm{kHz}$ \cite{ribeill2016} limiting its practical use.
    
    To remedy the limitations of current-biased JPMs, another design has been proposed and experimentally demonstrated; cf. Refs.~\cite{shnyrkov2018, opremcak2018, opremcak2021, shnyrkov2023}.
    In this design, a JJ is shunted by an inductor, which adds the quadratic term to the JJ cosine potential. 
    By adjusting the external flux through the JPM loop, the JPM potential can be set to a two-well configuration.
    This prevents the JPM from switching completely to the voltage state due to photon absorption and subsequent tunneling out of the local potential well.
    Instead, photon absorption results in a single tunneling event.
    By reading which well the particle is in, one can determine if the tunneling event has occurred.
    The inductively-shunted design provides fast readout and reset. 

    The macroscopic response of JPMs indicates absorption of a single photon with high detection efficiency.
    This feature sets them apart from most single-photon detectors operating in the optical domain, whose macroscopic response is indistinguishable for single- and multiphoton absorption.
    This brings us to the idea of using JPMs for fast repetition measurements.
    In this paper, we theoretically examine whether such a measurement sequence can give a result similar to that of ideal photon-number resolving (PNR) detectors.

    We focus on two counting techniques.
    In the first, we repeat the measurement a fixed number of times, regardless of the obtained outcomes.
    The second approach consists in counting photons until either the first no-click event or the end of the sequence.
    The number of photons is then associated with the number of clicks.

    Our theoretical methods imply an accurate derivation of the positive operator-valued measures (POVMs), see e.g., Refs.~\cite{nielsen2000book, vanenk2017}, for both counting techniques.
    This in fact means that we get photocounting formulas tailored to the considered scenarios.
    We introduce a measure of photon-number resolution, based on a recent observation in Ref.~\cite{Len2022}, that the probability of registering $n$ clicks given $n$ photons is a relevant characteristic of photon-number resolution.
    We optimize this measure with parameters of JPM and show that in practical situation such detectors can be considered as pseudo-PNR detectors, with high ability to resolve between adjacent photon numbers.

    We use our theory to show how JPMs can be employed to test nonclassicality of electromagnetic radiation.
    As widely accepted in quantum optics \cite{titulaer1965, mandel1986, sperling2018a, *sperling2018b, sperling2020, mandelwolf1995book, vogel2006book, agarwal2013book}, a quantum state of light is nonclassical if it cannot be considered as a statistical mixture of coherent states, i.e., if its Glauber-Sudarshan $P$ distribution \cite{glauber1963, sudarshan1963} cannot be interpreted as a probability density function.
    We utilize the methods recently developed in Refs.~\cite{semenov2021, kovtoniuk2023}, which enable us to test whether the photocounting statistics for the given quantum state can be reproduced with the classical electromagnetic radiation.
    
    The rest of the paper is organized as follows.
    Section~\ref{sec:jpm} provides a detailed description of the operation principle of the considered JPM.
    The Hamiltonian of the resonator mode--JPM system is derived in Sec.~\ref{sec:mdl}.
    The POVM of the JPM counter for two different photon counting techniques is derived in Sec.~\ref{sec:stat}.
    In Sec~\ref{sec:rsl}, we introduce a measure characterizing the ability of detectors to resolve between adjacent numbers of photons and use it for the optimization of the detector parameters.
    In Sec.~\ref{sec:exmpl}, we determine the counting statistics of the JPM for different quantum states of the resonator field.
    In Sec.~\ref{sec:noncl}, we apply our theory to develop a method for testing nonclassicality of photocounting statistics with the JPM. 
    Our results are summarized in Sec.~\ref{sec:summ}.
    The derivation of the classical circuit Hamiltonian is given in Appendix~\ref{sec:ham_der}.

\section{Josephson photomultiplier operation principle} \label{sec:jpm}

\begin{figure*}[t!]
	\centering
	\includegraphics{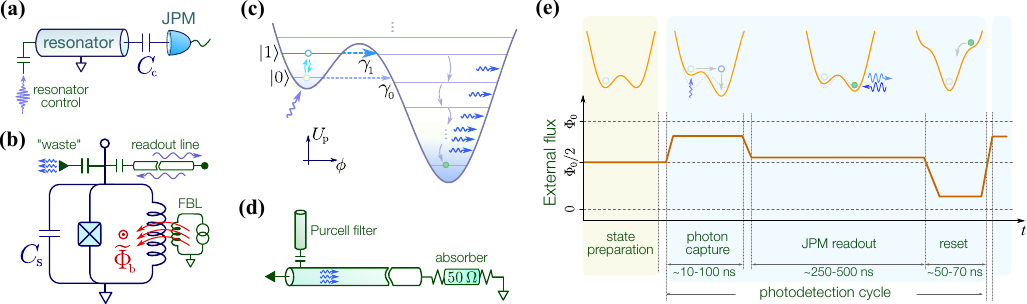}
	\caption{Setup and JPM details. (a) General scheme of the considered system. The JPM is coupled to the resonator via a static capacitance $\textstyle C_\mathrm{c}$.
		(b) Schematics of the JPM realization we consider.
		(c) Illustration of the JPM potential featuring a two-well structure: the shallow left well accommodates just two levels ($\textstyle |0\rangle$ and $\textstyle |1\rangle$), while the deep right well accommodates multiple levels.
		The metastable levels $\textstyle |0\rangle$ and $\textstyle |1\rangle$ in the shallow well have tunneling rates $\textstyle \gamma_0$ and $\textstyle \gamma_1$, correspondingly.
		(d) Potential implementation of the ``waste" mode.
		The transmission line, terminated by the absorbing element, is supplemented by the Purcell filter, which rejects the photon frequencies in the vicinity of the JPM working frequency $\textstyle \omega_\pd$.
		(e) Considered photodetection procedure.
         We start with preparing a quantum state in the resonator mode. Then, to capture the photon, the JPM potential is adjusted to a configuration with two levels in the left well. Next, the JPM readout is performed to determine whether the tunneling has occurred. After the readout, the JPM is reset. For the next measurement, the JPM is again adjusted for the photon capture.
		\label{fig:fig_1}}
\end{figure*}

We consider a circuit QED setup consisting of a Josephson photomultiplier (JPM) coupled to a mode of a microwave resonator. 
The scheme of the considered system is shown in Fig.~\ref{fig:fig_1}(a).
The resonator is represented by either a coplanar waveguide resonator (CPW) \cite{goppl2008, megr2012, bruno2015}, a lumped-element resonator \cite{allman2010, collodo2019, vrajitoarea2020}, coaxial resonator \cite{reagor2016, heidler2021}, or a three-dimensional (3D) microwave cavity \cite{paik2011, reagor2013, romanenko2020, milul2023}.
All these types of resonators are routinely employed in the state-of-the-art circuit QED systems.
In the considered setup, the coupling between the resonator mode and the JPM is mediated by a capacitor $\textstyle C_\cpl$.

The JPM design we consider is based on a flux-biased phase qubit~\cite{steff2006} constituted by the JJ shunted by a gradiometric inductance $\textstyle \LG$ and a large capacitance $\textstyle \Cs \gg C_\JJ$, where $\textstyle C_\JJ$ is the capacitance of the junction.
The JJ is characterized by the critical current $\textstyle I_0$ and the Josephson energy $\textstyle E_\JJ = I_0 \Phi_0/(2\pi)$, where $\textstyle \Phi_0 = h/(2e)$ denotes the magnetic flux quantum.
The JJ and the gradiometric inductance $\textstyle \LG$ comprise together an rf-SQUID loop threaded by the external flux $\textstyle \Phib$ generated by the external flux-bias line (FBL).
Such a JPM design was experimentally demonstrated by Opremcak \textit{et al.}~\cite{opremcak2018, opremcak2021}.
For gaining more control over the JPM operation frequency, one can replace the single JJ in the JPM circuit with a dc SQUID controlled via an individual FBL \cite{shnyrkov2018}.
However, for conciseness, here we consider a JPM design with a single JJ similar to that demonstrated in Refs.~\cite{opremcak2018, opremcak2021}.
The JPM schematics is demonstrated in Fig.~\ref{fig:fig_1}(b).

With the proper choice of the Josephson energy $\textstyle \EJ$ and the gradiometric inductance $\textstyle \LG$, the JPM potential can feature either one or two wells depending on the value of the external flux $\textstyle \Phib$.
For probing the resonator field, one tunes the flux $\textstyle \Phib$ in such a way that the JPM potential has an asymmetric two-well configuration with a deep well, accommodating multiple energy states, and a shallow well hosting only two metastable states (denoted as $\textstyle |0\rangle$ and $\textstyle |1\rangle$) as shown in Fig.~\ref{fig:fig_1}(c).
For ensuring the efficient excitation exchange between the resonator and the JPM, the resonator frequency and the frequency $\textstyle \omega_\pd$ of $\textstyle |0\rangle \leftrightarrow |1\rangle$ JPM transition are tuned to resonance.
The level $\textstyle |1\rangle$ lies near the top of the potential barrier separating the wells.
This provides the rapid tunneling of the excitation to the deep well.
Once the excitation has tunneled to the deep well, it undergoes the fast cascaded relaxation via multiple levels accommodated in the deep well and ultimately relaxes to the global minimum of the JPM potential.
This process is schematically illustrated in Fig.~\ref{fig:fig_1}(c).
The tunneling and relaxation to the deep potential well results in the change of the JPM classical fluxoid state, which is interpreted as a ``click".

To ensure the rapid relaxation to the bottom of the potential well, we propose to couple the JPM to an additional ``waste" mode.
The latter constitutes an engineered electromagnetic environment implemented by a long microwave transmission line terminated by an impedance-matched absorbing element \cite{cattaneo2021} and supplemented by a side-coupled resonator with frequency $\textstyle \omega_\pd$.
This resonator serves as a band-stop Purcell filter \cite{reed2010} rejecting the photons with frequencies close to the JPM working frequency.
The scheme of this JPM supplement is shown in Fig.~\ref{fig:fig_1}(d).
Besides, such a ``waste" mode coupled to the JPM efficiently channels the broadband cascade of microwave photons generated in the course of the JPM relaxation out of the resonator-JPM system.
This provides an additional suppression of the spurious population of the resonator by those photons.
The latter effect was observed in experiment in Ref.~\cite{opremcak2018}.

Now, let us outline the procedure of photodetection [see diagram in Fig.~\ref{fig:fig_1}(e)] in the setup we consider.
Here, we mainly follow Refs.~\cite{opremcak2018, opremcak2021}, and the considered sequence is essentially a simplified version of a detection procedure presented therein.
On the initial stage, we prepare a quantum state of the resonator mode.
In particular, Fock states \cite{hofheinz2008, premaratne2017, pfaff2017} and their arbitrary superpositions \cite{hofheinz2009}, squeezed vacuum \cite{moon2005, didier2014, kono2017, malnou2018, dass2021, eickbusch2022}, and Schr\"{o}dinger cat states \cite{leghtas2013, pfaff2017, ma2019,girvin2019} can be prepared for further analysis.
On the preparation stage, the JPM potential is set in a symmetric two-well configuration.
The frequencies of the JPM interlevel transitions are far detuned from the resonator frequency and there is no excitation exchange between the resonator mode and the JPM.

After the resonator field state being prepared, we need to efficiently capture the resonator photons by the JPM.
On the photon capture stage, the JPM potential is rapidly tilted in such a way, that the left well contains just two levels and the transition frequency between them is close to the resonator frequency, which enables excitation exchange between the resonator and the JPM.
As we mentioned earlier, in this regime, the resonator photon drives the JPM from the ground to the excited state.
The excitation then tunnels via the barrier and relaxes to the bottom of the right deep well.
We let the resonator and the JPM interact during time $\textstyle t_\mathrm{cpt}$ long enough to ensure high probability of photon absorption and subsequent interwell tunneling.

Next, we need to perform the JPM readout to determine whether the photon absorption and tunneling have occurred and the JPM has changed its fluxoid state.
On the readout stage, the JPM potential is tilted back to a slightly asymmetric two-well configuration that each well is characterized by a different plasma frequency.
By probing the JPM with a weak microwave signal on one of the plasma resonances, one can determine the JPM fluxoid state with high ($\textstyle >99.9\%$) fidelity in $\textstyle \sim 250\,\mathrm{ns}$ \cite{opremcak2021}.

Besides the JPM readout approach of Refs.~\cite{opremcak2018, opremcak2021} outlined above, there are other techniques for the JPM readout.
For example, the interwell tunneling induced by the absorption of the resonator photon results in the change of the JPM fluxoid state which can be detected by the rf SQUID weakly coupled to the JPM \cite{shnyrkov2023}.
Alternatively, the JPM fluxoid state can be detected by the ballistic fluxons propagating in the underdamped Josephson transmission line interfaced with the single-flux quantum logic digital circuitry \cite{howington2019}.

Finally, after the readout the JPM is reset to the single-well configuration and the current photodetection procedure is over.
For the next measurement, the JPM potential is again set to the photon capture regime and the whole detection procedure is repeated.

\section{The Model} \label{sec:mdl}

\subsection{The Hamiltonian} \label{sec:hmlt}
Let us start our analysis with a classical description of the considered system comprised by the resonator mode coupled to the JPM.
It is described by the classical Hamiltonian
\begin{equation} \label{eq:clham0}
	  H = H_\res + H_\pd + H_\mathrm{int},
\end{equation}
see Appendix~\ref{sec:ham_der} for details of its derivation.
Here $H_\res$, $H_\pd$, and $H_\mathrm{int}$ are terms describing the resonator mode, the JPM, and the interaction between them, respectively.

The first term in Eq.~\eqref{eq:clham0} is given by
\begin{equation} \label{eq:clham_res}
	H_\res = 4 E_{C,\res} n_\res^2 + E_{L,\res}\frac{\phi_\res^2}{2},
\end{equation}
where $\textstyle n_\res$ and $\textstyle \phi_\res$ denote the resonator Cooper-pair number and phase variables, respectively, which are proportional to field quadratures of the resonator mode.
In Eq.~\eqref{eq:clham_res}, parameters $E_{C,\res} = e^2/(2 C'_\res)$ and $E_{L,\res} = (\Phi_0/2\pi)^2/L_\res$ are the resonator capacitive and inductive energies, respectively.
The last term in Eq.~\eqref{eq:clham_res} is the resonator potential energy.

The second term in Eq.~(\ref{eq:clham0}) is the JPM Hamiltonian expressed as
\begin{equation} \label{eq:clham_pd}
	H_\pd = 4 E_{C,\pd} n_\pd^2 + \overbrace{E_{L,\pd} \frac{(\phi_\pd-\phib)^2}{2} - E_\JJ \cos \phi_\pd}^{U_\pd(\phi_\pd)},
\end{equation}
where $\textstyle n_\pd$ and $\textstyle \phi_\pd$ denote the JPM Cooper-pair number and phase variables, respectively.
The quantities $\textstyle E_{\cpc,\pd} = e^2/(2 C'_\pd)$ and $\textstyle E_{\ind,\pd} = (\Phi_0/2\pi)^2/\LG$ stand for the capacitive and inductive energies of the photodetector, respectively.
The last two terms in Eq.~\eqref{eq:clham_pd} constitute the JPM potential energy $\textstyle U_\pd(\phi_\pd)$, where $E_\JJ$ is the Josephson energy.

The third term in Eq.~\eqref{eq:clham0}, describing the interaction between the resonator mode and the JPM, reads as
\begin{equation} \label{eq:clham_cpl}
	H_\mathrm{int} = E_{\cpc,\cpl} \, n_\res n_\pd, 
\end{equation}
where $\textstyle E_{\cpc, \cpl} = e^2/(2C'_\cpl)$ stands for the energy of the capacitive coupling between the resonator and the JPM.
The definitions of the renormalized (loaded) capacitances $\textstyle C'_\res$, $\textstyle C'_\pd$, and $\textstyle C'_\cpl$ arising in Eqs.~\eqref{eq:clham_res}--\eqref{eq:clham_cpl} are given in Eq.~\eqref{eq:caps_ld} in Appendix~\ref{sec:ham_der}.

In the phase qubit regime, one has $\textstyle E_{\cpc,\pd} \ll E_{\JJ}, E_{\ind,\pd}$.
This implies that the JPM phase ``particle" is localized in the vicinity of the local minimum of the JPM potential resembling a behavior of a classical heavy particle oscillating near its equilibrium position in a potential well.
The positions of the local minima of the JPM potential $\textstyle U_\pd(\phi)$ are determined from the standard conditions
\begin{equation*}
	\frac{\dd}{\dd \phi} U_\pd(\phi) \big|_{\phi^\mnm_\pd} = 0, \quad
        \frac{\dd^2}{\dd\phi^2} U_\pd(\phi)\big|_{\phi^\mnm_\pd} > 0,
\end{equation*}
which yield
\begin{equation} \label{eq:eqmin}
   \begin{split}
   	& E_{\ind,\pd} (\phim_{\pd} - \phib) + E_\JJ \sin \phim_{\pd}  = 0, \\
   	& E_{\ind,\pd} + E_\JJ \cos\phim_{\pd} > 0.
   \end{split}
\end{equation}
Next, let us define $\textstyle \phi_\pd = \phim_{\pd} + \varphi_\pd$ with $\textstyle \varphi_\pd$ being a fluctuation from the equilibrium position of the JPM phase ``particle".
Similar arguments hold for the resonator phase ``particle" as well, since $\textstyle E_{\cpc,\res} \ll E_{\ind, \res}$.
Thus, we can formally write $\textstyle \phi_\res = \phim_{\res} + \varphi_\res$, where $\phim_{\res}$ is the minimum of the resonator potential. However, since $\textstyle \phim_{\res} = 0$, we just replace $\textstyle \phi_\res$ with $\textstyle \varphi_\res$ for consistency of notations.

To proceed, we expand the cosine term in the JPM potential $\textstyle U_\pd(\varphi_\pd+\phim_\pd)$ up to the fourth order in $\textstyle \varphi_\pd$.
Accounting for the first line in Eq.~\eqref{eq:eqmin} and dropping the constant terms, one obtains
\begin{equation} \label{eq:U_pd}
	U_\pd(\varphi_\pd) = E'_{\ind,\pd} \, \frac{\varphi_\pd^2}{2} - E_\JJ \sin\phim_\pd \, \frac{\varphi_\pd^3}{6} - E_\JJ\cos\phim_\pd \, \frac{\varphi_\pd^4}{24}.
\end{equation}
Here $\textstyle E'_{\ind,\pd} = E_{\ind,\pd} + E_\JJ\cos\phim_\pd$ can be interpreted as the total inductive energy of the JPM.

Now, we proceed to a quantum description of the considered resonator mode--JPM system.
For obtaining the quantum version of the circuit Hamiltonian, we follow the canonical quantization approach \cite{devoret1997, vool2017, rasmussen2021}.
We promote the classical variables of the JPM and the resonator mode to the quantum operators obeying the commutation relations $\textstyle [\hat{\varphi}_\pd,\hat{n}_\pd] = [\hat{\varphi}_\res,\hat{n}_\res] =\ii$.
Then, we express these operators in terms of the ladder operators.
For the resonator phase and Cooper-pair number variables $\textstyle \hat{\varphi}_\res$ and $\textstyle \hat{n}_\res$, one has
\begin{equation} \label{eq:ops_res}
	\hat{\varphi}_\res = \varphi_{\zpf,\res} (\hat{a}^\dag + \hat{a}), \quad \hat{n}_\res = -\ii  \, n_{\zpf,\res} (\hat{a}^\dag - \hat{a}),
\end{equation}
where $\textstyle \hat{a}$ ($\textstyle \hat{a}^\dag$) is the annihilation (creation) operator of an excitation (photon) in the resonator mode.
Parameters $\textstyle \varphi_{\zpf,\res} = (2E_{C,\res}/E_{L,\res})^{1/4}$ and $\textstyle n_{\zpf,\res} = [E_{L,\res}/(32E_{C,\res})]^{1/4}$ correspond to the zero-point fluctuations of the phase and the Cooper-pair number of the resonator, respectively.
Analogously, the JPM phase and the Cooper-pair number variables are expressed as
\begin{equation} \label{eq:ops_jpm}
	\hat{\varphi}_\pd = \varphi_{\zpf,\pd} (\hat{b}^\dag + \hat{b}), \quad \hat{n}_\pd = -\ii  \, n_{\zpf,\pd} (\hat{b}^\dag - \hat{b}).
\end{equation}
Here $\textstyle \hat{b}$ ($\textstyle \hat{b}^\dag$) is an annihilation (creation) operator of an excitation in the JPM, $\textstyle \varphi_{\zpf,\pd} = (2E_{\cpc,\pd}/E'_{\ind,\pd})^{1/4}$ and $\textstyle n_{\zpf,\pd} = [E'_{\ind,\pd}/(32E_{\cpc,\pd})]^{1/4}$ are the zero-point fluctuations of the JPM phase and Cooper-pair number.

Substitution of the expressions given by Eqs.~\eqref{eq:ops_res} and \eqref{eq:ops_jpm} in the quantum counterparts of Eqs.~\eqref {eq:clham0}, \eqref{eq:clham_res}, \eqref{eq:clham_pd}, and \eqref{eq:U_pd} yields the quantum Hamiltonian of the resonator mode--JPM circuit,
\begin{equation} \label{eq:qham1}
	\begin{split}
	  \frac{\opH}{\hbar} = & \, \overbrace{\omegar \hat{a}^\dag \hat{a}}^{\opH_\res} + \overbrace{\varpi_\pd \hat{b}^\dag \hat{b} - \varXi_3 (\hat{b}^\dag + \hat{b})^3 - \varXi_4 (\hat{b}^\dag + \hat{b})^4}^{\opH_\pd} \\
	  & \, \underbrace{- g(\hat{a}^\dag - \hat{a})(\hat{b}^\dag - \hat{b})}_{\opH_\mathrm{int}}.
	\end{split}
\end{equation}
Here the first term is the Hamiltonian of the quantum harmonic oscillator (mode of the resonator) with frequency $\textstyle \omega_\res = 1/\sqrt{L_\res C'_\res}$.
The second term in the above equation is the Hamiltonian of the JPM, where $\textstyle \varpi_\pd$ is the JPM plasma frequency determined as
\begin{equation}
	\varpi_\pd = \frac{1}{\hbar}\sqrt{8 E_{\cpc,\pd} E'_{\ind,\pd}}.
\end{equation}
Note that $\textstyle \varpi_\pd$ depends on the external flux $\textstyle \Phib$.
Parameters $\textstyle \varXi_3$ and $\textstyle \varXi_4$ in Eq.~\eqref{eq:qham1} are defined as
\begin{equation}
	\varXi_3 = \frac{E_\JJ}{6\hbar} \varphi_{\zpf,\pd}^3 \sin\phim_{\pd}, \quad
	\varXi_4 = \frac{E_\JJ}{24\hbar} \varphi_{\zpf,\pd}^4 \cos\phim_{\pd}.
\end{equation}
The last term in Eq.~\eqref{eq:qham1} describes the coupling between the resonator mode and the JPM, where
\begin{equation} \label{eq:def_g}
	g = \frac{E_{\cpc,\cpl}}{\hbar} \, n_{\zpf, \res} n_{\zpf, \pd}
\end{equation}
is the strength of this coupling.

\begin{figure}[t!]
	\centering
	\includegraphics{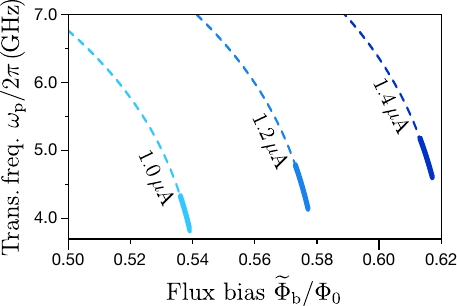}
	\caption{Dependence of the transition frequency $\textstyle \omega_\pd$ of the uncoupled ($\textstyle C_\cpl = 0$) JPM  on the external flux $\textstyle \Phib$ for various values of the JJ critical current $\textstyle I_0$ (indicated near the corresponding lines).
		The rest of the JPM parameters used for computations are $\textstyle \LG = 0.5\,\rm{nH}$ and $C_\pd = 1.0\,\rm{pF}$.
		Thick solid parts of lines indicate that the shallow well of the JPM potential hosts only two levels, while the thin dashed parts correspond to more than two levels existing in the shallow well.
		\label{fig:fig_2}
	}
\end{figure}

The JPM Hamiltonian in Eq.~\eqref{eq:qham1} can be approximately diagonalized using the Schrieffer-Wolff (SW) unitary transformation \cite{bravyi2011, klimov2000}:
\begin{equation}
	\opH_\pd \rightarrow \ee^{\lambda\opS} \opH_\pd \ee^{-\lambda\opS},
\end{equation}
where $\textstyle \lambda = \varXi_3/\varpi_\pd$ and the anti-Hermitian operator $\textstyle \opS$ is given by
\begin{equation} \label{eq:unit}
	\opS = \frac{1}{3}(\hat{b}^{\dag 3} - \hat{b}^3) + 3(\hat{b}^{\dag 2} \hat{b} - \hat{b}^\dag \hat{b}^2) + 3(\hat{b}^\dag - \hat{b}).
\end{equation}
For the typical circuit parameters we work with, one has $\textstyle \lambda^2 \ll 1$.
Applying the SW transformation to the Hamiltonian in Eq.~\eqref{eq:qham1}, keeping the terms only up to the first order in $\textstyle \lambda$, and making the rotating-wave approximation by dropping the fast-oscillating terms, one arrives at the effective circuit Hamiltonian as follows
\begin{equation} \label{eq:qham2}
	\begin{split}
		\frac{\opH}{\hbar} = & \, \omega_\res \hat{a}^\dag \hat{a} + \omega_\pd \hat{b}^\dag \hat{b} - \frac{\xi}{2} \hat{b}^{\dag 2} \hat{b}^2 \\
		& \, + g (\hat{b}^\dag \hat{a} + \hat{a}^\dag \hat{b}) - g_2 (\hat{b}^{\dag 2} \hat{a} + \hat{a}^\dag \hat{b}^2),
	\end{split}
\end{equation}
where $\textstyle \omega_\pd = \varpi_\pd - \xi$ is the JPM transition frequency and $\textstyle \xi = 60\lambda \varXi_3 + 12 \varXi_4$ is the anharmonicity of the JPM levels.
Figure~\ref{fig:fig_2} demonstrates the dependence of the JPM transition frequency $\textstyle \omega_\pd$ on the bias flux $\textstyle \Phib$ and the JJ critical current $\textstyle I_0$.
The terms in the bottom line in Eq.~\eqref{eq:qham2} describe couplings between the resonator mode and the JPM.
The first term in the bottom line describes the coupling between the resonator mode and the JPM giving rise to the coherent single-excitation exchange between them.
The second term in the bottom line in Eq.~\eqref{eq:qham2} describes the resonator mode-JPM coupling with strength $\textstyle g_2 = 4\lambda g$, which leads to conversion of a single excitation (photon) in the resonator into a pair of excitations in the JPM.
This nondipolar interaction arises due to the asymmetry of the shallow well of the JPM potential resulting in that the transitions between the JPM states with the same parity of their wavefunction become allowed.

On the photon capture stage, for providing the fastest excitation exchange, the JPM frequency is tuned close to the resonator frequency $\textstyle |\Delta_\pd| \ll g$, where $\textstyle \Delta_\pd = \omega_\pd - \omega_\res$ is the detuning between the frequencies of the JPM and the resonator.
Besides, for the circuit parameters we work with, one has $\textstyle \xi \gg g$.
Under these conditions, the JPM acts effectively as a two-level emitter (2LE) even when the shallow well accommodates more than two levels.
In this case, the nondipolar interaction described by the last term in Eq.~\eqref{eq:qham2} is strongly suppressed and this term can be neglected.
Thus, the Hamiltonian \eqref{eq:qham2} reduces to the standard Hamiltonian of the Jaynes-Cummings model describing a mode of a resonator coupled to a single 2LE \cite{jc1963, shore1993}:
\begin{equation} \label{eq:qham3}
	\frac{\opH}{\hbar} = \omegar \hat{a}^\dag \hat{a} + (\omegar + \Delta_\pd) \hat{\sigma}_{11} + g (\hat{\sigma}_{10} \hat{a} + \hat{a}^\dag \hat{\sigma}_{01}).
\end{equation}
Here we introduced the 2LE operator $\textstyle \hat{\sigma}_{jj'} \equiv |j\rangle\langle j'|$, where $\textstyle j,j' \in \{0,1\}$, obeying the spin-$1/2$ commutator algebra $\textstyle \left[\hat\sigma_{jj'},\hat\sigma_{ll'}\right] = \hat\sigma_{jl'}\delta_{j'l} - \hat\sigma_{lj'}\delta_{l'j}$.

To proceed, it is convenient to move to a rotating frame via a transformation 
\begin{equation}
 \opH \rightarrow \opR(t) \opH \opR^\dag(t) + \ii \hbar \frac{\partial \opR(t)}{\partial t} \, \opR^\dag(t),
\end{equation}
where $\textstyle \opR(t) = \exp(-\ii \omega_\res t \, \opN)$ with $\textstyle \opN = \hat a^\dag \hat a + \hat \sigma_{11}$ being the operator of the total number of excitations in the system.
In the rotating frame, the Hamiltonian of the resonator mode-JPM system reads as
\begin{equation} \label{eq:ham_cpt}
	\frac{\opH}{\hbar} = \Delta_\pd \hat\sigma_{11} + g (\hat a^\dag \hat \sigma_{01} + \hat \sigma_{10} \hat a).
\end{equation}
This is the Hamiltonian we use in what follows for describing the coupled resonator mode--JPM system on the photon capture stage. 

Contrary to the resonant regime of the interaction between the JPM and the resonator mode on the capture stage, on the JPM readout stage the potential is tilted to a slightly asymmetric configuration \cite{opremcak2018} and the JPM transition $\textstyle \bar{\omega}_\pd$ frequency becomes strongly detuned from the resonator frequency $\textstyle |\bar{\Delta}_\pd| \gg g$, where $\textstyle \bar{\Delta}_\pd = \bar{\omega}_\pd - \omega_\res$.
Thus, the excitation exchange between the resonator and the JPM is essentially suppressed on the readout stage.
The same holds also for the reset stage.
Thus, on these stages, we can treat the resonator mode and the JPM as decoupled subsystems.

\subsection{Master equation} \label{sec:meq}

We describe the evolution of the system state $\textstyle \hat\rho(t)$ by the master equation \cite{breuer2002book}
\begin{equation} \label{eq:me}
	\partial_t \hat{\rho}(t) = \hat{\mathcal{L}}(t) \hat\rho(t).
\end{equation}
Here, assuming that the switching between the regimes of photon capture and the JPM readout occurs much faster than the durations of these regimes, we represent the time-dependent Liouvillian superoperator $\textstyle \hat{\mathcal{L}}(t)$ as
\begin{equation} \label{eq:lvln1}
	\begin{split}
		\hat{\mathcal{L}}(t) = & \, \left[\Theta(t) - \Theta(t-t_\mathrm{cpt})\right]\hat{\mathcal{L}}_\mathrm{cpt} \\
		& \, + \left[\Theta(t-t_\mathrm{cpt}) - \Theta(t-t_\mathrm{msr})\right] \hat{\mathcal{L}}_\mathrm{rdt},
	\end{split}
\end{equation}
where $\textstyle t_\mathrm{msr} = t_\mathrm{cpt} + t_\mathrm{rdt} + t_\mathrm{rst}$ is the total duration of a single photodetection cycle comprised by the times of the photon capture $\textstyle t_\mathrm{cpt}$, readout $\textstyle t_\mathrm{rdt}$, and reset $\textstyle t_\mathrm{rst}$; $\textstyle \Theta(t)$ denotes the Heaviside step-function.
The first term in the above expression describes the evolution of the system during the photon capture stage, while the second term corresponds to the JPM readout and reset stages.

The Liouvillian $\textstyle \hat{\mathcal{L}}_\mathrm{cpt}$ has the Lindblad form \cite{manzano2020} and reads, cf. Refs.~\cite{govia2012, sokolov2020},
\begin{equation} \label{eq:lvln2}
	\begin{split}
		\hat{\mathcal{L}}_\mathrm{cpt}\hat\rho = & \,  - \frac{\ii}{\hbar}\left[\opH, \hat\rho\right] + \sum_{j=0,1 }\gamma_j \opD\left[|\mathrm{c}\rangle\langle j|\right]\hat\rho \\ & \, + \Gamma_{10}\opD[\hat \sigma_{01}]\hat\rho + \Gamma_{11} \opD[\hat \sigma_{11}]\hat\rho + \kappa \opD[\hat a]\hat\rho,
	\end{split}
\end{equation}
where $\textstyle \opD[\hat\bullet] \hat{\rho} = \hat\bullet \hat{\rho} \hat\bullet^\dag - (\hat\bullet^\dag \hat\bullet \hat{\rho} + \hat{\rho} \hat\bullet^\dag \hat \bullet)/2$.
The first term in Eq.~\eqref{eq:lvln2} describes the unitary quantum evolution during the photon capture stage described by the Hamiltonian $\textstyle \opH$ given by Eq.~\eqref{eq:ham_cpt}.
The second term describes the interwell tunneling of the JPM metastable states $\textstyle |0\rangle$ and $\textstyle |1\rangle$ from the shallow well to the deep well.
Here, similarly to Ref.~\cite{govia2012}, we replaced the cascade of multiple states in the deep well with a single state $\textstyle |c\rangle$ into which the metastable states $\textstyle |0\rangle$ and $\textstyle |1\rangle$ can irreversibly transit to.
We assume that the relaxation of the top levels in the deep well occurs much faster than the interwell tunneling.
That implies that once the particle tunneled from the shallow well, it does not tunnel back but rapidly dissipates its energy, eventually falling to the bottom level of the deep potential well.
Such a simplified description significantly reduces the Hilbert space of the JPM states, abating the numerical complexity of the problem while retaining the essential general features of quantum evolution of the considered system.
The third and fourth terms in Eq.~\eqref{eq:lvln2} describe the processes of the JPM relaxation with rate $\textstyle \Gamma_{10}$ and the pure dephasing with rate $\textstyle \Gamma_{11}$, respectively.
The last term corresponds to the resonator mode relaxation with rate $\textstyle \kappa$.

\begin{figure}[t!]
	\includegraphics{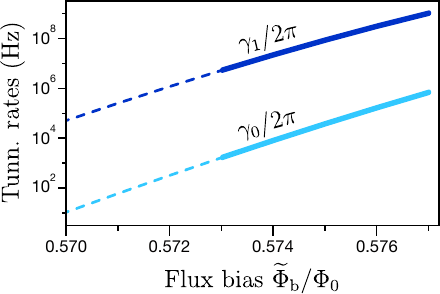}
	\caption{Dependence of the tunneling rates $\textstyle \gamma_1$ and $\textstyle \gamma_0$ on the external flux $\textstyle \Phib$ threading via the JPM.
		Thick solid and thin dashed parts of lines have the same meaning as in Fig.~\ref{fig:fig_2}.
		Computations are performed for $\textstyle I_0 = 1.2\,\rm{\mu A}$, while the rest of the circuit parameters are the same as in Fig.~\ref{fig:fig_2}. \label{fig:fig_3}}
\end{figure}

The tunneling rates of the metastable levels $\textstyle |0\rangle$ and $\textstyle |1\rangle$ are evaluated using the WKB approximation \cite{strauch2004}
\begin{equation} \label{eq:tunn_rate}
	\begin{split}
		\gamma_j = & \, \frac{\Omega_j}{j! \sqrt{2\pi}} \left(\frac{j + 1/2}{\ee}\right)^{j+1/2} \\
		& \, \times \exp\left(-\frac{2}{\hbar} \int^{\phi_2}_{\phi_1} \dd\phi \, \sqrt{2C'_\pd[E_j - U_\pd(\phi)]}\right),
	\end{split}
\end{equation}
where $\textstyle \Omega_j = E_j/\hbar$ with $\textstyle E_j$ being the eigenenergy of the $j$-th level in the shallow well.
Figure~\ref{fig:fig_3} shows the dependence of the tunneling rates of the levels $|0\rangle$ and $|1\rangle$ on the external flux $\textstyle \Phib$.
Calculations demonstrate that $\gamma_1/\gamma_0 \gtrsim 10^3$ for the typical JPM parameters we use.
A similar estimate was obtained, e.g., in Ref.~\cite{shnyrkov2023}.

As we mentioned earlier, on the readout and reset stages, the JPM transitions are effectively decoupled from the resonator mode due to strong detuning.
Since the JPM dynamics on these stages does not affect the resonator field and the detailed description of the JPM dynamics is not required, we can take partial trace with respect to the JPM states and consider the evolution of the resonator mode $\textstyle \hat\rho_\res(t)$ alone, where $\textstyle t_\mathrm{cpt} < t < t_\mathrm{msr}$.
Here $\textstyle \hat\rho_\res(t) \equiv \Tr_\pd\left[\hat\rho(t)\right]$, where the subscript $\mathrm{p}$ of the trace operator implies that only the JPM states are traced out.
The master equation governing the evolution of the state of the resonator mode reads as
\begin{equation} \label{eq:me_res}
    \partial_t \hat{\rho}_\res(t) = \hat{\mathcal{L}}_\mathrm{rdt} \hat\rho_\res(t), \quad t_\mathrm{cpt} < t < t_\mathrm{msr},
\end{equation}
where the Lindbladian
\begin{equation} \label{eq:lvln3}
  \hat{\mathcal{L}}_\mathrm{rdt}\hat\rho_\res = - \frac{\ii}{\hbar}\left[\opH_\res, \hat\rho_\res \right] + \kappa \opD [\hat a] \hat\rho_\res
\end{equation}
describes the free evolution and the relaxation of the resonator mode.

The Lindblad master equation given by Eqs.~\eqref{eq:me}--\eqref{eq:lvln2}, \eqref{eq:me_res}, and \eqref{eq:lvln3} is solved numerically using \textsf{QuantumOptics.jl} -- Julia library offering versatile toolkit for simulating various quantum optics problems \cite{kramer2018}.
For simulations, we assume that at the beginning of the photodetection (we set this moment as $\textstyle t=0$) the state of the system $\textstyle \hat\rho(0)$ is separable with the JPM residing in its lowest metastable state $|0\rangle$, i.e., $\textstyle \hat\rho(0) = \hat\varrho \otimes|0\rangle\langle 0|$ with $\textstyle \hat\varrho$ being the initial state of the resonator mode.

In our analysis, we neglect the effects of the thermal excitations on the system evolution.
The working temperatures of the circuit QED setups are usually lie in the range $\textstyle T_\mathrm{sys} \sim 10 \textrm{---} 30\,\rm{m K}$.
For the typical operational frequencies $\textstyle \omega_\mathrm{sys}/(2\pi) \sim 4 \textrm{---} 10\,\mathrm{GHz}$, assuming the Bose-Einstein distribution of the thermal excitations, one obtains the upper estimate for its average number $\textstyle n_\mathrm{th} \lesssim 10^{-2}$ which justifies our approximation.

\section{Photocounting statistics} \label{sec:stat}

Having developed the model describing the resonator mode--JPM system, we proceed to considering the process of counting photons in the resonator mode.
Here we aim to determine the statistics of the photocounts obtained as a result of the repeated measurements of the resonator mode by the single JPM.
As mentioned in introduction, the photocounting statistics is a prominent example of quantum measurements.
Its outcome probabilities are described by Born's rule \cite{born1926}.
The photocounting distribution for the resonator mode in the state $\textstyle \hat\varrho$ is given by
    \begin{equation} \label{eq:pn}
    	P(k) = \Tr\{\hat\varrho \, \hat\Pi_k\},
    \end{equation}
where $\textstyle \hat\Pi_k$ is the POVM element corresponding to the outcome with $\textstyle k$ clicks.
Thus, as it follows from Eq.~\eqref{eq:pn}, also known as the photocounting formula, calculation of photocounting statistics requires knowledge of the POVM.

The POVM of the JPM-based photocounter we study depends on the counting technique.
Here we consider two approaches to counting resonator photons.
The first approach is to count photons by performing a fixed number of measurement (photodetection) iterations, $\textstyle M$, regardless of the obtained results.
In the second approach, the measurement is iterated until either the first no-click event occurs or the maximum number of iterations, $\textstyle M$, is reached.
In both cases, the measurement outcome $k$ is given by the total number of clicks. 

The POVM is convenient to consider in the Fock-state basis, 
    \begin{align} \label{eq:Wkn}
        P(k|n)= \vphantom{\langle}_\res\langle n|\hat \Pi_k|n\rangle_\res.
    \end{align}
Here $P(k|n)$ are diagonal matrix elements of the POVM elements, defining the probability of obtaining $\textstyle k$ clicks given the $\textstyle n$-photon Fock state.
Importantly, since photocounting is a phase-insensitive measurement, the nondiagonal matrix elements of the POVM vanish, $\textstyle \vphantom{\langle}_\res\langle n|\hat\Pi_k|m\rangle_{\res} = 0$ for $\textstyle n \neq m$.
On the other hand, the spectral decomposition theorem yields
    \begin{align} \label{eq:Pik}
        \hat{\Pi}_k=\sum\limits_{n=0}^{+\infty} P(k|n)\left|n\right\rangle_{\res}\!\left\langle n \right|.
    \end{align}
This implies that the POVM in Eq.~(\ref{eq:pn}) is uniquely defined by its diagonal matrix elements.

\subsection{Statistics of subsequent events} \label{sec:subs}

Our initial objective is to determine the probability of obtaining a series of measurement outcomes $\mathcal{I}_M = \{i_M,\ldots,i_1\}$ given an $n$-photon Fock state.
Each event within this sequence, $l=1,\ldots,M$, can have a binary result: $i_l=0$ and $i_l=1$, corresponding to ``no-click'' and ``click'', respectively. 
The entire measurement procedure could be explained by Bernoulli's process, if all events were independent and described by the same binary probability distribution.
In such a scenario, the two measurement techniques mentioned in the preamble of this section would be described by binomial and geometric distributions, respectively.  
However, in our case, every event modifies the state of the resonator mode and, consequently, their probabilities are not independent.

Let us define the probability $\textstyle W(m, i|n)$ to get $m$ photons in the resonator mode and the measurement result $i=0,1$ given $n$ photons before the iteration event.
In the sequence of measurements, the result of the subsequent event depends solely on the number of photons that remain after the previous measurement event, and not on any prior states of the resonator mode.
This indicates a typical Markov process.
Thus, we can calculate the joint probabilities to obtain the outcome sequence $\mathcal{I}$ and the sequence of remaining photon numbers $\{m_M,\ldots,m_1\}$.
Averaging this probability with respect to the latter results in an expression for the probability of obtaining a series of measurement outcomes, $\mathcal{I}_M$, from the $n$-photon Fock state at the beginning,
\begin{equation} \label{eq:prob_M}
    P(\mathcal{I}_M|n) = \sum_{m_M=0}^{+\infty} \cdots \sum_{m_1=0}^{+\infty}\prod_{l=1}^M W(m_l, i_l|m_{l-1}),
\end{equation}
where we set $m_0=n$. 

Using Bayes' theorem, we can decompose the probabilities in the right-hand side of Eq.~(\ref{eq:prob_M}),
    \begin{equation}\label{eq:prob_decomp}
        W(m_l, i_l|m_{l-1}) = W(i_l|m_{l-1}) W(m_l|i_l, m_{l-1}).
    \end{equation}
Here $W(i_l|m_{l-1})$ is the probability to get the measurement result $i_l$ given $m_{l-1}$ photons before the iteration event and $W(m_l|i_l, m_{l-1})$ is the probability that the state of the resonator mode contains $m_l$ photons given the measurement result $i_{l}$ and also $m_{l-1}$ photons before the iteration event. 
Both these probabilities can be calculated with the model considered above.

\subsection{Positive operator-valued measures} \label{sec:povm}

The probabilities $P(\mathcal{I}_M|n)$ enable us to obtain the POVM in the Fock basis.
The first technique, which involves counting the total number of click events in the sequence $\mathcal{I}$, yields a probability distribution of clicks that can be considered a generalization of the binomial distribution.
In this scenario, the POVM in the Fock basis is given by
    \begin{equation}
    	P_{\mathrm{b}}(k|n) = \sum_{\mathcal{I}_M \in \boldsymbol{\mathcal{I}}_k} P(\mathcal{I}_M|n).
    \end{equation}
Here 
    \begin{align}
        \boldsymbol{\mathcal{I}}_k = \left\{\mathcal{I}_M\left|\sum\limits_{l=1}^M i_l=k\right.\right\}.
    \end{align}
is the set of all measurement sequences with $k$ click events.
We refer to this counting technique as generalized binomial.

The second technique involves counting clicks until the first no-click event or reaching the end of the sequence.
This yields the probability distributions of clicks, which is a generalization of the geometric distribution.
Thus, the POVM in the Fock representation for this scenario reads
    \begin{equation}
        P_{\mathrm{g}}(k|n) =P(\mathcal{I}_k^{(1)}|n).
    \end{equation}
Here
    \begin{equation}
       \mathcal{I}_k^{(1)}=\left\{
       \begin{array}{ll}
        \{i_{k+1}=0,i_k=1,\ldots,i_1=1\},& k<M\\
        \{i_M=1,\ldots,i_1=1\},& k=M
       \end{array}
       \right. 
    \end{equation}
is the sequence of measurement outcomes with $k$ consecutive click events. 
This counting technique will be referred to as generalized geometric.

\subsection{Click probability} \label{sec:click}

As mentioned, the probabilities $W(i|n)$ in Eq.~(\ref{eq:prob_decomp}) can be calculated with the model considered above.
In this context, we first note that $W(0|n)=1-W(1|n)$.
Hence, it is sufficient to determine only the probability $W(1|n)$ to get a click given $n$ photons.

The probability $\textstyle W(1|n)$ of the JPM delivering a click is given by
    \begin{equation} \label{eq:pcl_n}
        W(1|n) = \beta \PP_c(n; t_\mathrm{cpt}).
    \end{equation}
Here $\textstyle \PP_c(n; t) = \Tr\left[\hat\rho(n;t)|c\rangle\langle c|\right]$ stands for the population of the state $\textstyle |c\rangle$ at an instant $\textstyle t$ for the resonator mode initially in the $n$-photon Fock state.
Here we introduced a notation $\textstyle \hat\rho(n; t)$ for the system density operator at a moment of time $t$ given that at the beginning of the photon capture stage the resonator mode is in an $n$-photon Fock state $\textstyle \hat\varrho = |n\rangle_\res\langle n|$.
The phenomenological parameter $\textstyle 0\leq\beta\leq 1$ introduced in Eq.~\eqref{eq:pcl_n} models the efficiency of the JPM readout procedure outlined in Sec.~\ref{sec:jpm}.
As we mentioned therein, the efficiency of the reflectometry-based JPM readout can exceed $99.9\%$.

\subsection{Conditional photon-number distribution} \label{sec:back}

Next, we determine the probability $W(m|i, n)$ to get $m$ photons given the measurement result $i$ and $n$ photons before the measurement iteration.
For this purpose, we need to determine the postmeasurement state $\textstyle \hat\varrho(i, n)$ of the resonator mode, i.e., the state after obtaining the measurement result $i$ under the Fock state with $n$ photons.
This yields the probability to find $m$ photons in the resonator mode given $i$ and $n$,
\begin{equation}
    W(m|i,n) = \vphantom{\langle}_\res \langle m|\hat\varrho(i,n)|m\rangle_\res.
\end{equation}
Here partial averaging is taken over the Fock state of the resonator mode, $|m\rangle_\res$. 

The resonator mode postmeasurement state $\textstyle \hat\varrho(i, n)$ is determined as \cite{govia2012}
\begin{equation} \label{eq:res_post}
	\hat\varrho(i, n) = \frac{\Tr_\pd[\hat\rho(n; t_\mathrm{msr}) \, \hat\pi_i]}{W(i|n)},
\end{equation}
where $\textstyle \hat\rho(n; t_\mathrm{msr})$ is the state of the system at the end of the measurement sequence given the initial $\textstyle n$-photon Fock state of the resonator mode.
As above, the subscript $\textrm{p}$ of the trace operator indicates that the trace is taken over the JPM states only.
In Eq.~\eqref{eq:res_post}$, \textstyle \hat\pi_i$ stands for operators in the space of the JPM states,
\begin{equation}
        \hat\pi_1 = \beta |c\rangle\langle c|, \quad \hat\pi_0 = \hat{\mathbb{1}} - \hat\pi_1,
\end{equation}
where $i=0,1$ denotes the measurement result.
In this case, the unity operator reads as $\textstyle \hat{\mathbb{1}} = |0\rangle\langle 0| + |1\rangle\langle 1| + |c\rangle\langle c|$.

\begin{figure*}[t!]
    \centering
    \includegraphics{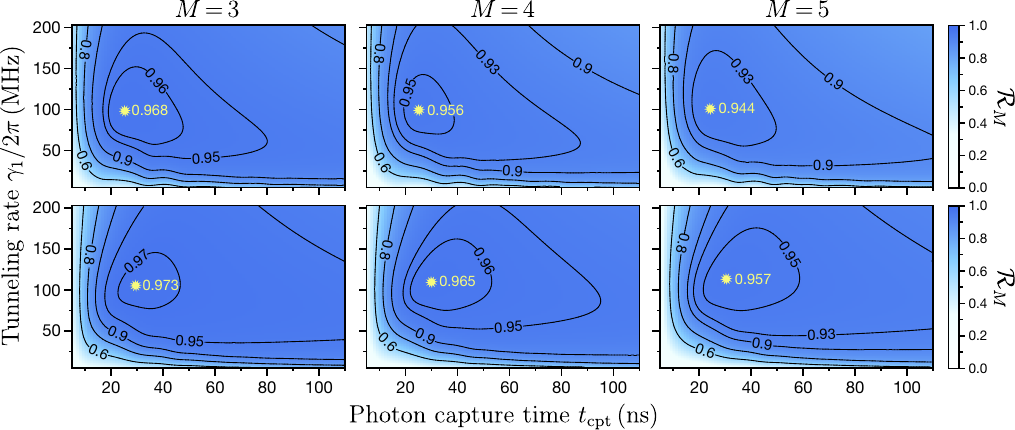}
    \caption{Dependence of the photocounter resolution $\textstyle \mathcal{R}_M$ (for $\textstyle M = 3, 4, 5$) on the photon capture time $\textstyle t_\mathrm{cpt}$ and the JPM upper level tunneling rate $\textstyle \gamma_1$ for the generalized binomial (upper row) and for the generalized geometric (lower row) counting technique.
    Parameters of the system used for computations are $\textstyle g/2\pi = 30\,\mathrm{MHz}$, $\textstyle \Gamma_{10}/2\pi = 1\,\mathrm{MHz}$, $\textstyle \Gamma_{11} = 5\Gamma_{10}$, and $\textstyle \kappa/2\pi = 1\,\mathrm{kHz}$.
    For computations, we assume the resonant regime of the resonator mode-JPM coupling $\textstyle \Delta = 0$ and take $\textstyle t_\mathrm{rdt} + t_\mathrm{rst} = 300\,\mathrm{ns}$.
    \label{fig:fig_4}}
\end{figure*}

\section{Photon-number resolution} \label{sec:rsl}

As mentioned above, our goal is to tune the parameters of the JPM detectors to make such counters as close as possible to the ideal PNR detectors.
Such an optimization requires a quantity (measure) that characterizes the ability of detectors to resolve between adjacent numbers of photons.
As discussed in Ref.~\cite{Len2022}, the probability $\textstyle P(n|n)$ of obtaining $\textstyle n$ clicks given $n$ photons can be considered as a relevant characteristic of the photon-number resolution for a fixed photon number $n$.

We generalize this characteristic to define a measure of the photon-number resolution between $M$ photons.
Let us assume that (i) the resonator mode does not contain more than $M$ photons and (ii) our \textit{a priori} information about the photon number is minimal for $n=0,\ldots,M$.
In other words, this means that the prior photon-number distribution is uniform for these $M+1$ values.
Therefore, the quantity
    \begin{equation}
    	\mathcal{R}_M = \frac{1}{M+1} \sum_{n=0}^M P(n|n).
    \end{equation}
can be considered as the probability to get the correct value for the number of photons in the aforementioned domain, given the minimal \textit{a priori} information and assuming that the state of the resonator mode is restricted by $M$ photons.
Obviously, this quantity can be considered as a suitable measure of the photon-number resolution, which we will refer to as the photocounter resolution.
This measure takes values in the interval $\mathcal{R}_M\in[0,1]$.
For $\mathcal{R}_M=1$ we get the ideal resolution between $M$ photons, for $\mathcal{R}_M=0$ it vanishes at all.
Note that even for PNR detectors with losses this measure is less than one.

Figure~\ref{fig:fig_4} demonstrates the dependence of the photocounter resolution $\textstyle \mathcal{R}_M$ (for $\textstyle M = 3, 4, 5$) on the photon capture time $\textstyle t_\mathrm{cpt}$ and the tunneling rate of the JPM excited state $\textstyle \gamma_1$ for the different resonator mode--JPM couplings $\textstyle g$.
Computations reveal that for the given value of the coupling $\textstyle g$ and the JPM relaxation rate $\textstyle \Gamma_{10}$ there is a combination of $\textstyle \gamma_1$ and $\textstyle t_\mathrm{cpt}$ for which the resolution attains its maximal value $\textstyle \mathcal{R}^\mathrm{max}_M$.

This feature of the resolution emerges due to an interplay between different competing processes occurring in the considered system.
The obtained results show that the resolution $\textstyle \mathcal{R}_M$ rapidly grows with the increase of $\textstyle \gamma_1/g$ attaining its global maximum for $\textstyle \gamma_1/g \sim 3\ldots 4$ and then starts to slowly diminish with the further increase of $\textstyle \gamma_1/g$.
Similar interplay between the coherent excitation exchange and the relaxation is characteristic for noise-assisted transport \cite{plenio2008, rebentrost2009}.
It was also observed in the experimental study of simplified models of light-harvesting complexes in circuit QED \cite{potocnik2018}.

\begin{figure*}[t!]
        \centering
	\includegraphics{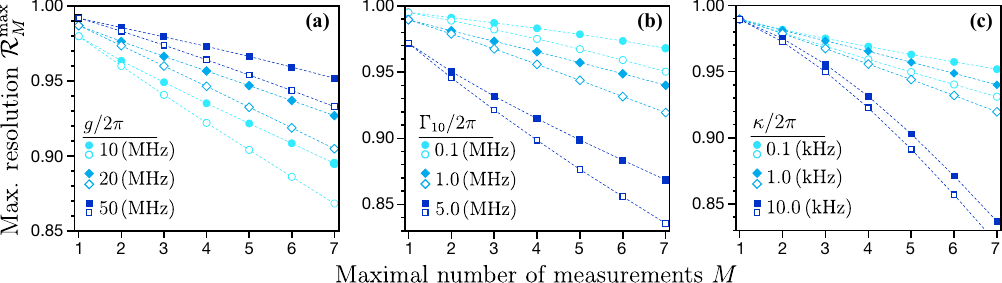}
	\caption{Dependence of the maximal photocounter resolution $\textstyle \mathcal{R}^\mathrm{max}_M$ on $M$ for (a) various coupling strengths $\textstyle g$ for $\textstyle \Gamma_{10}/2\pi = 1.0\,\mathrm{MHz}$ and $\textstyle \kappa/2\pi = 1.0\,\mathrm{kHz}$, (b) various relaxation rates $\textstyle \Gamma_{10}$ for $\textstyle g/2\pi = 30\,\mathrm{MHz}$ and $\textstyle \kappa/2\pi = 1.0\,\mathrm{kHz}$, and (c) various resonator loss rates $\textstyle \kappa$ for $\textstyle g/2\pi = 30\,\mathrm{MHz}$ and $\textstyle \Gamma_{10}/2\pi = 1.0\,\mathrm{MHz}$.
	The rest of the system parameters are the same as in Fig.~\ref{fig:fig_4}.
	Empty (filled) markers correspond to the generalized binomial (geometric) counting techniques.
	Thin dashed lines are added for the better perception.
	\label{fig:fig_5}}
\end{figure*}

The existence of the optimal capture time $\textstyle t^\mathrm{opt}_\mathrm{cpt}$ agrees with the earlier results of Ref.~\cite{poudel2012}.
On the one hand, the prolongation of the photon capture stage increases the
probability of the resonator photon to be absorbed by the JPM and induce a click.
On the other hand, the longer capture time increases the probability of the false click due to tunneling from the lower metastable level $\textstyle |0\rangle$ in the case of the vacuum-state input, which reduces the probability $\textstyle P(n|n)$.

Figure~\ref{fig:fig_5} shows the effect of the system parameters on the dependence of the maximal resolution $\textstyle \mathcal{R}^\mathrm{max}_M$ on $\textstyle M$.
Computations demonstrate that $\textstyle \mathcal{R}^\mathrm{max}_M$ decreases with the increase of $M$ for both generalized geometric and binomial photocounting techniques.
The increase of the coupling between the resonator mode and the JPM leads to the increase of the maximal resolution $\textstyle \mathcal{R}^\mathrm{max}_M$ as illustrated in Fig.~\ref{fig:fig_5}(a).
As expected, stronger JPM and resonator mode relaxation reduces the resolution which is shown in Figs.~\ref{fig:fig_5}(b) and \ref{fig:fig_5}(c).
Stronger losses deteriorate the resolution with more pronounced impact for larger $M$.

It is worth noting that the setup parameters we use for computations are accessible for the current or near-term circuit QED technologies.
Frequencies of the resonator mode and JPM transition as well as the coupling parameters we work with are typical for the circuit QED setups \cite{gu2017, *blais2021rmp}.
The internal Q-factors of the CPW resonators can exceed $\textstyle 10^6$ \cite{megr2012, bruno2015} which corresponds to the photon loss rates $\kappa/2\pi \lesssim 10\,\mathrm{kHz}$.
Comparable Q-factors $\textstyle \sim 10^6$ were achieved in the lumped-element resonators \cite{shi2022}.
Even higher internal Q-factors up to $\textstyle 10^8$ can be attained in the coaxial resonators \cite{reagor2016, heidler2021} and the 3D microwave cavities can demonstrate Q-factors exceeding $10^{10}$ \cite{paik2011, reagor2013, romanenko2020, milul2023}.
The experimentally reported relaxation rate of the flux-biased JPM is $\textstyle \Gamma_{10}/2\pi \sim 15\,\mathrm{MHz}$ \cite{opremcak2018}.
However, we anticipate that by using different materials and fabrication techniques the JPM relaxation will be reduced in the near-term devices. 

The obtained results show that for the same coupling $\textstyle g$ and the JPM relaxation $\textstyle \Gamma_{10}$, the generalized geometric counting technique provides better resolution than the generalized binomial one.
We attribute this behavior to the effect of the false clicks arising due to superfluous tunneling from the lower JPM level.
In the case of the generalized binomial counting technique, we perform the fixed number of measurement iterations which results in the higher probability (for the same photon capture time $\textstyle t_\mathrm{cpt}$) of obtaining a false click compared with the generalized geometric technique when we stop counting if one obtains no click.
For reducing the probability of obtaining a false click, one can shorten the photon capture time.
That, however, results in the reduction of the probability of absorbing a photon.
As a result of the trade off between these competing effects, for the generalized binomial counting technique, the optimal photon capture time $\textstyle t_\mathrm{cpt}$ is shorter and the maximal resolution is lower than those for the generalized geometric technique.
These qualitative considerations agree with the numerical results presented in Fig.~\ref{fig:fig_4}, demonstrating that the optimal photon capture time $\textstyle t_\mathrm{cpt}$ is indeed shorter for the generalized binomial counting technique.

\section{Examples of photocounting statistics} \label{sec:exmpl}

\begin{figure}[t!]
    \centering
    \includegraphics{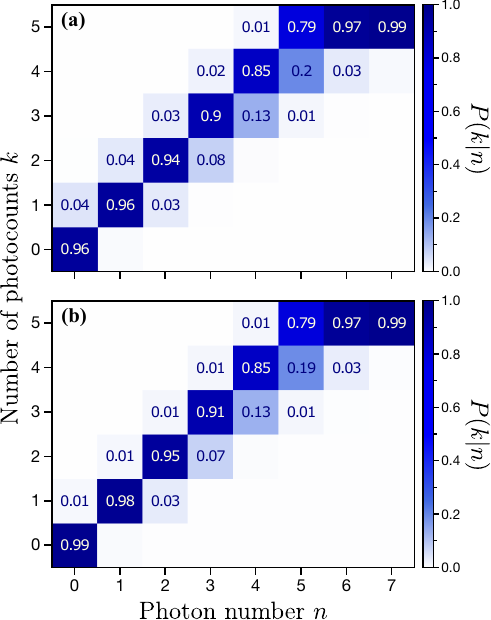}
    \caption{The POVM for the case of (a) the generalized binomial and (b) the generalized geometric counting techniques for the maximal number of iterations $M=5$.
    Parameters of the system used for computations are:
    (a) $\textstyle \gamma_1/2\pi = 162.3\,\mathrm{MHz}$, and $\textstyle t_\mathrm{cpt} = 17.4\,\mathrm{ns}$; (b) $\textstyle \gamma_1/2\pi = 173.6\,\mathrm{MHz}$, and $\textstyle t_\mathrm{cpt} = 19.2\,\mathrm{ns}$.
    The values of the tunneling rate $\textstyle \gamma_1$ and the capture time $\textstyle t_\mathrm{cpt}$ are taken to maximize the photodetector resolution $\textstyle \mathcal{R}_M$.
    The rest of the parameters are as follows: $\textstyle g/2\pi = 30\,\mathrm{MHz}$, $\textstyle \Gamma_{10}/2\pi = 1 \,\mathrm{MHz}$, $\textstyle \Gamma_{11} = 5\Gamma_{10}$, and $\textstyle \kappa/2\pi = 10\,\mathrm{kHz}$.
    As earlier, we assume the resonant coupling regime $\textstyle \Delta = 0$ and set $\textstyle t_\mathrm{rdt} + t_\mathrm{rst} = 300\,\mathrm{ns}$.
    \label{fig:fig_6}}
\end{figure}

\begin{figure}[t!]
    \centering
    \includegraphics{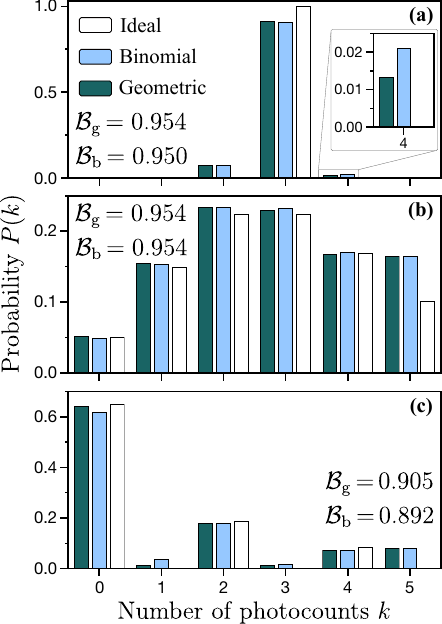}
    \caption{Photocounting statistics obtained for the resonator mode prepared in (a) the Fock state $\textstyle |n\rangle_\res$ with $n=3$, (b) the coherent state $\textstyle |\alpha\rangle_\res$ with the amplitude $\textstyle \alpha = \sqrt{3}$, and (c) the squeezed vacuum state $\textstyle |r\rangle_\res$ with the squeezing parameter $\textstyle r = 1$.
    The corresponding POVMs and the parameters of the system used for computations are given in Fig.~\ref{fig:fig_6} and its caption.
    \label{fig:fig_7}}
\end{figure}

In this section, we apply the description of the JPM-based photocounter developed in Sec.~\ref{sec:stat} to obtain photocounting statistics for the typical quantum states for which the protocols of preparation are already implemented in the circuit QED architecture.
The examples of the POVMs we use for evaluation of the photocounting statistics are shown in Fig.~\ref{fig:fig_6}(a) for the generalized binomial technique and Fig.~\ref{fig:fig_6}(b) for the generalized geometric technique.

As a measure of an overlap between the obtained photocounting statistics $\textstyle P(k)$  of the considered JPM-based photocounter and the counting statistics $\textstyle P_\mathrm{i}(k)$ of the ideal PNR detector, we use the Bhattacharyya coefficient $\textstyle 0 \leq \mathcal{B} \leq 1$ defined as
\begin{equation} \label{eq:bhatt}
    \mathcal{B} = \sum\limits_{k=0}^{M} \sqrt{P(k) P_\mathrm{i}(k)}.
\end{equation}
The counting statistics obtained by the ideal PNR detector is equivalent to the photon-number distribution of the initial state of the resonator mode $\textstyle P_\mathrm{i}(k) = \vphantom{\langle}_\res\langle k|\hat\varrho|k\rangle_\res$.
The coefficient $\textstyle \mathcal{B}$ is related to the distance between the probability distributions and shows how similar the obtained photocounting statistics of the JPM photocounter is to the genuine photon-number distribution.
The limiting case of $\mathcal{B} = 1$ implies that $\textstyle P(k) = P_\mathrm{i}(k)$, while $\textstyle \mathcal{B} = 0$ indicates no overlap between distributions $\textstyle P(k)$ and $P_\mathrm{i}(k)$.

The first example we consider is the initial Fock state of the resonator field $\textstyle |n\rangle_\res$.
The obtained photocounting statistics and the corresponding Bhattacharyya coefficients for $\textstyle n=3$ are shown in Fig.~\ref{fig:fig_7}(a).
The tunneling from the lower metastable level $\textstyle |0\rangle$ leads to false counts manifesting themselves as nonzero probability of obtaining more counts than there were initially photons in the resonator mode.
This contribution is lower for the geometric technique as demonstrated in the inset in Fig.~\ref{fig:fig_7}(a).
Resonator and JPM losses are the primary source of imperfect resolution of the Fock states in the considered photocounting setup.

Next, we consider the resonator mode prepared in the coherent state $\textstyle |\alpha\rangle_\res$ with the photon average number $\textstyle |\alpha|^2 = 3$.
The corresponding photocounting statistics are shown in Fig.~\ref{fig:fig_7}(b).
For the chosen amplitude $\textstyle \alpha$, both counting techniques give almost identical statistics.
Here, the discrepancy from the statistics of the ideal detector stems from the lack of resolution for the photon numbers $n \geq 5$.

\begin{figure}[t!]
    \centering
    \includegraphics{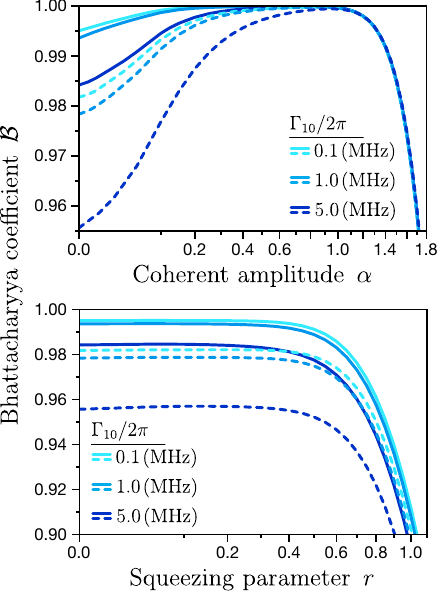}
    \caption{Dependence of the Bhattacharyya coefficient on (a) the amplitude $\textstyle \alpha$ for the initial coherent state $\textstyle |\alpha\rangle_\res$ of the resonator field and (b) the squeezing parameter $r$ in the case of the resonator field prepared in the squeezed vacuum state $\textstyle |r\rangle_\res$ for $\textstyle M=5$ and the different values of the JPM relaxation rate $\textstyle \Gamma_{10}$.
    Solid lines correspond to the generalized geometric counting technique while the dashed lines correspond to the generalized binomial technique.
    The parameters of the setup used for computations are $\textstyle g/2\pi = 30\,\mathrm{MHz}$, $\textstyle \Gamma_{11} = 5\Gamma_{10}$, and $\textstyle \kappa/2\pi = 1\,\mathrm{kHz}$.
    \label{fig:fig_8}}
\end{figure}

Finally, we study the case of the resonator mode prepared in the squeezed vacuum state
\begin{equation}\label{eq:sqzd}
    |r\rangle_\res = \frac{1}{\sqrt{\cosh r}} \sum_{k=0}^{+\infty} \binom{2n}{n}^{1/2} \left(\frac{\tanh r}{2}\right)^2 |2n\rangle_\res,
\end{equation}
where $\textstyle r$ is the squeezing parameter.
Figure~\ref{fig:fig_7}(c) shows the photocounting statistics for $\textstyle r=1$.
In contrast with the ideal photodetector, the JPM photocounter can deliver an odd number of clicks.
These erroneous outcomes arise mainly due to the JPM relaxation.
The generalized binomial counting technique delivers higher deviation from the statistics of the ideal detector than the generalized geometric technique.

Figure~\ref{fig:fig_8} shows the effect of the JPM relaxation on the dependence of the Bhattacharyya coefficient $\mathcal{B}$ on the coherent amplitude $\alpha$ for the case of the resonator mode prepared in the coherent state and the squeezing parameter $r$ for the resonator mode prepared in the squeezed vacuum state.
The results demonstrate that the photocounting statistics are closer to the statistics of the ideal PNR detector when the average number of photons (determined as $\textstyle |\alpha|^2$ for the coherent states and $\sinh^2 (r)$ for the squeezed vacuum) is much less than the number of measurement iterations $M$.
Degradation of $\textstyle \mathcal{B}$ for the weak coherent states ($|\alpha|^2 \ll 1$) is explained by the excess probability of getting no clicks due to JPM relaxation.
With the increase of the average photon number, the coefficient $\textstyle \mathcal{B}$ starts to rapidly decrease due to the lack of resolution for higher photon numbers.
Similarly to the results obtained in Sec.~\ref{sec:rsl}, computations of the Bhattacharyya coefficient suggest that the generalized geometric technique provides better results than the binomial technique, especially for the stronger JPM relaxation.

\section{Nonclassicality of photocounting statistics} \label{sec:noncl}

The aim of this section is to analyze the principal ability of the JPMs to serve as a playground for testing nonclassical properties of electromagnetic radiation in a microwave resonator.
Based on the results of the previous section, we anticipate that the considered JPM detector will provide sufficiently high quality photocounting statistics to solve this task efficiently.
It is also important to realize how test results for JPM detectors differ from results for the ideal PNR detectors.

Any measurement conducted under the coherent states or their statistical mixtures can be explained by classical electrodynamics; see, e.g., Refs.~\cite{titulaer1965, mandel1986, sperling2018a, *sperling2018b, sperling2020, mandelwolf1995book, vogel2006book, agarwal2013book}.
The density operator of any quantum state can be expanded by coherent states as
    \begin{equation} \label{eq:rho_decomp}
    	\hat\varrho = \int_{\alpha \in \CC} \dd^2\alpha \, P(\alpha)|\alpha\rangle_\res \langle\alpha|,
    \end{equation}
where $P(\alpha)$ is the Glauber-Sudarshan $P$ function \cite{glauber1963,sudarshan1963}.
Hence, nonclassical states are characterized by $P(\alpha) \ngeq 0$, which cannot be interpreted as probability distributions.

Photocounting measurements, even if they are ideal, do not provide complete information about the quantum state.
Consequently, any test with such measurements represents only a sufficient condition for nonclassicality of quantum states.
On the other hand, such tests address another important issue: can the given photocounting statistics be replicated with a statistical mixture of classical electromagnetic fields? 
For instance, this question is addressed in the test of the sub-Poissonian statistics of photocounting based on the Mandel $Q$ parameter \cite{mandel1979}, its higher-order generalization \cite{agarwal92}, the Klyshko test \cite{klyshko1996} and its generalization \cite{innocenti2022}, etc.
An issue with these tests is that they are formulated for ideal PNR detectors.
When applied directly to JPM detectors, even with high values of the photocounter resolution $\mathcal{R}_M$, they may lead to incorrect conclusions.
Other tests adapted for arrays of on-off detectors (see, for example, Refs.~\cite{sperling12c,bartley13,sperling13b}) are not applicable in our case since they are based on a significantly different POVM.

The nonclassicality test adapted to arbitrary POVMs has been proposed in Refs.~\cite{semenov2021, kovtoniuk2023,rivas2009}.
This states that the photocounting statistics can be explained classically if and only if, for any $\lambda(n)$, the inequality
    \begin{equation} \label{eq:ineq}
    	\sum_{n=0}^{M-1} \lambda(n) P(n) \leq \sup_{\alpha \in \CC} \sum_{n=0}^{M-1} \lambda(n) \Pi(n|\alpha)
    \end{equation}
holds.
Therefore, if there exists such $\lambda(n)$ that this inequality is violated, then photocounting statistics is nonclassical.
In Eq.~(\ref{eq:ineq})
    \begin{equation}
       \Pi(n|\alpha) = \vphantom{\langle}_\res\langle\alpha|\hat\Pi_n|\alpha\rangle_\res=\ee^{-|\alpha|^{2}} 
       \sum\limits_{k=0}^{+\infty} \frac{|\alpha|^{2k}}{k!} P(n|k)
    \end{equation}
is the $Q$ symbol of the POVM, describing the probability of obtaining $n$ clicks of the photocounter given the resonator mode prepared in the coherent state $\textstyle |\alpha\rangle_\res$.
Both sides of this inequality can be estimated experimentally. The left-hand side can be estimated with the tested nonclassical state, while the right-hand side can be estimated with a set of coherent states.
Hence, in experiments one should not even trust the POVM model to make a conclusion about nonclassicality of photocounting statistics.

The squeezed vacuum states described in Eq.~(\ref{eq:sqzd}) serve as archetype examples of nonclassical states, and they can be prepared well beyond the 3~dB limit of squeezing in circuit QED \cite{malnou2018, dass2021, eickbusch2022}.
These states are used here as a test case.
Interestingly, the Mandel $Q$ parameter associated with these states is positive.
However, other tests show a nonclassical character of the photocounting statistics for squeezed vacuum states; see, e.g., Refs.~\cite{semenov2021,kovtoniuk2023}.

\begin{figure}[t!]
    \centering
    \includegraphics{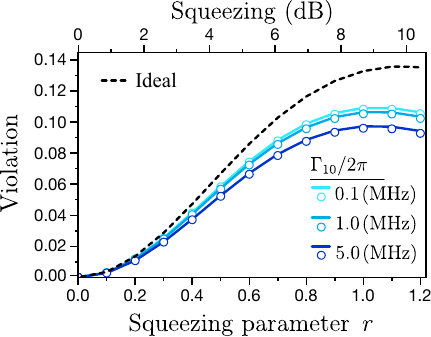}
    \caption{Dependence of the maximal violation of inequality~\eqref{eq:ineq} (difference of its left- and right-hand sides) on the squeezing parameter $r$ given the resonator mode is prepared in the squeezed vacuum state $\textstyle |r\rangle_\res$ for $M = 3$ and the various values of the JPM relaxation rate $\textstyle \Gamma_{10}$.
    Solid lines correspond to the generalized geometric technique.
    Empty circles represents the data for the generalized binomial technique.
    Dashed line corresponds to the ideal PNR detector.
    Here we take $\textstyle g/2\pi = 50\,\mathrm{MHz}$, while the rest of the system parameters are the same as in Fig.~\ref{fig:fig_8}.
    \label{fig:fig_9}}
\end{figure}

We have considered the case of $M=3$.
Inequality~\eqref{eq:ineq} has been optimized numerically with respect to $\textstyle \lambda(n)$ to find its maximum violation.
We define the violation as the difference of the left- and right-hand sides of inequality~\eqref{eq:ineq}.
For the numerical optimization, we used the adaptive differential evolution algorithm \cite{Zhang2009}.
This means that for each quantum state (the squeezing parameter $r$ for the considered example) we search for such a function $\lambda(n)$ that the violation is maximized.
This procedure has been performed for both generalized binomial and geometric counting techniques, as well as for different values of the detector parameters.
We also applied this procedure to the ideal PNR detectors, considering only $n=0,\ldots,M{-}1$.
The results are shown in Fig.~\ref{fig:fig_9}.
Evidently, counting techniques with the JPM detectors can be efficiently considered to test nonclassicality of squeezed vacuum states.
Moreover, for small squeezing ($\lesssim 2$ dB) the results for the JPMs are almost the same as for the ideal PNR detectors.
Therefore, for small squeezing parameters, JPMs can be used for nonclassicality tests as proper PNR detectors.
It is worth noting that the described nonclassicality test does not necessarily require optimization of the detector parameters and can be applied even with state-of-the-art devices.

\section{Summary} \label{sec:summ}

In this paper, we have theoretically investigated the possibility of using repeated measurements of microwave electromagnetic radiation confined in a resonator mode by a JPM detector.
Our goal was to understand if this method can be used to obtain the maximum possible resolution between adjacent numbers of photons---a key feature of photocounters required for various fundamental and applied problems.
We have considered two counting techniques referred to as generalized binomial and geometric.
The first involves counting the number of detector clicks in a given range of the predetermined measurement sequence.
The second assumes counting up to either the first no-click event or the end of the measurement sequence. 

Our theoretical methods involve three steps.
First, we have developed a description of a JPM photocounter coupled to a resonator mode with the Linblad master equation, which we then solve numerically.
Second, this solution has been applied to derive the POVMs for both counting techniques.
Third, the POVM is used in the proposed measure of photon-number resolution, which is then optimized by detector parameters.

We conclude that the currently available or near-term circuit QED devices can be used to achieve a high degree of photon-number resolution up to about seven photons.
Interestingly, the generalized geometric technique shows better results compared with the generalized binomial technique.
That could be explained by the fact that in the latter technique, we perform a fixed number of measurement iterations. This leads to a higher probability of obtaining erroneous results compared with the generalized geometric technique, where we stop counting once a no-click result is obtained. The erroneous result can arise due to photon loss in the JPM or the resonator mode or the superfluous tunneling from the lower metastable level of the JPM.

We have applied our theory to show that the JPM detectors can be used to experimentally solve a typical problem in quantum optics---testing nonclassicality of photocounting statistics.
Using the example of the squeezed vacuum state, it is shown that for relatively small squeezing parameters, the results are indistinguishable from those obtained with ideal PNR detectors. 
Therefore, nonclassical properties of low-intensity radiation analyzed by JPM detectors can be assumed to be analyzed by the ideal PNR detectors with a high degree of confidence.
We hope that our theoretical results will be useful in fundamental and applied research requiring detectors with high photon-number resolution. 

\newpage
\begin{acknowledgements}
 The authors thank A. Sokolov, R. Baskov, and V. Shnyrkov for fruitful discussions.
 This work was supported by the National Research Foundation of Ukraine through the Project No. 2020.02/0111, Nonclassical and hybrid correlations of quantum systems under realistic conditions.
\end{acknowledgements}

\appendix

\section{Derivation of the circuit Hamiltonian} \label{sec:ham_der}

\begin{figure}[b!]
	\centering
	\includegraphics{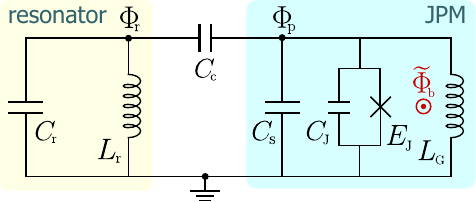}
	\caption{The lumped-element circuit diagram of the system comprised by the resonator coupled to the JPM. \label{fig:fig_10}}
\end{figure}

Figure~\ref{fig:fig_10} illustrates the circuit diagram of the considered resonator-JPM system. 
Following the standard procedure \cite{devoret1997, vool2017}, for deriving the classical Hamiltonian of the circuit shown in Fig.~\ref{fig:fig_10}, we start with the circuit Largrangian $\textstyle \Lgr$ given by
\begin{equation} \label{eq:lgrn}
	\begin{split}
		\Lgr = & \, \frac{C_\pd \dot{\Phi}^2_\pd}{2} + \EJ \cos\left(2\pi\frac{\Phi_\pd}{\Phi_0}\right) - \frac{(\Phi_\pd - \Phib)^2}{2\LG} \\
		& \, + \frac{C_\res \dot{\Phi}_\res^2}{2} - \frac{\Phi_\res^2}{2 L_\res} + \frac{C_\cpl (\dot{\Phi}_\res - \dot{\Phi}_\pd)^2}{2},
	\end{split}
\end{equation}
where $\textstyle C_\pd = \Cs + C_\JJ$ is the total capacitance of the JPM. $\textstyle \Phir$ and $\textstyle \Phi_\pd$ denote the resonator and JPM node fluxes, respectively.
Using the Legendre transform \cite{goldstein1980}, one derives the Hamiltonian:
\begin{equation} \label{eq:lgndr}
		H = Q_\pd \dot{\Phi}_\pd + Q_\res \dot{\Phi}_\res - \Lgr,
\end{equation}
where $\textstyle Q_\pd = \partial \Lgr/\partial \dot{\Phi}_\pd$ and $\textstyle Q_\res = \partial \Lgr/\partial \dot{\Phi}_\res$ are the generalized momenta corresponding to the charges on the resonator and JPM capacitances, respectively.
Using Eq.~\eqref{eq:lgrn}, one obtains
\begin{equation} \label{eq:Qs}
	\left(\!
	\begin{array}{l}
		Q_\pd \\
		Q_\res
	\end{array}
	\!\right)
	=
	\begin{pmatrix}
		C_\pd + C_\cpl && -C_\cpl \\
		-C_\cpl && C_\res + C_\cpl
	\end{pmatrix}
	\left(\!
	\begin{array}{l}
		\dot{\Phi}_\pd \\
		\dot{\Phi}_\res
	\end{array}
	\!\right)
\end{equation}

Plugging Eqs.~\eqref{eq:Qs} and \eqref{eq:lgrn} into Eq.~\eqref{eq:lgndr} leads to

\begin{equation} \label{eq:ham_1}
        \begin{split}
          H = & \, \frac{(C_\pd + C_\cpl)\dot{\Phi}^2_\pd}{2} + \frac{(C_\mathrm{r} + C_\cpl)\dot{\Phi}^2_\res}{2}
          - C_\cpl \dot{\Phi}_\pd \dot{\Phi}_\mathrm{r} \\
          & \, + \frac{\Phi^2_\res}{2L_\res} + \frac{(\Phi_\pd - \Phib)^2}{2\LG} - \EJ \cos\left(2\pi \frac{\Phi_\pd}{\Phi_0}\right).
        \end{split}
\end{equation}

Expressing $\textstyle \dot{\Phi}_\pd$ and $\textstyle \dot{\Phi}_\res$ via $\textstyle Q_\pd$ and $\textstyle Q_\res$ using Eq.~\eqref{eq:Qs} gives
\begin{subequations}
      \begin{equation}
	 \begin{split}
	   \dot{\Phi}_\pd = & \, \frac{C_\res + C_\cpl}{C_\pd (C_\res + C_\cpl) + C_\res C_\cpl} \, Q_\pd \\
          & \, + \frac{C_\cpl}{C_\pd (C_\res + C_\cpl) + C_\res C_\cpl} \, Q_\res,
	 \end{split}
      \end{equation}
      \begin{equation}
	 \begin{split}
	   \dot{\Phi}_\res = & \, \frac{C_\cpl}{C_\pd (C_\res + C_\cpl) + C_\res C_\cpl} Q_\pd \\
          & \, + \frac{C_\pd + C_\cpl}{C_\res (C_\pd + C_\cpl) + C_\pd C_\cpl} Q_\res.
	 \end{split}
      \end{equation}
\end{subequations}
Substituting the above expressions into Eq.~\eqref{eq:ham_1}, one arrives at the circuit Hamiltonian expressed as
\begin{equation} \label{eq:ham_cl}
	\begin{split}
		H = & \, \frac{Q^2_\res}{2C'_\res} + \frac{\Phi^2_\mathrm{r}}{2L_\res} + \frac{Q_\pd Q_\res}{C'_\cpl} \\
		& \, + \frac{Q_\pd^2}{2C'_\pd} + \frac{(\Phi_\pd - \Phib)^2}{2\LG} - \EJ\cos\left(2\pi\frac{\Phi_\pd}{\Phi_0}\right),
	\end{split}
\end{equation}
where the renormalized capacitances $\textstyle C'_\pd$, $\textstyle C'_\res$, and $\textstyle C'_\cpl$ are given by
\begin{equation} \label{eq:caps_ld}
	\begin{split}
		& C'_\pd = C_\pd + \frac{C_\res C_\cpl}{C_\res + C_\cpl},
		\quad C'_\res = C_\res + \frac{C_\pd C_\cpl}{C_\pd + C_\cpl},
	\end{split}
\end{equation}
and
\begin{equation}
	C'_\cpl = C_\res + C_\pd + \frac{C_\res C_\pd}{C_\cpl}.
\end{equation}

Next, we define the capacitive and inductive energies of the corresponding circuit elements and introduce the phases $\textstyle \phi_{\res} = 2\pi \Phi_{\res}/\Phi_0$ and $\textstyle \phi_\pd = 2\pi \Phi_{\pd}/\Phi_0$ across the resonator and the JPM, respectively.
We also introduce the numbers of Cooper pairs $\textstyle n_{\res} = - Q_{\res}/(2e)$ and $\textstyle n_{\pd} = - Q_{\pd}/(2e)$ on the resonator and JPM capacitors, correspondingly.
Note that variables $\textstyle n_\res$ and $\textstyle n_\pd$ can take both positive and negative values, corresponding to an excess or deficiency of the Cooper pairs.
Using these notations in Eq.~\eqref{eq:ham_cl} yields the Hamiltonian~\eqref{eq:clham0}.


\bibliography{bibliography}

\begin{thebibliography}{140}%
\makeatletter
\providecommand \@ifxundefined [1]{%
 \@ifx{#1\undefined}
}%
\providecommand \@ifnum [1]{%
 \ifnum #1\expandafter \@firstoftwo
 \else \expandafter \@secondoftwo
 \fi
}%
\providecommand \@ifx [1]{%
 \ifx #1\expandafter \@firstoftwo
 \else \expandafter \@secondoftwo
 \fi
}%
\providecommand \natexlab [1]{#1}%
\providecommand \enquote  [1]{``#1''}%
\providecommand \bibnamefont  [1]{#1}%
\providecommand \bibfnamefont [1]{#1}%
\providecommand \citenamefont [1]{#1}%
\providecommand \href@noop [0]{\@secondoftwo}%
\providecommand \href [0]{\begingroup \@sanitize@url \@href}%
\providecommand \@href[1]{\@@startlink{#1}\@@href}%
\providecommand \@@href[1]{\endgroup#1\@@endlink}%
\providecommand \@sanitize@url [0]{\catcode `\\12\catcode `\$12\catcode
  `\&12\catcode `\#12\catcode `\^12\catcode `\_12\catcode `\%12\relax}%
\providecommand \@@startlink[1]{}%
\providecommand \@@endlink[0]{}%
\providecommand \url  [0]{\begingroup\@sanitize@url \@url }%
\providecommand \@url [1]{\endgroup\@href {#1}{\urlprefix }}%
\providecommand \urlprefix  [0]{URL }%
\providecommand \Eprint [0]{\href }%
\providecommand \doibase [0]{https://doi.org/}%
\providecommand \selectlanguage [0]{\@gobble}%
\providecommand \bibinfo  [0]{\@secondoftwo}%
\providecommand \bibfield  [0]{\@secondoftwo}%
\providecommand \translation [1]{[#1]}%
\providecommand \BibitemOpen [0]{}%
\providecommand \bibitemStop [0]{}%
\providecommand \bibitemNoStop [0]{.\EOS\space}%
\providecommand \EOS [0]{\spacefactor3000\relax}%
\providecommand \BibitemShut  [1]{\csname bibitem#1\endcsname}%
\let\auto@bib@innerbib\@empty
\bibitem [{\citenamefont {Busch}\ \emph {et~al.}(2016)\citenamefont {Busch},
  \citenamefont {Lahti}, \citenamefont {Pellonp\"{a}\"{a}},\ and\ \citenamefont
  {Ylinen}}]{busch2016}%
  \BibitemOpen
  \bibfield  {author} {\bibinfo {author} {\bibfnamefont {P.}~\bibnamefont
  {Busch}}, \bibinfo {author} {\bibfnamefont {P.}~\bibnamefont {Lahti}},
  \bibinfo {author} {\bibfnamefont {J.-P.}\ \bibnamefont {Pellonp\"{a}\"{a}}},\
  and\ \bibinfo {author} {\bibfnamefont {K.}~\bibnamefont {Ylinen}},\ }\href
  {https://doi.org/0.1007/978-3-319-43389-9} {\emph {\bibinfo {title} {Quantum
  Measurement}}}\ (\bibinfo  {publisher} {Springer Cham},\ \bibinfo {year}
  {2016})\BibitemShut {NoStop}%
\bibitem [{\citenamefont {Mandel}\ \emph {et~al.}(1964)\citenamefont {Mandel},
  \citenamefont {Sudarshan},\ and\ \citenamefont {Wolf}}]{mandel1964}%
  \BibitemOpen
  \bibfield  {author} {\bibinfo {author} {\bibfnamefont {L.}~\bibnamefont
  {Mandel}}, \bibinfo {author} {\bibfnamefont {E.~C.~G.}\ \bibnamefont
  {Sudarshan}},\ and\ \bibinfo {author} {\bibfnamefont {E.}~\bibnamefont
  {Wolf}},\ }\bibfield  {title} {\bibinfo {title} {Theory of photoelectric
  detection of light fluctuations},\ }\href
  {https://doi.org/10.1088/0370-1328/84/3/313} {\bibfield  {journal} {\bibinfo
  {journal} {Proceedings of the Physical Society}\ }\textbf {\bibinfo {volume}
  {84}},\ \bibinfo {pages} {435} (\bibinfo {year} {1964})}\BibitemShut
  {NoStop}%
\bibitem [{\citenamefont {Kelley}\ and\ \citenamefont
  {Kleiner}(1964)}]{kelley1964}%
  \BibitemOpen
  \bibfield  {author} {\bibinfo {author} {\bibfnamefont {P.~L.}\ \bibnamefont
  {Kelley}}\ and\ \bibinfo {author} {\bibfnamefont {W.~H.}\ \bibnamefont
  {Kleiner}},\ }\bibfield  {title} {\bibinfo {title} {Theory of electromagnetic
  field measurement and photoelectron counting},\ }\href
  {https://doi.org/10.1103/PhysRev.136.A316} {\bibfield  {journal} {\bibinfo
  {journal} {Phys. Rev.}\ }\textbf {\bibinfo {volume} {136}},\ \bibinfo {pages}
  {A316} (\bibinfo {year} {1964})}\BibitemShut {NoStop}%
\bibitem [{\citenamefont {Lamb~Jr}\ and\ \citenamefont
  {Scully}(1969)}]{lamb1969}%
  \BibitemOpen
  \bibfield  {author} {\bibinfo {author} {\bibfnamefont {W.~E.}\ \bibnamefont
  {Lamb~Jr}}\ and\ \bibinfo {author} {\bibfnamefont {M.~O.}\ \bibnamefont
  {Scully}},\ }\bibfield  {title} {\bibinfo {title} {The photoelectric effect
  without photons},\ }in\ \href@noop {} {\emph {\bibinfo {booktitle}
  {Polarization, Matter and Radiation}}}\ (\bibinfo  {publisher} {Press Univ.
  de France},\ \bibinfo {address} {Paris},\ \bibinfo {year} {1969})\BibitemShut
  {NoStop}%
\bibitem [{\citenamefont {Mandel}\ and\ \citenamefont
  {Wolf}(1995)}]{mandelwolf1995book}%
  \BibitemOpen
  \bibfield  {author} {\bibinfo {author} {\bibfnamefont {L.}~\bibnamefont
  {Mandel}}\ and\ \bibinfo {author} {\bibfnamefont {E.}~\bibnamefont {Wolf}},\
  }\href {https://doi.org/10.1017/CBO9781139644105} {\emph {\bibinfo {title}
  {Optical Coherence and Quantum Optics}}}\ (\bibinfo  {publisher} {Cambridge
  University Press},\ \bibinfo {year} {1995})\BibitemShut {NoStop}%
\bibitem [{\citenamefont {Knill}\ \emph {et~al.}(2001)\citenamefont {Knill},
  \citenamefont {Laflamme},\ and\ \citenamefont {Milburn}}]{knill2001}%
  \BibitemOpen
  \bibfield  {author} {\bibinfo {author} {\bibfnamefont {E.}~\bibnamefont
  {Knill}}, \bibinfo {author} {\bibfnamefont {R.}~\bibnamefont {Laflamme}},\
  and\ \bibinfo {author} {\bibfnamefont {G.~J.}\ \bibnamefont {Milburn}},\
  }\bibfield  {title} {\bibinfo {title} {A scheme for efficient quantum
  computation with linear optics},\ }\href {https://doi.org/10.1038/35051009}
  {\bibfield  {journal} {\bibinfo  {journal} {Nature}\ }\textbf {\bibinfo
  {volume} {409}},\ \bibinfo {pages} {46} (\bibinfo {year} {2001})}\BibitemShut
  {NoStop}%
\bibitem [{\citenamefont {Kok}\ \emph {et~al.}(2007)\citenamefont {Kok},
  \citenamefont {Munro}, \citenamefont {Nemoto}, \citenamefont {Ralph},
  \citenamefont {Dowling},\ and\ \citenamefont {Milburn}}]{kok2007}%
  \BibitemOpen
  \bibfield  {author} {\bibinfo {author} {\bibfnamefont {P.}~\bibnamefont
  {Kok}}, \bibinfo {author} {\bibfnamefont {W.~J.}\ \bibnamefont {Munro}},
  \bibinfo {author} {\bibfnamefont {K.}~\bibnamefont {Nemoto}}, \bibinfo
  {author} {\bibfnamefont {T.~C.}\ \bibnamefont {Ralph}}, \bibinfo {author}
  {\bibfnamefont {J.~P.}\ \bibnamefont {Dowling}},\ and\ \bibinfo {author}
  {\bibfnamefont {G.~J.}\ \bibnamefont {Milburn}},\ }\bibfield  {title}
  {\bibinfo {title} {Linear optical quantum computing with photonic qubits},\
  }\href {https://doi.org/10.1103/RevModPhys.79.135} {\bibfield  {journal}
  {\bibinfo  {journal} {Rev. Mod. Phys.}\ }\textbf {\bibinfo {volume} {79}},\
  \bibinfo {pages} {135} (\bibinfo {year} {2007})}\BibitemShut {NoStop}%
\bibitem [{\citenamefont {Gisin}\ and\ \citenamefont {Thew}(2007)}]{gisin2007}%
  \BibitemOpen
  \bibfield  {author} {\bibinfo {author} {\bibfnamefont {N.}~\bibnamefont
  {Gisin}}\ and\ \bibinfo {author} {\bibfnamefont {R.}~\bibnamefont {Thew}},\
  }\bibfield  {title} {\bibinfo {title} {Quantum communication},\ }\href
  {https://doi.org/10.1038/nphoton.2007.22} {\bibfield  {journal} {\bibinfo
  {journal} {Nat. Photonics}\ }\textbf {\bibinfo {volume} {1}},\ \bibinfo
  {pages} {165} (\bibinfo {year} {2007})}\BibitemShut {NoStop}%
\bibitem [{\citenamefont {Hadfield}(2009)}]{hadfield2009}%
  \BibitemOpen
  \bibfield  {author} {\bibinfo {author} {\bibfnamefont {R.~H.}\ \bibnamefont
  {Hadfield}},\ }\bibfield  {title} {\bibinfo {title} {Single-photon detectors
  for optical quantum information applications},\ }\href
  {https://doi.org/10.1038/nphoton.2009.230} {\bibfield  {journal} {\bibinfo
  {journal} {Nature Photonics}\ }\textbf {\bibinfo {volume} {3}},\ \bibinfo
  {pages} {696} (\bibinfo {year} {2009})}\BibitemShut {NoStop}%
\bibitem [{\citenamefont {Natarajan}\ \emph {et~al.}(2012)\citenamefont
  {Natarajan}, \citenamefont {Tanner},\ and\ \citenamefont
  {Hadfield}}]{natarajan2012}%
  \BibitemOpen
  \bibfield  {author} {\bibinfo {author} {\bibfnamefont {C.~M.}\ \bibnamefont
  {Natarajan}}, \bibinfo {author} {\bibfnamefont {M.~G.}\ \bibnamefont
  {Tanner}},\ and\ \bibinfo {author} {\bibfnamefont {R.~H.}\ \bibnamefont
  {Hadfield}},\ }\bibfield  {title} {\bibinfo {title} {Superconducting nanowire
  single-photon detectors: physics and applications},\ }\href
  {http://stacks.iop.org/0953-2048/25/i=6/a=063001} {\bibfield  {journal}
  {\bibinfo  {journal} {Supercond. Sci. Technol.}\ }\textbf {\bibinfo {volume}
  {25}},\ \bibinfo {pages} {063001} (\bibinfo {year} {2012})}\BibitemShut
  {NoStop}%
\bibitem [{\citenamefont {You}(2020)}]{you2020}%
  \BibitemOpen
  \bibfield  {author} {\bibinfo {author} {\bibfnamefont {L.}~\bibnamefont
  {You}},\ }\bibfield  {title} {\bibinfo {title} {Superconducting nanowire
  single-photon detectors for quantum information},\ }\href
  {https://doi.org/doi:10.1515/nanoph-2020-0186} {\bibfield  {journal}
  {\bibinfo  {journal} {Nanophotonics}\ }\textbf {\bibinfo {volume} {9}},\
  \bibinfo {pages} {2673} (\bibinfo {year} {2020})}\BibitemShut {NoStop}%
\bibitem [{\citenamefont {Giovannetti}\ \emph {et~al.}(2011)\citenamefont
  {Giovannetti}, \citenamefont {Lloyd},\ and\ \citenamefont
  {Maccone}}]{giovannetti2011}%
  \BibitemOpen
  \bibfield  {author} {\bibinfo {author} {\bibfnamefont {V.}~\bibnamefont
  {Giovannetti}}, \bibinfo {author} {\bibfnamefont {S.}~\bibnamefont {Lloyd}},\
  and\ \bibinfo {author} {\bibfnamefont {L.}~\bibnamefont {Maccone}},\
  }\bibfield  {title} {\bibinfo {title} {Advances in quantum metrology},\
  }\href {https://doi.org/10.1038/nphoton.2011.35} {\bibfield  {journal}
  {\bibinfo  {journal} {Nat. Photonics}\ }\textbf {\bibinfo {volume} {5}},\
  \bibinfo {pages} {222} (\bibinfo {year} {2011})}\BibitemShut {NoStop}%
\bibitem [{\citenamefont {Degen}\ \emph {et~al.}(2017)\citenamefont {Degen},
  \citenamefont {Reinhard},\ and\ \citenamefont {Cappellaro}}]{degen2017}%
  \BibitemOpen
  \bibfield  {author} {\bibinfo {author} {\bibfnamefont {C.~L.}\ \bibnamefont
  {Degen}}, \bibinfo {author} {\bibfnamefont {F.}~\bibnamefont {Reinhard}},\
  and\ \bibinfo {author} {\bibfnamefont {P.}~\bibnamefont {Cappellaro}},\
  }\bibfield  {title} {\bibinfo {title} {Quantum sensing},\ }\href
  {https://doi.org/10.1103/RevModPhys.89.035002} {\bibfield  {journal}
  {\bibinfo  {journal} {Rev. Mod. Phys.}\ }\textbf {\bibinfo {volume} {89}},\
  \bibinfo {pages} {035002} (\bibinfo {year} {2017})}\BibitemShut {NoStop}%
\bibitem [{\citenamefont {Pirandola}\ \emph {et~al.}(2020)\citenamefont
  {Pirandola}, \citenamefont {Andersen}, \citenamefont {Banchi}, \citenamefont
  {Berta}, \citenamefont {Bunandar}, \citenamefont {Colbeck}, \citenamefont
  {Englund}, \citenamefont {Gehring}, \citenamefont {Lupo}, \citenamefont
  {Ottaviani}, \citenamefont {Pereira}, \citenamefont {Razavi}, \citenamefont
  {Shaari}, \citenamefont {Tomamichel}, \citenamefont {Usenko}, \citenamefont
  {Vallone}, \citenamefont {Villoresi},\ and\ \citenamefont
  {Wallden}}]{pirandola2020}%
  \BibitemOpen
  \bibfield  {author} {\bibinfo {author} {\bibfnamefont {S.}~\bibnamefont
  {Pirandola}}, \bibinfo {author} {\bibfnamefont {U.~L.}\ \bibnamefont
  {Andersen}}, \bibinfo {author} {\bibfnamefont {L.}~\bibnamefont {Banchi}},
  \bibinfo {author} {\bibfnamefont {M.}~\bibnamefont {Berta}}, \bibinfo
  {author} {\bibfnamefont {D.}~\bibnamefont {Bunandar}}, \bibinfo {author}
  {\bibfnamefont {R.}~\bibnamefont {Colbeck}}, \bibinfo {author} {\bibfnamefont
  {D.}~\bibnamefont {Englund}}, \bibinfo {author} {\bibfnamefont
  {T.}~\bibnamefont {Gehring}}, \bibinfo {author} {\bibfnamefont
  {C.}~\bibnamefont {Lupo}}, \bibinfo {author} {\bibfnamefont {C.}~\bibnamefont
  {Ottaviani}}, \bibinfo {author} {\bibfnamefont {J.~L.}\ \bibnamefont
  {Pereira}}, \bibinfo {author} {\bibfnamefont {M.}~\bibnamefont {Razavi}},
  \bibinfo {author} {\bibfnamefont {J.~S.}\ \bibnamefont {Shaari}}, \bibinfo
  {author} {\bibfnamefont {M.}~\bibnamefont {Tomamichel}}, \bibinfo {author}
  {\bibfnamefont {V.~C.}\ \bibnamefont {Usenko}}, \bibinfo {author}
  {\bibfnamefont {G.}~\bibnamefont {Vallone}}, \bibinfo {author} {\bibfnamefont
  {P.}~\bibnamefont {Villoresi}},\ and\ \bibinfo {author} {\bibfnamefont
  {P.}~\bibnamefont {Wallden}},\ }\bibfield  {title} {\bibinfo {title}
  {Advances in quantum cryptography},\ }\href
  {https://doi.org/10.1364/AOP.361502} {\bibfield  {journal} {\bibinfo
  {journal} {Adv. Opt. Photon.}\ }\textbf {\bibinfo {volume} {12}},\ \bibinfo
  {pages} {1012} (\bibinfo {year} {2020})}\BibitemShut {NoStop}%
\bibitem [{\citenamefont {Polino}\ \emph {et~al.}(2020)\citenamefont {Polino},
  \citenamefont {Valeri}, \citenamefont {Spagnolo},\ and\ \citenamefont
  {Sciarrino}}]{polino2020}%
  \BibitemOpen
  \bibfield  {author} {\bibinfo {author} {\bibfnamefont {E.}~\bibnamefont
  {Polino}}, \bibinfo {author} {\bibfnamefont {M.}~\bibnamefont {Valeri}},
  \bibinfo {author} {\bibfnamefont {N.}~\bibnamefont {Spagnolo}},\ and\
  \bibinfo {author} {\bibfnamefont {F.}~\bibnamefont {Sciarrino}},\ }\bibfield
  {title} {\bibinfo {title} {{Photonic quantum metrology}},\ }\href
  {https://doi.org/10.1116/5.0007577} {\bibfield  {journal} {\bibinfo
  {journal} {AVS Quantum Science}\ }\textbf {\bibinfo {volume} {2}},\ \bibinfo
  {pages} {024703} (\bibinfo {year} {2020})}\BibitemShut {NoStop}%
\bibitem [{\citenamefont {Xu}\ \emph {et~al.}(2020)\citenamefont {Xu},
  \citenamefont {Ma}, \citenamefont {Zhang}, \citenamefont {Lo},\ and\
  \citenamefont {Pan}}]{xu2020}%
  \BibitemOpen
  \bibfield  {author} {\bibinfo {author} {\bibfnamefont {F.}~\bibnamefont
  {Xu}}, \bibinfo {author} {\bibfnamefont {X.}~\bibnamefont {Ma}}, \bibinfo
  {author} {\bibfnamefont {Q.}~\bibnamefont {Zhang}}, \bibinfo {author}
  {\bibfnamefont {H.-K.}\ \bibnamefont {Lo}},\ and\ \bibinfo {author}
  {\bibfnamefont {J.-W.}\ \bibnamefont {Pan}},\ }\bibfield  {title} {\bibinfo
  {title} {Secure quantum key distribution with realistic devices},\ }\href
  {https://doi.org/10.1103/RevModPhys.92.025002} {\bibfield  {journal}
  {\bibinfo  {journal} {Rev. Mod. Phys.}\ }\textbf {\bibinfo {volume} {92}},\
  \bibinfo {pages} {025002} (\bibinfo {year} {2020})}\BibitemShut {NoStop}%
\bibitem [{\citenamefont {Shalm}\ \emph {et~al.}(2015)\citenamefont {Shalm},
  \citenamefont {Meyer-Scott}, \citenamefont {Christensen}, \citenamefont
  {Bierhorst}, \citenamefont {Wayne}, \citenamefont {Stevens}, \citenamefont
  {Gerrits}, \citenamefont {Glancy}, \citenamefont {Hamel}, \citenamefont
  {Allman}, \citenamefont {Coakley}, \citenamefont {Dyer}, \citenamefont
  {Hodge}, \citenamefont {Lita}, \citenamefont {Verma}, \citenamefont
  {Lambrocco}, \citenamefont {Tortorici}, \citenamefont {Migdall},
  \citenamefont {Zhang}, \citenamefont {Kumor}, \citenamefont {Farr},
  \citenamefont {Marsili}, \citenamefont {Shaw}, \citenamefont {Stern},
  \citenamefont {Abell\'an}, \citenamefont {Amaya}, \citenamefont {Pruneri},
  \citenamefont {Jennewein}, \citenamefont {Mitchell}, \citenamefont {Kwiat},
  \citenamefont {Bienfang}, \citenamefont {Mirin}, \citenamefont {Knill},\ and\
  \citenamefont {Nam}}]{shalm2015}%
  \BibitemOpen
  \bibfield  {author} {\bibinfo {author} {\bibfnamefont {L.~K.}\ \bibnamefont
  {Shalm}}, \bibinfo {author} {\bibfnamefont {E.}~\bibnamefont {Meyer-Scott}},
  \bibinfo {author} {\bibfnamefont {B.~G.}\ \bibnamefont {Christensen}},
  \bibinfo {author} {\bibfnamefont {P.}~\bibnamefont {Bierhorst}}, \bibinfo
  {author} {\bibfnamefont {M.~A.}\ \bibnamefont {Wayne}}, \bibinfo {author}
  {\bibfnamefont {M.~J.}\ \bibnamefont {Stevens}}, \bibinfo {author}
  {\bibfnamefont {T.}~\bibnamefont {Gerrits}}, \bibinfo {author} {\bibfnamefont
  {S.}~\bibnamefont {Glancy}}, \bibinfo {author} {\bibfnamefont {D.~R.}\
  \bibnamefont {Hamel}}, \bibinfo {author} {\bibfnamefont {M.~S.}\ \bibnamefont
  {Allman}}, \bibinfo {author} {\bibfnamefont {K.~J.}\ \bibnamefont {Coakley}},
  \bibinfo {author} {\bibfnamefont {S.~D.}\ \bibnamefont {Dyer}}, \bibinfo
  {author} {\bibfnamefont {C.}~\bibnamefont {Hodge}}, \bibinfo {author}
  {\bibfnamefont {A.~E.}\ \bibnamefont {Lita}}, \bibinfo {author}
  {\bibfnamefont {V.~B.}\ \bibnamefont {Verma}}, \bibinfo {author}
  {\bibfnamefont {C.}~\bibnamefont {Lambrocco}}, \bibinfo {author}
  {\bibfnamefont {E.}~\bibnamefont {Tortorici}}, \bibinfo {author}
  {\bibfnamefont {A.~L.}\ \bibnamefont {Migdall}}, \bibinfo {author}
  {\bibfnamefont {Y.}~\bibnamefont {Zhang}}, \bibinfo {author} {\bibfnamefont
  {D.~R.}\ \bibnamefont {Kumor}}, \bibinfo {author} {\bibfnamefont {W.~H.}\
  \bibnamefont {Farr}}, \bibinfo {author} {\bibfnamefont {F.}~\bibnamefont
  {Marsili}}, \bibinfo {author} {\bibfnamefont {M.~D.}\ \bibnamefont {Shaw}},
  \bibinfo {author} {\bibfnamefont {J.~A.}\ \bibnamefont {Stern}}, \bibinfo
  {author} {\bibfnamefont {C.}~\bibnamefont {Abell\'an}}, \bibinfo {author}
  {\bibfnamefont {W.}~\bibnamefont {Amaya}}, \bibinfo {author} {\bibfnamefont
  {V.}~\bibnamefont {Pruneri}}, \bibinfo {author} {\bibfnamefont
  {T.}~\bibnamefont {Jennewein}}, \bibinfo {author} {\bibfnamefont {M.~W.}\
  \bibnamefont {Mitchell}}, \bibinfo {author} {\bibfnamefont {P.~G.}\
  \bibnamefont {Kwiat}}, \bibinfo {author} {\bibfnamefont {J.~C.}\ \bibnamefont
  {Bienfang}}, \bibinfo {author} {\bibfnamefont {R.~P.}\ \bibnamefont {Mirin}},
  \bibinfo {author} {\bibfnamefont {E.}~\bibnamefont {Knill}},\ and\ \bibinfo
  {author} {\bibfnamefont {S.~W.}\ \bibnamefont {Nam}},\ }\bibfield  {title}
  {\bibinfo {title} {Strong loophole-free test of local realism},\ }\href
  {https://doi.org/10.1103/PhysRevLett.115.250402} {\bibfield  {journal}
  {\bibinfo  {journal} {Phys. Rev. Lett.}\ }\textbf {\bibinfo {volume} {115}},\
  \bibinfo {pages} {250402} (\bibinfo {year} {2015})}\BibitemShut {NoStop}%
\bibitem [{\citenamefont {Bohmann}\ \emph {et~al.}(2018)\citenamefont
  {Bohmann}, \citenamefont {Tiedau}, \citenamefont {Bartley}, \citenamefont
  {Sperling}, \citenamefont {Silberhorn},\ and\ \citenamefont
  {Vogel}}]{bohmann2018}%
  \BibitemOpen
  \bibfield  {author} {\bibinfo {author} {\bibfnamefont {M.}~\bibnamefont
  {Bohmann}}, \bibinfo {author} {\bibfnamefont {J.}~\bibnamefont {Tiedau}},
  \bibinfo {author} {\bibfnamefont {T.}~\bibnamefont {Bartley}}, \bibinfo
  {author} {\bibfnamefont {J.}~\bibnamefont {Sperling}}, \bibinfo {author}
  {\bibfnamefont {C.}~\bibnamefont {Silberhorn}},\ and\ \bibinfo {author}
  {\bibfnamefont {W.}~\bibnamefont {Vogel}},\ }\bibfield  {title} {\bibinfo
  {title} {Incomplete detection of nonclassical phase-space distributions},\
  }\href {https://doi.org/10.1103/PhysRevLett.120.063607} {\bibfield  {journal}
  {\bibinfo  {journal} {Phys. Rev. Lett.}\ }\textbf {\bibinfo {volume} {120}},\
  \bibinfo {pages} {063607} (\bibinfo {year} {2018})}\BibitemShut {NoStop}%
\bibitem [{\citenamefont {Herzog}(1996)}]{herzog1996}%
  \BibitemOpen
  \bibfield  {author} {\bibinfo {author} {\bibfnamefont {U.}~\bibnamefont
  {Herzog}},\ }\bibfield  {title} {\bibinfo {title} {Loss-error compensation in
  quantum-state measurements and the solution of the time-reversed damping
  equation},\ }\href {https://doi.org/10.1103/PhysRevA.53.1245} {\bibfield
  {journal} {\bibinfo  {journal} {Phys. Rev. A}\ }\textbf {\bibinfo {volume}
  {53}},\ \bibinfo {pages} {1245} (\bibinfo {year} {1996})}\BibitemShut
  {NoStop}%
\bibitem [{\citenamefont {Karp}\ \emph {et~al.}(1970)\citenamefont {Karp},
  \citenamefont {O'Neill},\ and\ \citenamefont {Gagliardi}}]{karp1970}%
  \BibitemOpen
  \bibfield  {author} {\bibinfo {author} {\bibfnamefont {S.}~\bibnamefont
  {Karp}}, \bibinfo {author} {\bibfnamefont {E.}~\bibnamefont {O'Neill}},\ and\
  \bibinfo {author} {\bibfnamefont {R.}~\bibnamefont {Gagliardi}},\ }\bibfield
  {title} {\bibinfo {title} {Communication theory for the free-space optical
  channel},\ }\href {https://doi.org/10.1109/PROC.1970.7985} {\bibfield
  {journal} {\bibinfo  {journal} {Proceedings of the IEEE}\ }\textbf {\bibinfo
  {volume} {58}},\ \bibinfo {pages} {1611} (\bibinfo {year}
  {1970})}\BibitemShut {NoStop}%
\bibitem [{\citenamefont {Lee}\ \emph {et~al.}(2004)\citenamefont {Lee},
  \citenamefont {Yurtsever}, \citenamefont {Kok}, \citenamefont {Hockney},
  \citenamefont {Adami}, \citenamefont {Braunstein},\ and\ \citenamefont
  {Dowling}}]{lee2004}%
  \BibitemOpen
  \bibfield  {author} {\bibinfo {author} {\bibfnamefont {H.}~\bibnamefont
  {Lee}}, \bibinfo {author} {\bibfnamefont {U.}~\bibnamefont {Yurtsever}},
  \bibinfo {author} {\bibfnamefont {P.}~\bibnamefont {Kok}}, \bibinfo {author}
  {\bibfnamefont {G.~M.}\ \bibnamefont {Hockney}}, \bibinfo {author}
  {\bibfnamefont {C.}~\bibnamefont {Adami}}, \bibinfo {author} {\bibfnamefont
  {S.~L.}\ \bibnamefont {Braunstein}},\ and\ \bibinfo {author} {\bibfnamefont
  {J.~P.}\ \bibnamefont {Dowling}},\ }\bibfield  {title} {\bibinfo {title}
  {Towards photostatistics from photon-number discriminating detectors},\
  }\href {https://doi.org/10.1080/09500340408235289} {\bibfield  {journal}
  {\bibinfo  {journal} {Journal of Modern Optics}\ }\textbf {\bibinfo {volume}
  {51}},\ \bibinfo {pages} {1517} (\bibinfo {year} {2004})}\BibitemShut
  {NoStop}%
\bibitem [{\citenamefont {Semenov}\ \emph {et~al.}(2008)\citenamefont
  {Semenov}, \citenamefont {Turchin},\ and\ \citenamefont
  {Gomonay}}]{semenov2008}%
  \BibitemOpen
  \bibfield  {author} {\bibinfo {author} {\bibfnamefont {A.~A.}\ \bibnamefont
  {Semenov}}, \bibinfo {author} {\bibfnamefont {A.~V.}\ \bibnamefont
  {Turchin}},\ and\ \bibinfo {author} {\bibfnamefont {H.~V.}\ \bibnamefont
  {Gomonay}},\ }\bibfield  {title} {\bibinfo {title} {Detection of quantum
  light in the presence of noise},\ }\href
  {https://doi.org/10.1103/PhysRevA.78.055803} {\bibfield  {journal} {\bibinfo
  {journal} {Phys. Rev. A}\ }\textbf {\bibinfo {volume} {78}},\ \bibinfo
  {pages} {055803} (\bibinfo {year} {2008})}\BibitemShut {NoStop}%
\bibitem [{\citenamefont {Semenov}\ \emph {et~al.}(2009)\citenamefont
  {Semenov}, \citenamefont {Turchin},\ and\ \citenamefont
  {Gomonay}}]{semenov2008err}%
  \BibitemOpen
  \bibfield  {author} {\bibinfo {author} {\bibfnamefont {A.~A.}\ \bibnamefont
  {Semenov}}, \bibinfo {author} {\bibfnamefont {A.~V.}\ \bibnamefont
  {Turchin}},\ and\ \bibinfo {author} {\bibfnamefont {H.~V.}\ \bibnamefont
  {Gomonay}},\ }\bibfield  {title} {\bibinfo {title} {Erratum: Detection of
  quantum light in the presence of noise [phys. rev. a 78, 055803 (2008)]},\
  }\href {https://doi.org/10.1103/PhysRevA.79.019902} {\bibfield  {journal}
  {\bibinfo  {journal} {Phys. Rev. A}\ }\textbf {\bibinfo {volume} {79}},\
  \bibinfo {pages} {019902(E)} (\bibinfo {year} {2009})}\BibitemShut {NoStop}%
\bibitem [{\citenamefont {Paul}\ \emph {et~al.}(1996)\citenamefont {Paul},
  \citenamefont {T\"orm\"a}, \citenamefont {Kiss},\ and\ \citenamefont
  {Jex}}]{paul1996}%
  \BibitemOpen
  \bibfield  {author} {\bibinfo {author} {\bibfnamefont {H.}~\bibnamefont
  {Paul}}, \bibinfo {author} {\bibfnamefont {P.}~\bibnamefont {T\"orm\"a}},
  \bibinfo {author} {\bibfnamefont {T.}~\bibnamefont {Kiss}},\ and\ \bibinfo
  {author} {\bibfnamefont {I.}~\bibnamefont {Jex}},\ }\bibfield  {title}
  {\bibinfo {title} {Photon chopping: New way to measure the quantum state of
  light},\ }\href {https://doi.org/10.1103/PhysRevLett.76.2464} {\bibfield
  {journal} {\bibinfo  {journal} {Phys. Rev. Lett.}\ }\textbf {\bibinfo
  {volume} {76}},\ \bibinfo {pages} {2464} (\bibinfo {year}
  {1996})}\BibitemShut {NoStop}%
\bibitem [{\citenamefont {Achilles}\ \emph {et~al.}(2003)\citenamefont
  {Achilles}, \citenamefont {Silberhorn}, \citenamefont {\'{S}liwa},
  \citenamefont {Banaszek},\ and\ \citenamefont {Walmsley}}]{achilles2003}%
  \BibitemOpen
  \bibfield  {author} {\bibinfo {author} {\bibfnamefont {D.}~\bibnamefont
  {Achilles}}, \bibinfo {author} {\bibfnamefont {C.}~\bibnamefont
  {Silberhorn}}, \bibinfo {author} {\bibfnamefont {C.}~\bibnamefont
  {\'{S}liwa}}, \bibinfo {author} {\bibfnamefont {K.}~\bibnamefont
  {Banaszek}},\ and\ \bibinfo {author} {\bibfnamefont {I.~A.}\ \bibnamefont
  {Walmsley}},\ }\bibfield  {title} {\bibinfo {title} {Fiber-assisted detection
  with photon number resolution},\ }\href
  {https://doi.org/10.1364/OL.28.002387} {\bibfield  {journal} {\bibinfo
  {journal} {Opt. Lett.}\ }\textbf {\bibinfo {volume} {28}},\ \bibinfo {pages}
  {2387} (\bibinfo {year} {2003})}\BibitemShut {NoStop}%
\bibitem [{\citenamefont {Fitch}\ \emph {et~al.}(2003)\citenamefont {Fitch},
  \citenamefont {Jacobs}, \citenamefont {Pittman},\ and\ \citenamefont
  {Franson}}]{fitch2003}%
  \BibitemOpen
  \bibfield  {author} {\bibinfo {author} {\bibfnamefont {M.~J.}\ \bibnamefont
  {Fitch}}, \bibinfo {author} {\bibfnamefont {B.~C.}\ \bibnamefont {Jacobs}},
  \bibinfo {author} {\bibfnamefont {T.~B.}\ \bibnamefont {Pittman}},\ and\
  \bibinfo {author} {\bibfnamefont {J.~D.}\ \bibnamefont {Franson}},\
  }\bibfield  {title} {\bibinfo {title} {Photon-number resolution using
  time-multiplexed single-photon detectors},\ }\href
  {https://doi.org/10.1103/PhysRevA.68.043814} {\bibfield  {journal} {\bibinfo
  {journal} {Phys. Rev. A}\ }\textbf {\bibinfo {volume} {68}},\ \bibinfo
  {pages} {043814} (\bibinfo {year} {2003})}\BibitemShut {NoStop}%
\bibitem [{\citenamefont {\ifmmode \check{R}\else
  \v{R}\fi{}eh\'a\ifmmode~\check{c}\else \v{c}\fi{}ek}\ \emph
  {et~al.}(2003)\citenamefont {\ifmmode \check{R}\else
  \v{R}\fi{}eh\'a\ifmmode~\check{c}\else \v{c}\fi{}ek}, \citenamefont {Hradil},
  \citenamefont {Haderka}, \citenamefont {Pe\ifmmode~\check{r}\else
  \v{r}\fi{}ina},\ and\ \citenamefont {Hamar}}]{rehacek2003}%
  \BibitemOpen
  \bibfield  {author} {\bibinfo {author} {\bibfnamefont {J.}~\bibnamefont
  {\ifmmode \check{R}\else \v{R}\fi{}eh\'a\ifmmode~\check{c}\else
  \v{c}\fi{}ek}}, \bibinfo {author} {\bibfnamefont {Z.}~\bibnamefont {Hradil}},
  \bibinfo {author} {\bibfnamefont {O.}~\bibnamefont {Haderka}}, \bibinfo
  {author} {\bibfnamefont {J.}~\bibnamefont {Pe\ifmmode~\check{r}\else
  \v{r}\fi{}ina}},\ and\ \bibinfo {author} {\bibfnamefont {M.}~\bibnamefont
  {Hamar}},\ }\bibfield  {title} {\bibinfo {title} {Multiple-photon resolving
  fiber-loop detector},\ }\href {https://doi.org/10.1103/PhysRevA.67.061801}
  {\bibfield  {journal} {\bibinfo  {journal} {Phys. Rev. A}\ }\textbf {\bibinfo
  {volume} {67}},\ \bibinfo {pages} {061801(R)} (\bibinfo {year}
  {2003})}\BibitemShut {NoStop}%
\bibitem [{\citenamefont {S.~A.~Castelletto}\ and\ \citenamefont
  {Migdall}(2007)}]{castellano2007}%
  \BibitemOpen
  \bibfield  {author} {\bibinfo {author} {\bibfnamefont {V.~S.}\ \bibnamefont
  {S.~A.~Castelletto}, \bibfnamefont {I.~P.~Degiovanni}}\ and\ \bibinfo
  {author} {\bibfnamefont {A.~L.}\ \bibnamefont {Migdall}},\ }\bibfield
  {title} {\bibinfo {title} {Reduced deadtime and higher rate photon-counting
  detection using a multiplexed detector array},\ }\href
  {https://doi.org/10.1080/09500340600779579} {\bibfield  {journal} {\bibinfo
  {journal} {Journal of Modern Optics}\ }\textbf {\bibinfo {volume} {54}},\
  \bibinfo {pages} {337} (\bibinfo {year} {2007})}\BibitemShut {NoStop}%
\bibitem [{\citenamefont {Schettini}\ \emph {et~al.}(2007)\citenamefont
  {Schettini}, \citenamefont {Polyakov}, \citenamefont {Degiovanni},
  \citenamefont {Brida}, \citenamefont {Castelletto},\ and\ \citenamefont
  {L.~Migdall}}]{schettini2007}%
  \BibitemOpen
  \bibfield  {author} {\bibinfo {author} {\bibfnamefont {V.}~\bibnamefont
  {Schettini}}, \bibinfo {author} {\bibfnamefont {S.~V.}\ \bibnamefont
  {Polyakov}}, \bibinfo {author} {\bibfnamefont {I.~P.}\ \bibnamefont
  {Degiovanni}}, \bibinfo {author} {\bibfnamefont {G.}~\bibnamefont {Brida}},
  \bibinfo {author} {\bibfnamefont {S.}~\bibnamefont {Castelletto}},\ and\
  \bibinfo {author} {\bibfnamefont {A.}~\bibnamefont {L.~Migdall}},\ }\bibfield
   {title} {\bibinfo {title} {Implementing a multiplexed system of detectors
  for higher photon counting rates},\ }\href
  {https://doi.org/10.1109/JSTQE.2007.902846} {\bibfield  {journal} {\bibinfo
  {journal} {IEEE Journal of Selected Topics in Quantum Electronics}\ }\textbf
  {\bibinfo {volume} {13}},\ \bibinfo {pages} {978} (\bibinfo {year}
  {2007})}\BibitemShut {NoStop}%
\bibitem [{\citenamefont {Blanchet}\ \emph {et~al.}(2008)\citenamefont
  {Blanchet}, \citenamefont {Devaux}, \citenamefont {Furfaro},\ and\
  \citenamefont {Lantz}}]{blanchet2008}%
  \BibitemOpen
  \bibfield  {author} {\bibinfo {author} {\bibfnamefont {J.-L.}\ \bibnamefont
  {Blanchet}}, \bibinfo {author} {\bibfnamefont {F.}~\bibnamefont {Devaux}},
  \bibinfo {author} {\bibfnamefont {L.}~\bibnamefont {Furfaro}},\ and\ \bibinfo
  {author} {\bibfnamefont {E.}~\bibnamefont {Lantz}},\ }\bibfield  {title}
  {\bibinfo {title} {Measurement of sub-shot-noise correlations of spatial
  fluctuations in the photon-counting regime},\ }\href
  {https://doi.org/10.1103/PhysRevLett.101.233604} {\bibfield  {journal}
  {\bibinfo  {journal} {Phys. Rev. Lett.}\ }\textbf {\bibinfo {volume} {101}},\
  \bibinfo {pages} {233604} (\bibinfo {year} {2008})}\BibitemShut {NoStop}%
\bibitem [{\citenamefont {Hlou\ifmmode~\check{s}\else \v{s}\fi{}ek}\ \emph
  {et~al.}(2019)\citenamefont {Hlou\ifmmode~\check{s}\else \v{s}\fi{}ek},
  \citenamefont {Dudka}, \citenamefont {Straka},\ and\ \citenamefont
  {Je\ifmmode~\check{z}\else \v{z}\fi{}ek}}]{hlousek2019}%
  \BibitemOpen
  \bibfield  {author} {\bibinfo {author} {\bibfnamefont {J.}~\bibnamefont
  {Hlou\ifmmode~\check{s}\else \v{s}\fi{}ek}}, \bibinfo {author} {\bibfnamefont
  {M.}~\bibnamefont {Dudka}}, \bibinfo {author} {\bibfnamefont
  {I.}~\bibnamefont {Straka}},\ and\ \bibinfo {author} {\bibfnamefont
  {M.}~\bibnamefont {Je\ifmmode~\check{z}\else \v{z}\fi{}ek}},\ }\bibfield
  {title} {\bibinfo {title} {Accurate detection of arbitrary photon
  statistics},\ }\href {https://doi.org/10.1103/PhysRevLett.123.153604}
  {\bibfield  {journal} {\bibinfo  {journal} {Phys. Rev. Lett.}\ }\textbf
  {\bibinfo {volume} {123}},\ \bibinfo {pages} {153604} (\bibinfo {year}
  {2019})}\BibitemShut {NoStop}%
\bibitem [{\citenamefont {Saleh}(1978)}]{saleh1978book}%
  \BibitemOpen
  \bibfield  {author} {\bibinfo {author} {\bibfnamefont {B.}~\bibnamefont
  {Saleh}},\ }\href@noop {} {\emph {\bibinfo {title} {Photoelectron
  Statistics}}}\ (\bibinfo  {publisher} {Springer Berlin},\ \bibinfo {year}
  {1978})\BibitemShut {NoStop}%
\bibitem [{\citenamefont {Sperling}\ \emph
  {et~al.}(2012{\natexlab{a}})\citenamefont {Sperling}, \citenamefont {Vogel},\
  and\ \citenamefont {Agarwal}}]{sperling2012}%
  \BibitemOpen
  \bibfield  {author} {\bibinfo {author} {\bibfnamefont {J.}~\bibnamefont
  {Sperling}}, \bibinfo {author} {\bibfnamefont {W.}~\bibnamefont {Vogel}},\
  and\ \bibinfo {author} {\bibfnamefont {G.~S.}\ \bibnamefont {Agarwal}},\
  }\bibfield  {title} {\bibinfo {title} {True photocounting statistics of
  multiple on-off detectors},\ }\href
  {https://doi.org/10.1103/PhysRevA.85.023820} {\bibfield  {journal} {\bibinfo
  {journal} {Phys. Rev. A}\ }\textbf {\bibinfo {volume} {85}},\ \bibinfo
  {pages} {023820} (\bibinfo {year} {2012}{\natexlab{a}})}\BibitemShut
  {NoStop}%
\bibitem [{\citenamefont {Kovalenko}\ \emph {et~al.}(2018)\citenamefont
  {Kovalenko}, \citenamefont {Sperling}, \citenamefont {Vogel},\ and\
  \citenamefont {Semenov}}]{kovalenko2018}%
  \BibitemOpen
  \bibfield  {author} {\bibinfo {author} {\bibfnamefont {O.~P.}\ \bibnamefont
  {Kovalenko}}, \bibinfo {author} {\bibfnamefont {J.}~\bibnamefont {Sperling}},
  \bibinfo {author} {\bibfnamefont {W.}~\bibnamefont {Vogel}},\ and\ \bibinfo
  {author} {\bibfnamefont {A.~A.}\ \bibnamefont {Semenov}},\ }\bibfield
  {title} {\bibinfo {title} {Geometrical picture of photocounting
  measurements},\ }\href {https://doi.org/10.1103/PhysRevA.97.023845}
  {\bibfield  {journal} {\bibinfo  {journal} {Phys. Rev. A}\ }\textbf {\bibinfo
  {volume} {97}},\ \bibinfo {pages} {023845} (\bibinfo {year}
  {2018})}\BibitemShut {NoStop}%
\bibitem [{\citenamefont {Uzunova}\ and\ \citenamefont
  {Semenov}(2022)}]{uzunova2022}%
  \BibitemOpen
  \bibfield  {author} {\bibinfo {author} {\bibfnamefont {V.~A.}\ \bibnamefont
  {Uzunova}}\ and\ \bibinfo {author} {\bibfnamefont {A.~A.}\ \bibnamefont
  {Semenov}},\ }\bibfield  {title} {\bibinfo {title} {Photocounting statistics
  of superconducting nanowire single-photon detectors},\ }\href
  {https://doi.org/10.1103/PhysRevA.105.063716} {\bibfield  {journal} {\bibinfo
   {journal} {Phys. Rev. A}\ }\textbf {\bibinfo {volume} {105}},\ \bibinfo
  {pages} {063716} (\bibinfo {year} {2022})}\BibitemShut {NoStop}%
\bibitem [{\citenamefont {Semenov}\ \emph {et~al.}(2023)\citenamefont
  {Semenov}, \citenamefont {Samelin}, \citenamefont {Boldt}, \citenamefont
  {Schünemann}, \citenamefont {Reiher}, \citenamefont {Vogel},\ and\
  \citenamefont {Hage}}]{semenov2023}%
  \BibitemOpen
  \bibfield  {author} {\bibinfo {author} {\bibfnamefont {A.~A.}\ \bibnamefont
  {Semenov}}, \bibinfo {author} {\bibfnamefont {J.}~\bibnamefont {Samelin}},
  \bibinfo {author} {\bibfnamefont {C.}~\bibnamefont {Boldt}}, \bibinfo
  {author} {\bibfnamefont {M.}~\bibnamefont {Schünemann}}, \bibinfo {author}
  {\bibfnamefont {C.}~\bibnamefont {Reiher}}, \bibinfo {author} {\bibfnamefont
  {W.}~\bibnamefont {Vogel}},\ and\ \bibinfo {author} {\bibfnamefont
  {B.}~\bibnamefont {Hage}},\ }\href@noop {} {\bibinfo {title} {Photocounting
  measurements with dead time and afterpulses in the continuous-wave regime}}
  (\bibinfo {year} {2023}),\ \Eprint {https://arxiv.org/abs/2303.14246}
  {arXiv:2303.14246 [quant-ph]} \BibitemShut {NoStop}%
\bibitem [{\citenamefont {Wendin}\ and\ \citenamefont
  {Shumeiko}(2007)}]{wendin2007ltp}%
  \BibitemOpen
  \bibfield  {author} {\bibinfo {author} {\bibfnamefont {G.}~\bibnamefont
  {Wendin}}\ and\ \bibinfo {author} {\bibfnamefont {V.~S.}\ \bibnamefont
  {Shumeiko}},\ }\bibfield  {title} {\bibinfo {title} {{Quantum bits with
  Josephson junctions (Review Article)}},\ }\href
  {https://doi.org/10.1063/1.2780165} {\bibfield  {journal} {\bibinfo
  {journal} {Low Temperature Physics}\ }\textbf {\bibinfo {volume} {33}},\
  \bibinfo {pages} {724} (\bibinfo {year} {2007})}\BibitemShut {NoStop}%
\bibitem [{\citenamefont {Clarke}\ and\ \citenamefont
  {Wilhelm}(2008)}]{clarke2008}%
  \BibitemOpen
  \bibfield  {author} {\bibinfo {author} {\bibfnamefont {J.}~\bibnamefont
  {Clarke}}\ and\ \bibinfo {author} {\bibfnamefont {F.~K.}\ \bibnamefont
  {Wilhelm}},\ }\bibfield  {title} {\bibinfo {title} {Superconducting quantum
  bits},\ }\href {https://doi.org/10.1038/nature07128} {\bibfield  {journal}
  {\bibinfo  {journal} {Nature}\ }\textbf {\bibinfo {volume} {453}},\ \bibinfo
  {pages} {1031} (\bibinfo {year} {2008})}\BibitemShut {NoStop}%
\bibitem [{\citenamefont {Gu}\ \emph {et~al.}(2017)\citenamefont {Gu},
  \citenamefont {Kockum}, \citenamefont {Miranowicz}, \citenamefont {xi~Liu},\
  and\ \citenamefont {Nori}}]{gu2017}%
  \BibitemOpen
  \bibfield  {author} {\bibinfo {author} {\bibfnamefont {X.}~\bibnamefont
  {Gu}}, \bibinfo {author} {\bibfnamefont {A.~F.}\ \bibnamefont {Kockum}},
  \bibinfo {author} {\bibfnamefont {A.}~\bibnamefont {Miranowicz}}, \bibinfo
  {author} {\bibfnamefont {Y.}~\bibnamefont {xi~Liu}},\ and\ \bibinfo {author}
  {\bibfnamefont {F.}~\bibnamefont {Nori}},\ }\bibfield  {title} {\bibinfo
  {title} {Microwave photonics with superconducting quantum circuits},\ }\href
  {https://doi.org/https://doi.org/10.1016/j.physrep.2017.10.002} {\bibfield
  {journal} {\bibinfo  {journal} {Physics Reports}\ }\textbf {\bibinfo {volume}
  {718-719}},\ \bibinfo {pages} {1 } (\bibinfo {year} {2017})}\BibitemShut
  {NoStop}%
\bibitem [{\citenamefont {Blais}\ \emph {et~al.}(2021)\citenamefont {Blais},
  \citenamefont {Grimsmo}, \citenamefont {Girvin},\ and\ \citenamefont
  {Wallraff}}]{blais2021rmp}%
  \BibitemOpen
  \bibfield  {author} {\bibinfo {author} {\bibfnamefont {A.}~\bibnamefont
  {Blais}}, \bibinfo {author} {\bibfnamefont {A.~L.}\ \bibnamefont {Grimsmo}},
  \bibinfo {author} {\bibfnamefont {S.~M.}\ \bibnamefont {Girvin}},\ and\
  \bibinfo {author} {\bibfnamefont {A.}~\bibnamefont {Wallraff}},\ }\bibfield
  {title} {\bibinfo {title} {Circuit quantum electrodynamics},\ }\href
  {https://doi.org/10.1103/RevModPhys.93.025005} {\bibfield  {journal}
  {\bibinfo  {journal} {Rev. Mod. Phys.}\ }\textbf {\bibinfo {volume} {93}},\
  \bibinfo {pages} {025005} (\bibinfo {year} {2021})}\BibitemShut {NoStop}%
\bibitem [{\citenamefont {Makhlin}\ \emph {et~al.}(2001)\citenamefont
  {Makhlin}, \citenamefont {Sch\"on},\ and\ \citenamefont
  {Shnirman}}]{makhlin2001rmp}%
  \BibitemOpen
  \bibfield  {author} {\bibinfo {author} {\bibfnamefont {Y.}~\bibnamefont
  {Makhlin}}, \bibinfo {author} {\bibfnamefont {G.}~\bibnamefont {Sch\"on}},\
  and\ \bibinfo {author} {\bibfnamefont {A.}~\bibnamefont {Shnirman}},\
  }\bibfield  {title} {\bibinfo {title} {Quantum-state engineering with
  {Josephson}-junction devices},\ }\href
  {https://doi.org/10.1103/RevModPhys.73.357} {\bibfield  {journal} {\bibinfo
  {journal} {Rev. Mod. Phys.}\ }\textbf {\bibinfo {volume} {73}},\ \bibinfo
  {pages} {357} (\bibinfo {year} {2001})}\BibitemShut {NoStop}%
\bibitem [{\citenamefont {Hofheinz}\ \emph {et~al.}(2008)\citenamefont
  {Hofheinz}, \citenamefont {Weig}, \citenamefont {Ansmann}, \citenamefont
  {Bialczak}, \citenamefont {Lucero}, \citenamefont {Neeley}, \citenamefont
  {O'Connell}, \citenamefont {Wang}, \citenamefont {Martinis},\ and\
  \citenamefont {Cleland}}]{hofheinz2008}%
  \BibitemOpen
  \bibfield  {author} {\bibinfo {author} {\bibfnamefont {M.}~\bibnamefont
  {Hofheinz}}, \bibinfo {author} {\bibfnamefont {E.~M.}\ \bibnamefont {Weig}},
  \bibinfo {author} {\bibfnamefont {M.}~\bibnamefont {Ansmann}}, \bibinfo
  {author} {\bibfnamefont {R.~C.}\ \bibnamefont {Bialczak}}, \bibinfo {author}
  {\bibfnamefont {E.}~\bibnamefont {Lucero}}, \bibinfo {author} {\bibfnamefont
  {M.}~\bibnamefont {Neeley}}, \bibinfo {author} {\bibfnamefont {A.~D.}\
  \bibnamefont {O'Connell}}, \bibinfo {author} {\bibfnamefont {H.}~\bibnamefont
  {Wang}}, \bibinfo {author} {\bibfnamefont {J.~M.}\ \bibnamefont {Martinis}},\
  and\ \bibinfo {author} {\bibfnamefont {A.~N.}\ \bibnamefont {Cleland}},\
  }\bibfield  {title} {\bibinfo {title} {Generation of {Fock} states in a
  superconducting quantum circuit},\ }\href
  {https://doi.org/10.1038/nature07136} {\bibfield  {journal} {\bibinfo
  {journal} {Nature}\ }\textbf {\bibinfo {volume} {454}},\ \bibinfo {pages}
  {310} (\bibinfo {year} {2008})}\BibitemShut {NoStop}%
\bibitem [{\citenamefont {Hofheinz}\ \emph {et~al.}(2009)\citenamefont
  {Hofheinz}, \citenamefont {Wang}, \citenamefont {Ansmann}, \citenamefont
  {Bialczak}, \citenamefont {Lucero}, \citenamefont {Neeley}, \citenamefont
  {O'Connell}, \citenamefont {Sank}, \citenamefont {Wenner}, \citenamefont
  {Martinis},\ and\ \citenamefont {Cleland}}]{hofheinz2009}%
  \BibitemOpen
  \bibfield  {author} {\bibinfo {author} {\bibfnamefont {M.}~\bibnamefont
  {Hofheinz}}, \bibinfo {author} {\bibfnamefont {H.}~\bibnamefont {Wang}},
  \bibinfo {author} {\bibfnamefont {M.}~\bibnamefont {Ansmann}}, \bibinfo
  {author} {\bibfnamefont {R.~C.}\ \bibnamefont {Bialczak}}, \bibinfo {author}
  {\bibfnamefont {E.}~\bibnamefont {Lucero}}, \bibinfo {author} {\bibfnamefont
  {M.}~\bibnamefont {Neeley}}, \bibinfo {author} {\bibfnamefont {A.~D.}\
  \bibnamefont {O'Connell}}, \bibinfo {author} {\bibfnamefont {D.}~\bibnamefont
  {Sank}}, \bibinfo {author} {\bibfnamefont {J.}~\bibnamefont {Wenner}},
  \bibinfo {author} {\bibfnamefont {J.~M.}\ \bibnamefont {Martinis}},\ and\
  \bibinfo {author} {\bibfnamefont {A.~N.}\ \bibnamefont {Cleland}},\
  }\bibfield  {title} {\bibinfo {title} {Synthesizing arbitrary quantum states
  in a superconducting resonator},\ }\href
  {https://doi.org/10.1038/nature08005} {\bibfield  {journal} {\bibinfo
  {journal} {Nature}\ }\textbf {\bibinfo {volume} {459}},\ \bibinfo {pages}
  {546} (\bibinfo {year} {2009})}\BibitemShut {NoStop}%
\bibitem [{\citenamefont {Premaratne}\ \emph {et~al.}(2017)\citenamefont
  {Premaratne}, \citenamefont {Wellstood},\ and\ \citenamefont
  {Palmer}}]{premaratne2017}%
  \BibitemOpen
  \bibfield  {author} {\bibinfo {author} {\bibfnamefont {S.~P.}\ \bibnamefont
  {Premaratne}}, \bibinfo {author} {\bibfnamefont {F.~C.}\ \bibnamefont
  {Wellstood}},\ and\ \bibinfo {author} {\bibfnamefont {B.~S.}\ \bibnamefont
  {Palmer}},\ }\bibfield  {title} {\bibinfo {title} {Microwave photon {Fock}
  state generation by stimulated raman adiabatic passage},\ }\href
  {https://doi.org/10.1038/ncomms14148} {\bibfield  {journal} {\bibinfo
  {journal} {Nature Communications}\ }\textbf {\bibinfo {volume} {8}},\
  \bibinfo {pages} {14148} (\bibinfo {year} {2017})}\BibitemShut {NoStop}%
\bibitem [{\citenamefont {Pfaff}\ \emph {et~al.}(2017)\citenamefont {Pfaff},
  \citenamefont {Axline}, \citenamefont {Burkhart}, \citenamefont {Vool},
  \citenamefont {Reinhold}, \citenamefont {Frunzio}, \citenamefont {Jiang},
  \citenamefont {Devoret},\ and\ \citenamefont {Schoelkopf}}]{pfaff2017}%
  \BibitemOpen
  \bibfield  {author} {\bibinfo {author} {\bibfnamefont {W.}~\bibnamefont
  {Pfaff}}, \bibinfo {author} {\bibfnamefont {C.~J.}\ \bibnamefont {Axline}},
  \bibinfo {author} {\bibfnamefont {L.~D.}\ \bibnamefont {Burkhart}}, \bibinfo
  {author} {\bibfnamefont {U.}~\bibnamefont {Vool}}, \bibinfo {author}
  {\bibfnamefont {P.}~\bibnamefont {Reinhold}}, \bibinfo {author}
  {\bibfnamefont {L.}~\bibnamefont {Frunzio}}, \bibinfo {author} {\bibfnamefont
  {L.}~\bibnamefont {Jiang}}, \bibinfo {author} {\bibfnamefont {M.~H.}\
  \bibnamefont {Devoret}},\ and\ \bibinfo {author} {\bibfnamefont {R.~J.}\
  \bibnamefont {Schoelkopf}},\ }\bibfield  {title} {\bibinfo {title}
  {Controlled release of multiphoton quantum states from a microwave cavity
  memory},\ }\href {https://doi.org/10.1038/nphys4143} {\bibfield  {journal}
  {\bibinfo  {journal} {Nature Physics}\ }\textbf {\bibinfo {volume} {13}},\
  \bibinfo {pages} {882} (\bibinfo {year} {2017})}\BibitemShut {NoStop}%
\bibitem [{\citenamefont {Leghtas}\ \emph {et~al.}(2013)\citenamefont
  {Leghtas}, \citenamefont {Kirchmair}, \citenamefont {Vlastakis},
  \citenamefont {Devoret}, \citenamefont {Schoelkopf},\ and\ \citenamefont
  {Mirrahimi}}]{leghtas2013}%
  \BibitemOpen
  \bibfield  {author} {\bibinfo {author} {\bibfnamefont {Z.}~\bibnamefont
  {Leghtas}}, \bibinfo {author} {\bibfnamefont {G.}~\bibnamefont {Kirchmair}},
  \bibinfo {author} {\bibfnamefont {B.}~\bibnamefont {Vlastakis}}, \bibinfo
  {author} {\bibfnamefont {M.~H.}\ \bibnamefont {Devoret}}, \bibinfo {author}
  {\bibfnamefont {R.~J.}\ \bibnamefont {Schoelkopf}},\ and\ \bibinfo {author}
  {\bibfnamefont {M.}~\bibnamefont {Mirrahimi}},\ }\bibfield  {title} {\bibinfo
  {title} {Deterministic protocol for mapping a qubit to coherent state
  superpositions in a cavity},\ }\href
  {https://doi.org/10.1103/PhysRevA.87.042315} {\bibfield  {journal} {\bibinfo
  {journal} {Phys. Rev. A}\ }\textbf {\bibinfo {volume} {87}},\ \bibinfo
  {pages} {042315} (\bibinfo {year} {2013})}\BibitemShut {NoStop}%
\bibitem [{\citenamefont {Eickbusch}\ \emph {et~al.}(2022)\citenamefont
  {Eickbusch}, \citenamefont {Sivak}, \citenamefont {Ding}, \citenamefont
  {Elder}, \citenamefont {Jha}, \citenamefont {Venkatraman}, \citenamefont
  {Royer}, \citenamefont {Girvin}, \citenamefont {Schoelkopf},\ and\
  \citenamefont {Devoret}}]{eickbusch2022}%
  \BibitemOpen
  \bibfield  {author} {\bibinfo {author} {\bibfnamefont {A.}~\bibnamefont
  {Eickbusch}}, \bibinfo {author} {\bibfnamefont {V.}~\bibnamefont {Sivak}},
  \bibinfo {author} {\bibfnamefont {A.~Z.}\ \bibnamefont {Ding}}, \bibinfo
  {author} {\bibfnamefont {S.~S.}\ \bibnamefont {Elder}}, \bibinfo {author}
  {\bibfnamefont {S.~R.}\ \bibnamefont {Jha}}, \bibinfo {author} {\bibfnamefont
  {J.}~\bibnamefont {Venkatraman}}, \bibinfo {author} {\bibfnamefont
  {B.}~\bibnamefont {Royer}}, \bibinfo {author} {\bibfnamefont {S.~M.}\
  \bibnamefont {Girvin}}, \bibinfo {author} {\bibfnamefont {R.~J.}\
  \bibnamefont {Schoelkopf}},\ and\ \bibinfo {author} {\bibfnamefont {M.~H.}\
  \bibnamefont {Devoret}},\ }\bibfield  {title} {\bibinfo {title} {Fast
  universal control of an oscillator with weak dispersive coupling to a
  qubit},\ }\href {https://doi.org/10.1038/s41567-022-01776-9} {\bibfield
  {journal} {\bibinfo  {journal} {Nature Physics}\ }\textbf {\bibinfo {volume}
  {18}},\ \bibinfo {pages} {1464} (\bibinfo {year} {2022})}\BibitemShut
  {NoStop}%
\bibitem [{\citenamefont {Devoret}\ and\ \citenamefont
  {Schoelkopf}(2013)}]{devoret2013}%
  \BibitemOpen
  \bibfield  {author} {\bibinfo {author} {\bibfnamefont {M.~H.}\ \bibnamefont
  {Devoret}}\ and\ \bibinfo {author} {\bibfnamefont {R.~J.}\ \bibnamefont
  {Schoelkopf}},\ }\bibfield  {title} {\bibinfo {title} {Superconducting
  circuits for quantum information: An outlook},\ }\href
  {https://doi.org/10.1126/science.1231930} {\bibfield  {journal} {\bibinfo
  {journal} {Science}\ }\textbf {\bibinfo {volume} {339}},\ \bibinfo {pages}
  {1169} (\bibinfo {year} {2013})}\BibitemShut {NoStop}%
\bibitem [{\citenamefont {Wendin}(2017)}]{wendin2017}%
  \BibitemOpen
  \bibfield  {author} {\bibinfo {author} {\bibfnamefont {G.}~\bibnamefont
  {Wendin}},\ }\bibfield  {title} {\bibinfo {title} {Quantum information
  processing with superconducting circuits: a review},\ }\href
  {https://doi.org/10.1088/1361-6633/aa7e1a} {\bibfield  {journal} {\bibinfo
  {journal} {Reports on Progress in Physics}\ }\textbf {\bibinfo {volume}
  {80}},\ \bibinfo {pages} {106001} (\bibinfo {year} {2017})}\BibitemShut
  {NoStop}%
\bibitem [{\citenamefont {Lamata}\ \emph {et~al.}(2018)\citenamefont {Lamata},
  \citenamefont {Parra-Rodriguez}, \citenamefont {Sanz},\ and\ \citenamefont
  {Solano}}]{lamata2018}%
  \BibitemOpen
  \bibfield  {author} {\bibinfo {author} {\bibfnamefont {L.}~\bibnamefont
  {Lamata}}, \bibinfo {author} {\bibfnamefont {A.}~\bibnamefont
  {Parra-Rodriguez}}, \bibinfo {author} {\bibfnamefont {M.}~\bibnamefont
  {Sanz}},\ and\ \bibinfo {author} {\bibfnamefont {E.}~\bibnamefont {Solano}},\
  }\bibfield  {title} {\bibinfo {title} {Digital-analog quantum simulations
  with superconducting circuits},\ }\href
  {https://doi.org/10.1080/23746149.2018.1457981} {\bibfield  {journal}
  {\bibinfo  {journal} {Advances in Physics: X}\ }\textbf {\bibinfo {volume}
  {3}},\ \bibinfo {pages} {1457981} (\bibinfo {year} {2018})}\BibitemShut
  {NoStop}%
\bibitem [{\citenamefont {Kreula}\ \emph {et~al.}(2016)\citenamefont {Kreula},
  \citenamefont {Garc\'{i}a-\'{A}lvarez}, \citenamefont {Lamata}, \citenamefont
  {Clark}, \citenamefont {Solano},\ and\ \citenamefont {Jaksch}}]{kreula2016}%
  \BibitemOpen
  \bibfield  {author} {\bibinfo {author} {\bibfnamefont {J.~M.}\ \bibnamefont
  {Kreula}}, \bibinfo {author} {\bibfnamefont {L.}~\bibnamefont
  {Garc\'{i}a-\'{A}lvarez}}, \bibinfo {author} {\bibfnamefont {L.}~\bibnamefont
  {Lamata}}, \bibinfo {author} {\bibfnamefont {S.~R.}\ \bibnamefont {Clark}},
  \bibinfo {author} {\bibfnamefont {E.}~\bibnamefont {Solano}},\ and\ \bibinfo
  {author} {\bibfnamefont {D.}~\bibnamefont {Jaksch}},\ }\bibfield  {title}
  {\bibinfo {title} {Few-qubit quantum-classical simulation of strongly
  correlated lattice fermions},\ }\href
  {https://doi.org/10.1140/epjqt/s40507-016-0049-1} {\bibfield  {journal}
  {\bibinfo  {journal} {EPJ Quantum Technology}\ }\textbf {\bibinfo {volume}
  {3}},\ \bibinfo {pages} {11} (\bibinfo {year} {2016})}\BibitemShut {NoStop}%
\bibitem [{\citenamefont {Poto{\v{c}}nik}\ \emph {et~al.}(2018)\citenamefont
  {Poto{\v{c}}nik}, \citenamefont {Bargerbos}, \citenamefont {Schr{\"o}der},
  \citenamefont {Khan}, \citenamefont {Collodo}, \citenamefont {Gasparinetti},
  \citenamefont {Salath{\'e}}, \citenamefont {Creatore}, \citenamefont
  {Eichler}, \citenamefont {T{\"u}reci}, \citenamefont {Chin},\ and\
  \citenamefont {Wallraff}}]{potocnik2018}%
  \BibitemOpen
  \bibfield  {author} {\bibinfo {author} {\bibfnamefont {A.}~\bibnamefont
  {Poto{\v{c}}nik}}, \bibinfo {author} {\bibfnamefont {A.}~\bibnamefont
  {Bargerbos}}, \bibinfo {author} {\bibfnamefont {F.~A. Y.~N.}\ \bibnamefont
  {Schr{\"o}der}}, \bibinfo {author} {\bibfnamefont {S.~A.}\ \bibnamefont
  {Khan}}, \bibinfo {author} {\bibfnamefont {M.~C.}\ \bibnamefont {Collodo}},
  \bibinfo {author} {\bibfnamefont {S.}~\bibnamefont {Gasparinetti}}, \bibinfo
  {author} {\bibfnamefont {Y.}~\bibnamefont {Salath{\'e}}}, \bibinfo {author}
  {\bibfnamefont {C.}~\bibnamefont {Creatore}}, \bibinfo {author}
  {\bibfnamefont {C.}~\bibnamefont {Eichler}}, \bibinfo {author} {\bibfnamefont
  {H.~E.}\ \bibnamefont {T{\"u}reci}}, \bibinfo {author} {\bibfnamefont
  {A.~W.}\ \bibnamefont {Chin}},\ and\ \bibinfo {author} {\bibfnamefont
  {A.}~\bibnamefont {Wallraff}},\ }\bibfield  {title} {\bibinfo {title}
  {Studying light-harvesting models with superconducting circuits},\ }\href
  {https://doi.org/10.1038/s41467-018-03312-x} {\bibfield  {journal} {\bibinfo
  {journal} {Nature Communications}\ }\textbf {\bibinfo {volume} {9}},\
  \bibinfo {pages} {904} (\bibinfo {year} {2018})}\BibitemShut {NoStop}%
\bibitem [{\citenamefont {Mezzacapo}\ \emph {et~al.}(2015)\citenamefont
  {Mezzacapo}, \citenamefont {Rico}, \citenamefont {Sab\'{\i}n}, \citenamefont
  {Egusquiza}, \citenamefont {Lamata},\ and\ \citenamefont
  {Solano}}]{mezzacapo2015}%
  \BibitemOpen
  \bibfield  {author} {\bibinfo {author} {\bibfnamefont {A.}~\bibnamefont
  {Mezzacapo}}, \bibinfo {author} {\bibfnamefont {E.}~\bibnamefont {Rico}},
  \bibinfo {author} {\bibfnamefont {C.}~\bibnamefont {Sab\'{\i}n}}, \bibinfo
  {author} {\bibfnamefont {I.~L.}\ \bibnamefont {Egusquiza}}, \bibinfo {author}
  {\bibfnamefont {L.}~\bibnamefont {Lamata}},\ and\ \bibinfo {author}
  {\bibfnamefont {E.}~\bibnamefont {Solano}},\ }\bibfield  {title} {\bibinfo
  {title} {Non-{A}belian {SU}(2) lattice gauge theories in superconducting
  circuits},\ }\href {https://doi.org/10.1103/PhysRevLett.115.240502}
  {\bibfield  {journal} {\bibinfo  {journal} {Phys. Rev. Lett.}\ }\textbf
  {\bibinfo {volume} {115}},\ \bibinfo {pages} {240502} (\bibinfo {year}
  {2015})}\BibitemShut {NoStop}%
\bibitem [{\citenamefont {Brennen}\ \emph {et~al.}(2016)\citenamefont
  {Brennen}, \citenamefont {Pupillo}, \citenamefont {Rico}, \citenamefont
  {Stace},\ and\ \citenamefont {Vodola}}]{brennen2016}%
  \BibitemOpen
  \bibfield  {author} {\bibinfo {author} {\bibfnamefont {G.~K.}\ \bibnamefont
  {Brennen}}, \bibinfo {author} {\bibfnamefont {G.}~\bibnamefont {Pupillo}},
  \bibinfo {author} {\bibfnamefont {E.}~\bibnamefont {Rico}}, \bibinfo {author}
  {\bibfnamefont {T.~M.}\ \bibnamefont {Stace}},\ and\ \bibinfo {author}
  {\bibfnamefont {D.}~\bibnamefont {Vodola}},\ }\bibfield  {title} {\bibinfo
  {title} {Loops and strings in a superconducting lattice gauge simulator},\
  }\href {https://doi.org/10.1103/PhysRevLett.117.240504} {\bibfield  {journal}
  {\bibinfo  {journal} {Phys. Rev. Lett.}\ }\textbf {\bibinfo {volume} {117}},\
  \bibinfo {pages} {240504} (\bibinfo {year} {2016})}\BibitemShut {NoStop}%
\bibitem [{\citenamefont {Chen}\ \emph {et~al.}(2011)\citenamefont {Chen},
  \citenamefont {Hover}, \citenamefont {Sendelbach}, \citenamefont {Maurer},
  \citenamefont {Merkel}, \citenamefont {Pritchett}, \citenamefont {Wilhelm},\
  and\ \citenamefont {McDermott}}]{yfchen2011}%
  \BibitemOpen
  \bibfield  {author} {\bibinfo {author} {\bibfnamefont {Y.-F.}\ \bibnamefont
  {Chen}}, \bibinfo {author} {\bibfnamefont {D.}~\bibnamefont {Hover}},
  \bibinfo {author} {\bibfnamefont {S.}~\bibnamefont {Sendelbach}}, \bibinfo
  {author} {\bibfnamefont {L.}~\bibnamefont {Maurer}}, \bibinfo {author}
  {\bibfnamefont {S.~T.}\ \bibnamefont {Merkel}}, \bibinfo {author}
  {\bibfnamefont {E.~J.}\ \bibnamefont {Pritchett}}, \bibinfo {author}
  {\bibfnamefont {F.~K.}\ \bibnamefont {Wilhelm}},\ and\ \bibinfo {author}
  {\bibfnamefont {R.}~\bibnamefont {McDermott}},\ }\bibfield  {title} {\bibinfo
  {title} {Microwave photon counter based on {J}osephson junctions},\ }\href
  {https://doi.org/10.1103/PhysRevLett.107.217401} {\bibfield  {journal}
  {\bibinfo  {journal} {Phys. Rev. Lett.}\ }\textbf {\bibinfo {volume} {107}},\
  \bibinfo {pages} {217401} (\bibinfo {year} {2011})}\BibitemShut {NoStop}%
\bibitem [{\citenamefont {Oelsner}\ \emph {et~al.}(2013)\citenamefont
  {Oelsner}, \citenamefont {Revin}, \citenamefont {Il'ichev}, \citenamefont
  {Pankratov}, \citenamefont {Meyer}, \citenamefont {Gr\"{o}nberg},
  \citenamefont {Hassel},\ and\ \citenamefont {Kuzmin}}]{oelsner2013}%
  \BibitemOpen
  \bibfield  {author} {\bibinfo {author} {\bibfnamefont {G.}~\bibnamefont
  {Oelsner}}, \bibinfo {author} {\bibfnamefont {L.~S.}\ \bibnamefont {Revin}},
  \bibinfo {author} {\bibfnamefont {E.}~\bibnamefont {Il'ichev}}, \bibinfo
  {author} {\bibfnamefont {A.~L.}\ \bibnamefont {Pankratov}}, \bibinfo {author}
  {\bibfnamefont {H.-G.}\ \bibnamefont {Meyer}}, \bibinfo {author}
  {\bibfnamefont {L.}~\bibnamefont {Gr\"{o}nberg}}, \bibinfo {author}
  {\bibfnamefont {J.}~\bibnamefont {Hassel}},\ and\ \bibinfo {author}
  {\bibfnamefont {L.~S.}\ \bibnamefont {Kuzmin}},\ }\bibfield  {title}
  {\bibinfo {title} {{Underdamped Josephson junction as a switching current
  detector}},\ }\href {https://doi.org/10.1063/1.4824308} {\bibfield  {journal}
  {\bibinfo  {journal} {Applied Physics Letters}\ }\textbf {\bibinfo {volume}
  {103}},\ \bibinfo {pages} {142605} (\bibinfo {year} {2013})}\BibitemShut
  {NoStop}%
\bibitem [{\citenamefont {Oelsner}\ \emph {et~al.}(2017)\citenamefont
  {Oelsner}, \citenamefont {Andersen}, \citenamefont {Reh\'ak}, \citenamefont
  {Schmelz}, \citenamefont {Anders}, \citenamefont {Grajcar}, \citenamefont
  {H\"ubner}, \citenamefont {M\o{}lmer},\ and\ \citenamefont
  {Il'ichev}}]{oelsner2017}%
  \BibitemOpen
  \bibfield  {author} {\bibinfo {author} {\bibfnamefont {G.}~\bibnamefont
  {Oelsner}}, \bibinfo {author} {\bibfnamefont {C.~K.}\ \bibnamefont
  {Andersen}}, \bibinfo {author} {\bibfnamefont {M.}~\bibnamefont {Reh\'ak}},
  \bibinfo {author} {\bibfnamefont {M.}~\bibnamefont {Schmelz}}, \bibinfo
  {author} {\bibfnamefont {S.}~\bibnamefont {Anders}}, \bibinfo {author}
  {\bibfnamefont {M.}~\bibnamefont {Grajcar}}, \bibinfo {author} {\bibfnamefont
  {U.}~\bibnamefont {H\"ubner}}, \bibinfo {author} {\bibfnamefont
  {K.}~\bibnamefont {M\o{}lmer}},\ and\ \bibinfo {author} {\bibfnamefont
  {E.}~\bibnamefont {Il'ichev}},\ }\bibfield  {title} {\bibinfo {title}
  {Detection of weak microwave fields with an underdamped {Josephson}
  junction},\ }\href {https://doi.org/10.1103/PhysRevApplied.7.014012}
  {\bibfield  {journal} {\bibinfo  {journal} {Phys. Rev. Appl.}\ }\textbf
  {\bibinfo {volume} {7}},\ \bibinfo {pages} {014012} (\bibinfo {year}
  {2017})}\BibitemShut {NoStop}%
\bibitem [{\citenamefont {Golubev}\ \emph {et~al.}(2021)\citenamefont
  {Golubev}, \citenamefont {Il'ichev},\ and\ \citenamefont
  {Kuzmin}}]{golubev2021}%
  \BibitemOpen
  \bibfield  {author} {\bibinfo {author} {\bibfnamefont {D.~S.}\ \bibnamefont
  {Golubev}}, \bibinfo {author} {\bibfnamefont {E.~V.}\ \bibnamefont
  {Il'ichev}},\ and\ \bibinfo {author} {\bibfnamefont {L.~S.}\ \bibnamefont
  {Kuzmin}},\ }\bibfield  {title} {\bibinfo {title} {Single-photon detection
  with a {Josephson} junction coupled to a resonator},\ }\href
  {https://doi.org/10.1103/PhysRevApplied.16.014025} {\bibfield  {journal}
  {\bibinfo  {journal} {Phys. Rev. Appl.}\ }\textbf {\bibinfo {volume} {16}},\
  \bibinfo {pages} {014025} (\bibinfo {year} {2021})}\BibitemShut {NoStop}%
\bibitem [{\citenamefont {Guarcello}\ \emph {et~al.}(2021)\citenamefont
  {Guarcello}, \citenamefont {Piedjou~Komnang}, \citenamefont {Barone},
  \citenamefont {Rettaroli}, \citenamefont {Gatti}, \citenamefont {Pagano},\
  and\ \citenamefont {Filatrella}}]{guarcello2021}%
  \BibitemOpen
  \bibfield  {author} {\bibinfo {author} {\bibfnamefont {C.}~\bibnamefont
  {Guarcello}}, \bibinfo {author} {\bibfnamefont {A.~S.}\ \bibnamefont
  {Piedjou~Komnang}}, \bibinfo {author} {\bibfnamefont {C.}~\bibnamefont
  {Barone}}, \bibinfo {author} {\bibfnamefont {A.}~\bibnamefont {Rettaroli}},
  \bibinfo {author} {\bibfnamefont {C.}~\bibnamefont {Gatti}}, \bibinfo
  {author} {\bibfnamefont {S.}~\bibnamefont {Pagano}},\ and\ \bibinfo {author}
  {\bibfnamefont {G.}~\bibnamefont {Filatrella}},\ }\bibfield  {title}
  {\bibinfo {title} {{Josephson}-based scheme for the detection of microwave
  photons},\ }\href {https://doi.org/10.1103/PhysRevApplied.16.054015}
  {\bibfield  {journal} {\bibinfo  {journal} {Phys. Rev. Appl.}\ }\textbf
  {\bibinfo {volume} {16}},\ \bibinfo {pages} {054015} (\bibinfo {year}
  {2021})}\BibitemShut {NoStop}%
\bibitem [{\citenamefont {Opremcak}\ \emph {et~al.}(2018)\citenamefont
  {Opremcak}, \citenamefont {Pechenezhskiy}, \citenamefont {Howington},
  \citenamefont {Christensen}, \citenamefont {Beck}, \citenamefont {Leonard},
  \citenamefont {Suttle}, \citenamefont {Wilen}, \citenamefont {Nesterov},
  \citenamefont {Ribeill}, \citenamefont {Thorbeck}, \citenamefont {Schlenker},
  \citenamefont {Vavilov}, \citenamefont {Plourde},\ and\ \citenamefont
  {McDermott}}]{opremcak2018}%
  \BibitemOpen
  \bibfield  {author} {\bibinfo {author} {\bibfnamefont {A.}~\bibnamefont
  {Opremcak}}, \bibinfo {author} {\bibfnamefont {I.~V.}\ \bibnamefont
  {Pechenezhskiy}}, \bibinfo {author} {\bibfnamefont {C.}~\bibnamefont
  {Howington}}, \bibinfo {author} {\bibfnamefont {B.~G.}\ \bibnamefont
  {Christensen}}, \bibinfo {author} {\bibfnamefont {M.~A.}\ \bibnamefont
  {Beck}}, \bibinfo {author} {\bibfnamefont {E.}~\bibnamefont {Leonard}},
  \bibinfo {author} {\bibfnamefont {J.}~\bibnamefont {Suttle}}, \bibinfo
  {author} {\bibfnamefont {C.}~\bibnamefont {Wilen}}, \bibinfo {author}
  {\bibfnamefont {K.~N.}\ \bibnamefont {Nesterov}}, \bibinfo {author}
  {\bibfnamefont {G.~J.}\ \bibnamefont {Ribeill}}, \bibinfo {author}
  {\bibfnamefont {T.}~\bibnamefont {Thorbeck}}, \bibinfo {author}
  {\bibfnamefont {F.}~\bibnamefont {Schlenker}}, \bibinfo {author}
  {\bibfnamefont {M.~G.}\ \bibnamefont {Vavilov}}, \bibinfo {author}
  {\bibfnamefont {B.~L.~T.}\ \bibnamefont {Plourde}},\ and\ \bibinfo {author}
  {\bibfnamefont {R.}~\bibnamefont {McDermott}},\ }\bibfield  {title} {\bibinfo
  {title} {Measurement of a superconducting qubit with a microwave photon
  counter},\ }\href {https://doi.org/10.1126/science.aat4625} {\bibfield
  {journal} {\bibinfo  {journal} {Science}\ }\textbf {\bibinfo {volume}
  {361}},\ \bibinfo {pages} {1239} (\bibinfo {year} {2018})}\BibitemShut
  {NoStop}%
\bibitem [{\citenamefont {Opremcak}\ \emph {et~al.}(2021)\citenamefont
  {Opremcak}, \citenamefont {Liu}, \citenamefont {Wilen}, \citenamefont
  {Okubo}, \citenamefont {Christensen}, \citenamefont {Sank}, \citenamefont
  {White}, \citenamefont {Vainsencher}, \citenamefont {Giustina}, \citenamefont
  {Megrant}, \citenamefont {Burkett}, \citenamefont {Plourde},\ and\
  \citenamefont {McDermott}}]{opremcak2021}%
  \BibitemOpen
  \bibfield  {author} {\bibinfo {author} {\bibfnamefont {A.}~\bibnamefont
  {Opremcak}}, \bibinfo {author} {\bibfnamefont {C.~H.}\ \bibnamefont {Liu}},
  \bibinfo {author} {\bibfnamefont {C.}~\bibnamefont {Wilen}}, \bibinfo
  {author} {\bibfnamefont {K.}~\bibnamefont {Okubo}}, \bibinfo {author}
  {\bibfnamefont {B.~G.}\ \bibnamefont {Christensen}}, \bibinfo {author}
  {\bibfnamefont {D.}~\bibnamefont {Sank}}, \bibinfo {author} {\bibfnamefont
  {T.~C.}\ \bibnamefont {White}}, \bibinfo {author} {\bibfnamefont
  {A.}~\bibnamefont {Vainsencher}}, \bibinfo {author} {\bibfnamefont
  {M.}~\bibnamefont {Giustina}}, \bibinfo {author} {\bibfnamefont
  {A.}~\bibnamefont {Megrant}}, \bibinfo {author} {\bibfnamefont
  {B.}~\bibnamefont {Burkett}}, \bibinfo {author} {\bibfnamefont {B.~L.~T.}\
  \bibnamefont {Plourde}},\ and\ \bibinfo {author} {\bibfnamefont
  {R.}~\bibnamefont {McDermott}},\ }\bibfield  {title} {\bibinfo {title}
  {High-fidelity measurement of a superconducting qubit using an on-chip
  microwave photon counter},\ }\href
  {https://doi.org/10.1103/PhysRevX.11.011027} {\bibfield  {journal} {\bibinfo
  {journal} {Phys. Rev. X}\ }\textbf {\bibinfo {volume} {11}},\ \bibinfo
  {pages} {011027} (\bibinfo {year} {2021})}\BibitemShut {NoStop}%
\bibitem [{\citenamefont {Poudel}\ \emph {et~al.}(2012)\citenamefont {Poudel},
  \citenamefont {McDermott},\ and\ \citenamefont {Vavilov}}]{poudel2012}%
  \BibitemOpen
  \bibfield  {author} {\bibinfo {author} {\bibfnamefont {A.}~\bibnamefont
  {Poudel}}, \bibinfo {author} {\bibfnamefont {R.}~\bibnamefont {McDermott}},\
  and\ \bibinfo {author} {\bibfnamefont {M.~G.}\ \bibnamefont {Vavilov}},\
  }\bibfield  {title} {\bibinfo {title} {Quantum efficiency of a microwave
  photon detector based on a current-biased {Josephson} junction},\ }\href
  {https://doi.org/10.1103/PhysRevB.86.174506} {\bibfield  {journal} {\bibinfo
  {journal} {Phys. Rev. B}\ }\textbf {\bibinfo {volume} {86}},\ \bibinfo
  {pages} {174506} (\bibinfo {year} {2012})}\BibitemShut {NoStop}%
\bibitem [{\citenamefont {Govia}\ \emph {et~al.}(2012)\citenamefont {Govia},
  \citenamefont {Pritchett}, \citenamefont {Merkel}, \citenamefont {Pineau},\
  and\ \citenamefont {Wilhelm}}]{govia2012}%
  \BibitemOpen
  \bibfield  {author} {\bibinfo {author} {\bibfnamefont {L.~C.~G.}\
  \bibnamefont {Govia}}, \bibinfo {author} {\bibfnamefont {E.~J.}\ \bibnamefont
  {Pritchett}}, \bibinfo {author} {\bibfnamefont {S.~T.}\ \bibnamefont
  {Merkel}}, \bibinfo {author} {\bibfnamefont {D.}~\bibnamefont {Pineau}},\
  and\ \bibinfo {author} {\bibfnamefont {F.~K.}\ \bibnamefont {Wilhelm}},\
  }\bibfield  {title} {\bibinfo {title} {Theory of {J}osephson
  photomultipliers: Optimal working conditions and back action},\ }\href
  {https://doi.org/10.1103/PhysRevA.86.032311} {\bibfield  {journal} {\bibinfo
  {journal} {Phys. Rev. A}\ }\textbf {\bibinfo {volume} {86}},\ \bibinfo
  {pages} {032311} (\bibinfo {year} {2012})}\BibitemShut {NoStop}%
\bibitem [{\citenamefont {Govia}\ \emph {et~al.}(2014)\citenamefont {Govia},
  \citenamefont {Pritchett}, \citenamefont {Xu}, \citenamefont {Plourde},
  \citenamefont {Vavilov}, \citenamefont {Wilhelm},\ and\ \citenamefont
  {McDermott}}]{govia2014}%
  \BibitemOpen
  \bibfield  {author} {\bibinfo {author} {\bibfnamefont {L.~C.~G.}\
  \bibnamefont {Govia}}, \bibinfo {author} {\bibfnamefont {E.~J.}\ \bibnamefont
  {Pritchett}}, \bibinfo {author} {\bibfnamefont {C.}~\bibnamefont {Xu}},
  \bibinfo {author} {\bibfnamefont {B.~L.~T.}\ \bibnamefont {Plourde}},
  \bibinfo {author} {\bibfnamefont {M.~G.}\ \bibnamefont {Vavilov}}, \bibinfo
  {author} {\bibfnamefont {F.~K.}\ \bibnamefont {Wilhelm}},\ and\ \bibinfo
  {author} {\bibfnamefont {R.}~\bibnamefont {McDermott}},\ }\bibfield  {title}
  {\bibinfo {title} {High-fidelity qubit measurement with a microwave-photon
  counter},\ }\href {https://doi.org/10.1103/PhysRevA.90.062307} {\bibfield
  {journal} {\bibinfo  {journal} {Phys. Rev. A}\ }\textbf {\bibinfo {volume}
  {90}},\ \bibinfo {pages} {062307} (\bibinfo {year} {2014})}\BibitemShut
  {NoStop}%
\bibitem [{\citenamefont {Sch\"ondorf}\ and\ \citenamefont
  {Wilhelm}(2018)}]{schondorf2018a}%
  \BibitemOpen
  \bibfield  {author} {\bibinfo {author} {\bibfnamefont {M.}~\bibnamefont
  {Sch\"ondorf}}\ and\ \bibinfo {author} {\bibfnamefont {F.~K.}\ \bibnamefont
  {Wilhelm}},\ }\bibfield  {title} {\bibinfo {title} {Nonlinear parity readout
  with a microwave photodetector},\ }\href
  {https://doi.org/10.1103/PhysRevA.97.043849} {\bibfield  {journal} {\bibinfo
  {journal} {Phys. Rev. A}\ }\textbf {\bibinfo {volume} {97}},\ \bibinfo
  {pages} {043849} (\bibinfo {year} {2018})}\BibitemShut {NoStop}%
\bibitem [{\citenamefont {Romero}\ \emph
  {et~al.}(2009{\natexlab{a}})\citenamefont {Romero}, \citenamefont
  {Garc\'{\i}a-Ripoll},\ and\ \citenamefont {Solano}}]{romero2009}%
  \BibitemOpen
  \bibfield  {author} {\bibinfo {author} {\bibfnamefont {G.}~\bibnamefont
  {Romero}}, \bibinfo {author} {\bibfnamefont {J.~J.}\ \bibnamefont
  {Garc\'{\i}a-Ripoll}},\ and\ \bibinfo {author} {\bibfnamefont
  {E.}~\bibnamefont {Solano}},\ }\bibfield  {title} {\bibinfo {title}
  {Microwave photon detector in circuit {QED}},\ }\href
  {https://doi.org/10.1103/PhysRevLett.102.173602} {\bibfield  {journal}
  {\bibinfo  {journal} {Phys. Rev. Lett.}\ }\textbf {\bibinfo {volume} {102}},\
  \bibinfo {pages} {173602} (\bibinfo {year} {2009}{\natexlab{a}})}\BibitemShut
  {NoStop}%
\bibitem [{\citenamefont {Romero}\ \emph
  {et~al.}(2009{\natexlab{b}})\citenamefont {Romero}, \citenamefont
  {Garc\'{\i}a-Ripoll},\ and\ \citenamefont {Solano}}]{romero2009ps}%
  \BibitemOpen
  \bibfield  {author} {\bibinfo {author} {\bibfnamefont {G.}~\bibnamefont
  {Romero}}, \bibinfo {author} {\bibfnamefont {J.~J.}\ \bibnamefont
  {Garc\'{\i}a-Ripoll}},\ and\ \bibinfo {author} {\bibfnamefont
  {E.}~\bibnamefont {Solano}},\ }\bibfield  {title} {\bibinfo {title}
  {Photodetection of propagating quantum microwaves in circuit {QED}},\ }\href
  {https://doi.org/10.1088/0031-8949/2009/T137/014004} {\bibfield  {journal}
  {\bibinfo  {journal} {Physica Scripta}\ }\textbf {\bibinfo {volume} {2009}},\
  \bibinfo {pages} {014004} (\bibinfo {year} {2009}{\natexlab{b}})}\BibitemShut
  {NoStop}%
\bibitem [{\citenamefont {Peropadre}\ \emph {et~al.}(2011)\citenamefont
  {Peropadre}, \citenamefont {Romero}, \citenamefont {Johansson}, \citenamefont
  {Wilson}, \citenamefont {Solano},\ and\ \citenamefont
  {Garc\'{\i}a-Ripoll}}]{peropadre2011}%
  \BibitemOpen
  \bibfield  {author} {\bibinfo {author} {\bibfnamefont {B.}~\bibnamefont
  {Peropadre}}, \bibinfo {author} {\bibfnamefont {G.}~\bibnamefont {Romero}},
  \bibinfo {author} {\bibfnamefont {G.}~\bibnamefont {Johansson}}, \bibinfo
  {author} {\bibfnamefont {C.~M.}\ \bibnamefont {Wilson}}, \bibinfo {author}
  {\bibfnamefont {E.}~\bibnamefont {Solano}},\ and\ \bibinfo {author}
  {\bibfnamefont {J.~J.}\ \bibnamefont {Garc\'{\i}a-Ripoll}},\ }\bibfield
  {title} {\bibinfo {title} {Approaching perfect microwave photodetection in
  circuit {QED}},\ }\href {https://doi.org/10.1103/PhysRevA.84.063834}
  {\bibfield  {journal} {\bibinfo  {journal} {Phys. Rev. A}\ }\textbf {\bibinfo
  {volume} {84}},\ \bibinfo {pages} {063834} (\bibinfo {year}
  {2011})}\BibitemShut {NoStop}%
\bibitem [{\citenamefont {Sch\"{o}ndorf}\ \emph {et~al.}(2018)\citenamefont
  {Sch\"{o}ndorf}, \citenamefont {Govia}, \citenamefont {Vavilov},
  \citenamefont {McDermott},\ and\ \citenamefont {Wilhelm}}]{schondorf2018b}%
  \BibitemOpen
  \bibfield  {author} {\bibinfo {author} {\bibfnamefont {M.}~\bibnamefont
  {Sch\"{o}ndorf}}, \bibinfo {author} {\bibfnamefont {L.~C.~G.}\ \bibnamefont
  {Govia}}, \bibinfo {author} {\bibfnamefont {M.~G.}\ \bibnamefont {Vavilov}},
  \bibinfo {author} {\bibfnamefont {R.}~\bibnamefont {McDermott}},\ and\
  \bibinfo {author} {\bibfnamefont {F.~K.}\ \bibnamefont {Wilhelm}},\
  }\bibfield  {title} {\bibinfo {title} {Optimizing microwave photodetection:
  input–output theory},\ }\href {https://doi.org/10.1088/2058-9565/aaa7f7}
  {\bibfield  {journal} {\bibinfo  {journal} {Quantum Science and Technology}\
  }\textbf {\bibinfo {volume} {3}},\ \bibinfo {pages} {024009} (\bibinfo {year}
  {2018})}\BibitemShut {NoStop}%
\bibitem [{\citenamefont {Sokolov}\ and\ \citenamefont
  {Stolyarov}(2020)}]{sok2020}%
  \BibitemOpen
  \bibfield  {author} {\bibinfo {author} {\bibfnamefont {A.~M.}\ \bibnamefont
  {Sokolov}}\ and\ \bibinfo {author} {\bibfnamefont {E.~V.}\ \bibnamefont
  {Stolyarov}},\ }\bibfield  {title} {\bibinfo {title} {Single-photon limit of
  dispersive readout of a qubit with a photodetector},\ }\href
  {https://doi.org/10.1103/PhysRevA.101.042306} {\bibfield  {journal} {\bibinfo
   {journal} {Phys. Rev. A}\ }\textbf {\bibinfo {volume} {101}},\ \bibinfo
  {pages} {042306} (\bibinfo {year} {2020})}\BibitemShut {NoStop}%
\bibitem [{\citenamefont {Martinis}(2009)}]{martinis2009}%
  \BibitemOpen
  \bibfield  {author} {\bibinfo {author} {\bibfnamefont {J.~M.}\ \bibnamefont
  {Martinis}},\ }\bibfield  {title} {\bibinfo {title} {Superconducting phase
  qubits},\ }\href {https://doi.org/10.1007/s11128-009-0105-1} {\bibfield
  {journal} {\bibinfo  {journal} {Quantum Information Processing}\ }\textbf
  {\bibinfo {volume} {8}},\ \bibinfo {pages} {81} (\bibinfo {year}
  {2009})}\BibitemShut {NoStop}%
\bibitem [{\citenamefont {Martinis}\ \emph {et~al.}(1985)\citenamefont
  {Martinis}, \citenamefont {Devoret},\ and\ \citenamefont
  {Clarke}}]{martinis1985}%
  \BibitemOpen
  \bibfield  {author} {\bibinfo {author} {\bibfnamefont {J.~M.}\ \bibnamefont
  {Martinis}}, \bibinfo {author} {\bibfnamefont {M.~H.}\ \bibnamefont
  {Devoret}},\ and\ \bibinfo {author} {\bibfnamefont {J.}~\bibnamefont
  {Clarke}},\ }\bibfield  {title} {\bibinfo {title} {Energy-level quantization
  in the zero-voltage state of a current-biased {Josephson} junction},\ }\href
  {https://doi.org/10.1103/PhysRevLett.55.1543} {\bibfield  {journal} {\bibinfo
   {journal} {Phys. Rev. Lett.}\ }\textbf {\bibinfo {volume} {55}},\ \bibinfo
  {pages} {1543} (\bibinfo {year} {1985})}\BibitemShut {NoStop}%
\bibitem [{\citenamefont {Lenander}\ \emph {et~al.}(2011)\citenamefont
  {Lenander}, \citenamefont {Wang}, \citenamefont {Bialczak}, \citenamefont
  {Lucero}, \citenamefont {Mariantoni}, \citenamefont {Neeley}, \citenamefont
  {O'Connell}, \citenamefont {Sank}, \citenamefont {Weides}, \citenamefont
  {Wenner}, \citenamefont {Yamamoto}, \citenamefont {Yin}, \citenamefont
  {Zhao}, \citenamefont {Cleland},\ and\ \citenamefont
  {Martinis}}]{lenander2011}%
  \BibitemOpen
  \bibfield  {author} {\bibinfo {author} {\bibfnamefont {M.}~\bibnamefont
  {Lenander}}, \bibinfo {author} {\bibfnamefont {H.}~\bibnamefont {Wang}},
  \bibinfo {author} {\bibfnamefont {R.~C.}\ \bibnamefont {Bialczak}}, \bibinfo
  {author} {\bibfnamefont {E.}~\bibnamefont {Lucero}}, \bibinfo {author}
  {\bibfnamefont {M.}~\bibnamefont {Mariantoni}}, \bibinfo {author}
  {\bibfnamefont {M.}~\bibnamefont {Neeley}}, \bibinfo {author} {\bibfnamefont
  {A.~D.}\ \bibnamefont {O'Connell}}, \bibinfo {author} {\bibfnamefont
  {D.}~\bibnamefont {Sank}}, \bibinfo {author} {\bibfnamefont {M.}~\bibnamefont
  {Weides}}, \bibinfo {author} {\bibfnamefont {J.}~\bibnamefont {Wenner}},
  \bibinfo {author} {\bibfnamefont {T.}~\bibnamefont {Yamamoto}}, \bibinfo
  {author} {\bibfnamefont {Y.}~\bibnamefont {Yin}}, \bibinfo {author}
  {\bibfnamefont {J.}~\bibnamefont {Zhao}}, \bibinfo {author} {\bibfnamefont
  {A.~N.}\ \bibnamefont {Cleland}},\ and\ \bibinfo {author} {\bibfnamefont
  {J.~M.}\ \bibnamefont {Martinis}},\ }\bibfield  {title} {\bibinfo {title}
  {Measurement of energy decay in superconducting qubits from nonequilibrium
  quasiparticles},\ }\href {https://doi.org/10.1103/PhysRevB.84.024501}
  {\bibfield  {journal} {\bibinfo  {journal} {Phys. Rev. B}\ }\textbf {\bibinfo
  {volume} {84}},\ \bibinfo {pages} {024501} (\bibinfo {year}
  {2011})}\BibitemShut {NoStop}%
\bibitem [{\citenamefont {Wang}\ \emph {et~al.}(2014)\citenamefont {Wang},
  \citenamefont {Gao}, \citenamefont {Pop}, \citenamefont {Vool}, \citenamefont
  {Axline}, \citenamefont {Brecht}, \citenamefont {Heeres}, \citenamefont
  {Frunzio}, \citenamefont {Devoret}, \citenamefont {Catelani}, \citenamefont
  {Glazman},\ and\ \citenamefont {Schoelkopf}}]{wang2014}%
  \BibitemOpen
  \bibfield  {author} {\bibinfo {author} {\bibfnamefont {C.}~\bibnamefont
  {Wang}}, \bibinfo {author} {\bibfnamefont {Y.~Y.}\ \bibnamefont {Gao}},
  \bibinfo {author} {\bibfnamefont {I.~M.}\ \bibnamefont {Pop}}, \bibinfo
  {author} {\bibfnamefont {U.}~\bibnamefont {Vool}}, \bibinfo {author}
  {\bibfnamefont {C.}~\bibnamefont {Axline}}, \bibinfo {author} {\bibfnamefont
  {T.}~\bibnamefont {Brecht}}, \bibinfo {author} {\bibfnamefont {R.~W.}\
  \bibnamefont {Heeres}}, \bibinfo {author} {\bibfnamefont {L.}~\bibnamefont
  {Frunzio}}, \bibinfo {author} {\bibfnamefont {M.~H.}\ \bibnamefont
  {Devoret}}, \bibinfo {author} {\bibfnamefont {G.}~\bibnamefont {Catelani}},
  \bibinfo {author} {\bibfnamefont {L.~I.}\ \bibnamefont {Glazman}},\ and\
  \bibinfo {author} {\bibfnamefont {R.~J.}\ \bibnamefont {Schoelkopf}},\
  }\bibfield  {title} {\bibinfo {title} {Measurement and control of
  quasiparticle dynamics in a superconducting qubit},\ }\href
  {https://doi.org/10.1038/ncomms6836} {\bibfield  {journal} {\bibinfo
  {journal} {Nature Communications}\ }\textbf {\bibinfo {volume} {5}},\
  \bibinfo {pages} {5836} (\bibinfo {year} {2014})}\BibitemShut {NoStop}%
\bibitem [{\citenamefont {Ribeill}(2016)}]{ribeill2016}%
  \BibitemOpen
  \bibfield  {author} {\bibinfo {author} {\bibfnamefont {G.}~\bibnamefont
  {Ribeill}},\ }\emph {\bibinfo {title} {Qubit Readout with the {Josephson}
  Photomultiplier}},\ \href@noop {} {Ph.D. thesis},\ \bibinfo  {school}
  {University of Wisconsin-Madison} (\bibinfo {year} {2016})\BibitemShut
  {NoStop}%
\bibitem [{\citenamefont {Shnyrkov}\ \emph {et~al.}(2018)\citenamefont
  {Shnyrkov}, \citenamefont {Yangcao}, \citenamefont {Soroka}, \citenamefont
  {Turutanov},\ and\ \citenamefont {Lyakhno}}]{shnyrkov2018}%
  \BibitemOpen
  \bibfield  {author} {\bibinfo {author} {\bibfnamefont {V.~I.}\ \bibnamefont
  {Shnyrkov}}, \bibinfo {author} {\bibfnamefont {W.}~\bibnamefont {Yangcao}},
  \bibinfo {author} {\bibfnamefont {A.~A.}\ \bibnamefont {Soroka}}, \bibinfo
  {author} {\bibfnamefont {O.~G.}\ \bibnamefont {Turutanov}},\ and\ \bibinfo
  {author} {\bibfnamefont {V.~Y.}\ \bibnamefont {Lyakhno}},\ }\bibfield
  {title} {\bibinfo {title} {{Frequency-tuned microwave photon counter based on
  a superconductive quantum interferometer}},\ }\href
  {https://doi.org/10.1063/1.5024538} {\bibfield  {journal} {\bibinfo
  {journal} {Low Temperature Physics}\ }\textbf {\bibinfo {volume} {44}},\
  \bibinfo {pages} {213} (\bibinfo {year} {2018})}\BibitemShut {NoStop}%
\bibitem [{\citenamefont {Shnyrkov}\ \emph {et~al.}(2023)\citenamefont
  {Shnyrkov}, \citenamefont {Shapovalov}, \citenamefont {Lyakhno},
  \citenamefont {Dumik}, \citenamefont {Kalenyuk},\ and\ \citenamefont
  {Febvre}}]{shnyrkov2023}%
  \BibitemOpen
  \bibfield  {author} {\bibinfo {author} {\bibfnamefont {V.~I.}\ \bibnamefont
  {Shnyrkov}}, \bibinfo {author} {\bibfnamefont {A.~P.}\ \bibnamefont
  {Shapovalov}}, \bibinfo {author} {\bibfnamefont {V.~Y.}\ \bibnamefont
  {Lyakhno}}, \bibinfo {author} {\bibfnamefont {A.~O.}\ \bibnamefont {Dumik}},
  \bibinfo {author} {\bibfnamefont {A.~A.}\ \bibnamefont {Kalenyuk}},\ and\
  \bibinfo {author} {\bibfnamefont {P.}~\bibnamefont {Febvre}},\ }\bibfield
  {title} {\bibinfo {title} {An {RF SQUID} readout for a flux qubit-based
  microwave single photon counter},\ }\href
  {https://doi.org/10.1088/1361-6668/acb10e} {\bibfield  {journal} {\bibinfo
  {journal} {Superconductor Science and Technology}\ }\textbf {\bibinfo
  {volume} {36}},\ \bibinfo {pages} {035005} (\bibinfo {year}
  {2023})}\BibitemShut {NoStop}%
\bibitem [{\citenamefont {Nielsen}\ and\ \citenamefont
  {Chuang}()}]{nielsen2000book}%
  \BibitemOpen
  \bibfield  {author} {\bibinfo {author} {\bibfnamefont {M.}~\bibnamefont
  {Nielsen}}\ and\ \bibinfo {author} {\bibfnamefont {I.}~\bibnamefont
  {Chuang}},\ }\href@noop {} {\emph {\bibinfo {title} {Quantum Computation and
  Quantum Information}}}\ (\bibinfo  {publisher} {Cambridge University
  Press})\BibitemShut {NoStop}%
\bibitem [{\citenamefont {van Enk}(2017)}]{vanenk2017}%
  \BibitemOpen
  \bibfield  {author} {\bibinfo {author} {\bibfnamefont {S.~J.}\ \bibnamefont
  {van Enk}},\ }\bibfield  {title} {\bibinfo {title} {Photodetector figures of
  merit in terms of {POVMs}},\ }\href
  {https://doi.org/10.1088/2399-6528/aa90ce} {\bibfield  {journal} {\bibinfo
  {journal} {J. Phys. Commun.}\ }\textbf {\bibinfo {volume} {1}},\ \bibinfo
  {pages} {045001} (\bibinfo {year} {2017})}\BibitemShut {NoStop}%
\bibitem [{\citenamefont {Len}\ \emph {et~al.}(2022)\citenamefont {Len},
  \citenamefont {Byelova}, \citenamefont {Uzunova},\ and\ \citenamefont
  {Semenov}}]{Len2022}%
  \BibitemOpen
  \bibfield  {author} {\bibinfo {author} {\bibfnamefont {V.~Y.}\ \bibnamefont
  {Len}}, \bibinfo {author} {\bibfnamefont {M.~M.}\ \bibnamefont {Byelova}},
  \bibinfo {author} {\bibfnamefont {V.~A.}\ \bibnamefont {Uzunova}},\ and\
  \bibinfo {author} {\bibfnamefont {A.~A.}\ \bibnamefont {Semenov}},\
  }\bibfield  {title} {\bibinfo {title} {Realistic photon-number resolution in
  generalized {H}ong-{O}u-{M}andel experiment},\ }\href
  {https://doi.org/10.1088/1402-4896/ac9095} {\bibfield  {journal} {\bibinfo
  {journal} {Phys. Scr.}\ }\textbf {\bibinfo {volume} {97}},\ \bibinfo {pages}
  {105102} (\bibinfo {year} {2022})}\BibitemShut {NoStop}%
\bibitem [{\citenamefont {Titulaer}\ and\ \citenamefont
  {Glauber}(1965)}]{titulaer1965}%
  \BibitemOpen
  \bibfield  {author} {\bibinfo {author} {\bibfnamefont {U.~M.}\ \bibnamefont
  {Titulaer}}\ and\ \bibinfo {author} {\bibfnamefont {R.~J.}\ \bibnamefont
  {Glauber}},\ }\bibfield  {title} {\bibinfo {title} {Correlation functions for
  coherent fields},\ }\href {https://doi.org/10.1103/PhysRev.140.B676}
  {\bibfield  {journal} {\bibinfo  {journal} {Phys. Rev.}\ }\textbf {\bibinfo
  {volume} {140}},\ \bibinfo {pages} {B676} (\bibinfo {year}
  {1965})}\BibitemShut {NoStop}%
\bibitem [{\citenamefont {Mandel}(1986)}]{mandel1986}%
  \BibitemOpen
  \bibfield  {author} {\bibinfo {author} {\bibfnamefont {L.}~\bibnamefont
  {Mandel}},\ }\bibfield  {title} {\bibinfo {title} {Non-classical states of
  the electromagnetic field},\ }\href
  {https://doi.org/10.1088/0031-8949/1986/t12/005} {\bibfield  {journal}
  {\bibinfo  {journal} {Phys. Scr.}\ }\textbf {\bibinfo {volume} {T12}},\
  \bibinfo {pages} {34} (\bibinfo {year} {1986})}\BibitemShut {NoStop}%
\bibitem [{\citenamefont {Sperling}\ and\ \citenamefont
  {Walmsley}(2018{\natexlab{a}})}]{sperling2018a}%
  \BibitemOpen
  \bibfield  {author} {\bibinfo {author} {\bibfnamefont {J.}~\bibnamefont
  {Sperling}}\ and\ \bibinfo {author} {\bibfnamefont {I.~A.}\ \bibnamefont
  {Walmsley}},\ }\bibfield  {title} {\bibinfo {title} {Quasiprobability
  representation of quantum coherence},\ }\href
  {https://doi.org/10.1103/PhysRevA.97.062327} {\bibfield  {journal} {\bibinfo
  {journal} {Phys. Rev. A}\ }\textbf {\bibinfo {volume} {97}},\ \bibinfo
  {pages} {062327} (\bibinfo {year} {2018}{\natexlab{a}})}\BibitemShut
  {NoStop}%
\bibitem [{\citenamefont {Sperling}\ and\ \citenamefont
  {Walmsley}(2018{\natexlab{b}})}]{sperling2018b}%
  \BibitemOpen
  \bibfield  {author} {\bibinfo {author} {\bibfnamefont {J.}~\bibnamefont
  {Sperling}}\ and\ \bibinfo {author} {\bibfnamefont {I.~A.}\ \bibnamefont
  {Walmsley}},\ }\bibfield  {title} {\bibinfo {title} {Quasistates and
  quasiprobabilities},\ }\href {https://doi.org/10.1103/PhysRevA.98.042122}
  {\bibfield  {journal} {\bibinfo  {journal} {Phys. Rev. A}\ }\textbf {\bibinfo
  {volume} {98}},\ \bibinfo {pages} {042122} (\bibinfo {year}
  {2018}{\natexlab{b}})}\BibitemShut {NoStop}%
\bibitem [{\citenamefont {Sperling}\ and\ \citenamefont
  {Vogel}(2020)}]{sperling2020}%
  \BibitemOpen
  \bibfield  {author} {\bibinfo {author} {\bibfnamefont {J.}~\bibnamefont
  {Sperling}}\ and\ \bibinfo {author} {\bibfnamefont {W.}~\bibnamefont
  {Vogel}},\ }\bibfield  {title} {\bibinfo {title} {Quasiprobability
  distributions for quantum-optical coherence and beyond},\ }\href
  {https://doi.org/10.1088/1402-4896/ab5501} {\bibfield  {journal} {\bibinfo
  {journal} {Phys. Scr.}\ }\textbf {\bibinfo {volume} {95}},\ \bibinfo {pages}
  {034007} (\bibinfo {year} {2020})}\BibitemShut {NoStop}%
\bibitem [{\citenamefont {Vogel}\ and\ \citenamefont
  {Welsch}(2006)}]{vogel2006book}%
  \BibitemOpen
  \bibfield  {author} {\bibinfo {author} {\bibfnamefont {W.}~\bibnamefont
  {Vogel}}\ and\ \bibinfo {author} {\bibfnamefont {D.-G.}\ \bibnamefont
  {Welsch}},\ }\href@noop {} {\emph {\bibinfo {title} {Quantum Optics}}}\
  (\bibinfo  {publisher} {Wiley-VCH},\ \bibinfo {year} {2006})\BibitemShut
  {NoStop}%
\bibitem [{\citenamefont {Agarwal}(2013)}]{agarwal2013book}%
  \BibitemOpen
  \bibfield  {author} {\bibinfo {author} {\bibfnamefont {G.~S.}\ \bibnamefont
  {Agarwal}},\ }\href@noop {} {\emph {\bibinfo {title} {Quantum Optics}}}\
  (\bibinfo  {publisher} {Cambridge University Press},\ \bibinfo {year}
  {2013})\BibitemShut {NoStop}%
\bibitem [{\citenamefont {Glauber}(1963)}]{glauber1963}%
  \BibitemOpen
  \bibfield  {author} {\bibinfo {author} {\bibfnamefont {R.~J.}\ \bibnamefont
  {Glauber}},\ }\bibfield  {title} {\bibinfo {title} {Coherent and incoherent
  states of the radiation field},\ }\href
  {https://doi.org/10.1103/PhysRev.131.2766} {\bibfield  {journal} {\bibinfo
  {journal} {Phys. Rev.}\ }\textbf {\bibinfo {volume} {131}},\ \bibinfo {pages}
  {2766} (\bibinfo {year} {1963})}\BibitemShut {NoStop}%
\bibitem [{\citenamefont {Sudarshan}(1963)}]{sudarshan1963}%
  \BibitemOpen
  \bibfield  {author} {\bibinfo {author} {\bibfnamefont {E.~C.~G.}\
  \bibnamefont {Sudarshan}},\ }\bibfield  {title} {\bibinfo {title}
  {Equivalence of semiclassical and quantum mechanical descriptions of
  statistical light beams},\ }\href
  {https://doi.org/10.1103/PhysRevLett.10.277} {\bibfield  {journal} {\bibinfo
  {journal} {Phys. Rev. Lett.}\ }\textbf {\bibinfo {volume} {10}},\ \bibinfo
  {pages} {277} (\bibinfo {year} {1963})}\BibitemShut {NoStop}%
\bibitem [{\citenamefont {Semenov}\ and\ \citenamefont
  {Klimov}(2021)}]{semenov2021}%
  \BibitemOpen
  \bibfield  {author} {\bibinfo {author} {\bibfnamefont {A.~A.}\ \bibnamefont
  {Semenov}}\ and\ \bibinfo {author} {\bibfnamefont {A.~B.}\ \bibnamefont
  {Klimov}},\ }\bibfield  {title} {\bibinfo {title} {Dual form of the
  phase-space classical simulation problem in quantum optics},\ }\href
  {https://doi.org/10.1088/1367-2630/ac40cc} {\bibfield  {journal} {\bibinfo
  {journal} {New J. Phys.}\ }\textbf {\bibinfo {volume} {23}},\ \bibinfo
  {pages} {123046} (\bibinfo {year} {2021})}\BibitemShut {NoStop}%
\bibitem [{\citenamefont {Kovtoniuk}\ \emph {et~al.}(2023)\citenamefont
  {Kovtoniuk}, \citenamefont {Stolyarov}, \citenamefont {Kliushnichenko},\ and\
  \citenamefont {Semenov}}]{kovtoniuk2023}%
  \BibitemOpen
  \bibfield  {author} {\bibinfo {author} {\bibfnamefont {V.~S.}\ \bibnamefont
  {Kovtoniuk}}, \bibinfo {author} {\bibfnamefont {E.~V.}\ \bibnamefont
  {Stolyarov}}, \bibinfo {author} {\bibfnamefont {O.~V.}\ \bibnamefont
  {Kliushnichenko}},\ and\ \bibinfo {author} {\bibfnamefont {A.~A.}\
  \bibnamefont {Semenov}},\ }\href@noop {} {\bibinfo {title} {Tight
  inequalities for nonclassicality of measurement statistics}} (\bibinfo {year}
  {2023}),\ \Eprint {https://arxiv.org/abs/2310.14263} {arXiv:2310.14263
  [quant-ph]} \BibitemShut {NoStop}%
\bibitem [{\citenamefont {G\"{o}ppl}\ \emph {et~al.}(2008)\citenamefont
  {G\"{o}ppl}, \citenamefont {Fragner}, \citenamefont {Baur}, \citenamefont
  {Bianchetti}, \citenamefont {Filipp}, \citenamefont {Fink}, \citenamefont
  {Leek}, \citenamefont {Puebla}, \citenamefont {Steffen},\ and\ \citenamefont
  {Wallraff}}]{goppl2008}%
  \BibitemOpen
  \bibfield  {author} {\bibinfo {author} {\bibfnamefont {M.}~\bibnamefont
  {G\"{o}ppl}}, \bibinfo {author} {\bibfnamefont {A.}~\bibnamefont {Fragner}},
  \bibinfo {author} {\bibfnamefont {M.}~\bibnamefont {Baur}}, \bibinfo {author}
  {\bibfnamefont {R.}~\bibnamefont {Bianchetti}}, \bibinfo {author}
  {\bibfnamefont {S.}~\bibnamefont {Filipp}}, \bibinfo {author} {\bibfnamefont
  {J.~M.}\ \bibnamefont {Fink}}, \bibinfo {author} {\bibfnamefont {P.~J.}\
  \bibnamefont {Leek}}, \bibinfo {author} {\bibfnamefont {G.}~\bibnamefont
  {Puebla}}, \bibinfo {author} {\bibfnamefont {L.}~\bibnamefont {Steffen}},\
  and\ \bibinfo {author} {\bibfnamefont {A.}~\bibnamefont {Wallraff}},\
  }\bibfield  {title} {\bibinfo {title} {Coplanar waveguide resonators for
  circuit quantum electrodynamics},\ }\href {https://doi.org/10.1063/1.3010859}
  {\bibfield  {journal} {\bibinfo  {journal} {Journal of Applied Physics}\
  }\textbf {\bibinfo {volume} {104}},\ \bibinfo {pages} {113904} (\bibinfo
  {year} {2008})}\BibitemShut {NoStop}%
\bibitem [{\citenamefont {Megrant}\ \emph {et~al.}(2012)\citenamefont
  {Megrant}, \citenamefont {Neill}, \citenamefont {Barends}, \citenamefont
  {Chiaro}, \citenamefont {Chen}, \citenamefont {Feigl}, \citenamefont {Kelly},
  \citenamefont {Lucero}, \citenamefont {Mariantoni}, \citenamefont
  {O’Malley}, \citenamefont {Sank}, \citenamefont {Vainsencher},
  \citenamefont {Wenner}, \citenamefont {White}, \citenamefont {Yin},
  \citenamefont {Zhao}, \citenamefont {Palmstrøm}, \citenamefont {Martinis},\
  and\ \citenamefont {Cleland}}]{megr2012}%
  \BibitemOpen
  \bibfield  {author} {\bibinfo {author} {\bibfnamefont {A.}~\bibnamefont
  {Megrant}}, \bibinfo {author} {\bibfnamefont {C.}~\bibnamefont {Neill}},
  \bibinfo {author} {\bibfnamefont {R.}~\bibnamefont {Barends}}, \bibinfo
  {author} {\bibfnamefont {B.}~\bibnamefont {Chiaro}}, \bibinfo {author}
  {\bibfnamefont {Y.}~\bibnamefont {Chen}}, \bibinfo {author} {\bibfnamefont
  {L.}~\bibnamefont {Feigl}}, \bibinfo {author} {\bibfnamefont
  {J.}~\bibnamefont {Kelly}}, \bibinfo {author} {\bibfnamefont
  {E.}~\bibnamefont {Lucero}}, \bibinfo {author} {\bibfnamefont
  {M.}~\bibnamefont {Mariantoni}}, \bibinfo {author} {\bibfnamefont {P.~J.~J.}\
  \bibnamefont {O’Malley}}, \bibinfo {author} {\bibfnamefont
  {D.}~\bibnamefont {Sank}}, \bibinfo {author} {\bibfnamefont {A.}~\bibnamefont
  {Vainsencher}}, \bibinfo {author} {\bibfnamefont {J.}~\bibnamefont {Wenner}},
  \bibinfo {author} {\bibfnamefont {T.~C.}\ \bibnamefont {White}}, \bibinfo
  {author} {\bibfnamefont {Y.}~\bibnamefont {Yin}}, \bibinfo {author}
  {\bibfnamefont {J.}~\bibnamefont {Zhao}}, \bibinfo {author} {\bibfnamefont
  {C.~J.}\ \bibnamefont {Palmstrøm}}, \bibinfo {author} {\bibfnamefont
  {J.~M.}\ \bibnamefont {Martinis}},\ and\ \bibinfo {author} {\bibfnamefont
  {A.~N.}\ \bibnamefont {Cleland}},\ }\bibfield  {title} {\bibinfo {title}
  {Planar superconducting resonators with internal quality factors above one
  million},\ }\href {https://doi.org/10.1063/1.3693409} {\bibfield  {journal}
  {\bibinfo  {journal} {Applied Physics Letters}\ }\textbf {\bibinfo {volume}
  {100}},\ \bibinfo {pages} {113510} (\bibinfo {year} {2012})}\BibitemShut
  {NoStop}%
\bibitem [{\citenamefont {Bruno}\ \emph {et~al.}(2015)\citenamefont {Bruno},
  \citenamefont {de~Lange}, \citenamefont {Asaad}, \citenamefont {van~der
  Enden}, \citenamefont {Langford},\ and\ \citenamefont {DiCarlo}}]{bruno2015}%
  \BibitemOpen
  \bibfield  {author} {\bibinfo {author} {\bibfnamefont {A.}~\bibnamefont
  {Bruno}}, \bibinfo {author} {\bibfnamefont {G.}~\bibnamefont {de~Lange}},
  \bibinfo {author} {\bibfnamefont {S.}~\bibnamefont {Asaad}}, \bibinfo
  {author} {\bibfnamefont {K.~L.}\ \bibnamefont {van~der Enden}}, \bibinfo
  {author} {\bibfnamefont {N.~K.}\ \bibnamefont {Langford}},\ and\ \bibinfo
  {author} {\bibfnamefont {L.}~\bibnamefont {DiCarlo}},\ }\bibfield  {title}
  {\bibinfo {title} {Reducing intrinsic loss in superconducting resonators by
  surface treatment and deep etching of silicon substrates},\ }\href
  {https://doi.org/10.1063/1.4919761} {\bibfield  {journal} {\bibinfo
  {journal} {Applied Physics Letters}\ }\textbf {\bibinfo {volume} {106}},\
  \bibinfo {pages} {182601} (\bibinfo {year} {2015})}\BibitemShut {NoStop}%
\bibitem [{\citenamefont {Allman}\ \emph {et~al.}(2010)\citenamefont {Allman},
  \citenamefont {Altomare}, \citenamefont {Whittaker}, \citenamefont {Cicak},
  \citenamefont {Li}, \citenamefont {Sirois}, \citenamefont {Strong},
  \citenamefont {Teufel},\ and\ \citenamefont {Simmonds}}]{allman2010}%
  \BibitemOpen
  \bibfield  {author} {\bibinfo {author} {\bibfnamefont {M.~S.}\ \bibnamefont
  {Allman}}, \bibinfo {author} {\bibfnamefont {F.}~\bibnamefont {Altomare}},
  \bibinfo {author} {\bibfnamefont {J.~D.}\ \bibnamefont {Whittaker}}, \bibinfo
  {author} {\bibfnamefont {K.}~\bibnamefont {Cicak}}, \bibinfo {author}
  {\bibfnamefont {D.}~\bibnamefont {Li}}, \bibinfo {author} {\bibfnamefont
  {A.}~\bibnamefont {Sirois}}, \bibinfo {author} {\bibfnamefont
  {J.}~\bibnamefont {Strong}}, \bibinfo {author} {\bibfnamefont {J.~D.}\
  \bibnamefont {Teufel}},\ and\ \bibinfo {author} {\bibfnamefont {R.~W.}\
  \bibnamefont {Simmonds}},\ }\bibfield  {title} {\bibinfo {title}
  {rf-{SQUID}-mediated coherent tunable coupling between a superconducting
  phase qubit and a lumped-element resonator},\ }\href
  {https://doi.org/10.1103/PhysRevLett.104.177004} {\bibfield  {journal}
  {\bibinfo  {journal} {Phys. Rev. Lett.}\ }\textbf {\bibinfo {volume} {104}},\
  \bibinfo {pages} {177004} (\bibinfo {year} {2010})}\BibitemShut {NoStop}%
\bibitem [{\citenamefont {Collodo}\ \emph {et~al.}(2019)\citenamefont
  {Collodo}, \citenamefont {Poto\ifmmode~\check{c}\else \v{c}\fi{}nik},
  \citenamefont {Gasparinetti}, \citenamefont {Besse}, \citenamefont {Pechal},
  \citenamefont {Sameti}, \citenamefont {Hartmann}, \citenamefont {Wallraff},\
  and\ \citenamefont {Eichler}}]{collodo2019}%
  \BibitemOpen
  \bibfield  {author} {\bibinfo {author} {\bibfnamefont {M.~C.}\ \bibnamefont
  {Collodo}}, \bibinfo {author} {\bibfnamefont {A.}~\bibnamefont
  {Poto\ifmmode~\check{c}\else \v{c}\fi{}nik}}, \bibinfo {author}
  {\bibfnamefont {S.}~\bibnamefont {Gasparinetti}}, \bibinfo {author}
  {\bibfnamefont {J.-C.}\ \bibnamefont {Besse}}, \bibinfo {author}
  {\bibfnamefont {M.}~\bibnamefont {Pechal}}, \bibinfo {author} {\bibfnamefont
  {M.}~\bibnamefont {Sameti}}, \bibinfo {author} {\bibfnamefont {M.~J.}\
  \bibnamefont {Hartmann}}, \bibinfo {author} {\bibfnamefont {A.}~\bibnamefont
  {Wallraff}},\ and\ \bibinfo {author} {\bibfnamefont {C.}~\bibnamefont
  {Eichler}},\ }\bibfield  {title} {\bibinfo {title} {Observation of the
  crossover from photon ordering to delocalization in tunably coupled
  resonators},\ }\href {https://doi.org/10.1103/PhysRevLett.122.183601}
  {\bibfield  {journal} {\bibinfo  {journal} {Phys. Rev. Lett.}\ }\textbf
  {\bibinfo {volume} {122}},\ \bibinfo {pages} {183601} (\bibinfo {year}
  {2019})}\BibitemShut {NoStop}%
\bibitem [{\citenamefont {Vrajitoarea}\ \emph {et~al.}(2020)\citenamefont
  {Vrajitoarea}, \citenamefont {Huang}, \citenamefont {Groszkowski},
  \citenamefont {Koch},\ and\ \citenamefont {Houck}}]{vrajitoarea2020}%
  \BibitemOpen
  \bibfield  {author} {\bibinfo {author} {\bibfnamefont {A.}~\bibnamefont
  {Vrajitoarea}}, \bibinfo {author} {\bibfnamefont {Z.}~\bibnamefont {Huang}},
  \bibinfo {author} {\bibfnamefont {P.}~\bibnamefont {Groszkowski}}, \bibinfo
  {author} {\bibfnamefont {J.}~\bibnamefont {Koch}},\ and\ \bibinfo {author}
  {\bibfnamefont {A.~A.}\ \bibnamefont {Houck}},\ }\bibfield  {title} {\bibinfo
  {title} {Quantum control of an oscillator using a stimulated {Josephson}
  nonlinearity},\ }\href {https://doi.org/10.1038/s41567-019-0703-5} {\bibfield
   {journal} {\bibinfo  {journal} {Nature Physics}\ }\textbf {\bibinfo {volume}
  {16}},\ \bibinfo {pages} {211} (\bibinfo {year} {2020})}\BibitemShut
  {NoStop}%
\bibitem [{\citenamefont {Reagor}\ \emph {et~al.}(2016)\citenamefont {Reagor},
  \citenamefont {Pfaff}, \citenamefont {Axline}, \citenamefont {Heeres},
  \citenamefont {Ofek}, \citenamefont {Sliwa}, \citenamefont {Holland},
  \citenamefont {Wang}, \citenamefont {Blumoff}, \citenamefont {Chou},
  \citenamefont {Hatridge}, \citenamefont {Frunzio}, \citenamefont {Devoret},
  \citenamefont {Jiang},\ and\ \citenamefont {Schoelkopf}}]{reagor2016}%
  \BibitemOpen
  \bibfield  {author} {\bibinfo {author} {\bibfnamefont {M.}~\bibnamefont
  {Reagor}}, \bibinfo {author} {\bibfnamefont {W.}~\bibnamefont {Pfaff}},
  \bibinfo {author} {\bibfnamefont {C.}~\bibnamefont {Axline}}, \bibinfo
  {author} {\bibfnamefont {R.~W.}\ \bibnamefont {Heeres}}, \bibinfo {author}
  {\bibfnamefont {N.}~\bibnamefont {Ofek}}, \bibinfo {author} {\bibfnamefont
  {K.}~\bibnamefont {Sliwa}}, \bibinfo {author} {\bibfnamefont
  {E.}~\bibnamefont {Holland}}, \bibinfo {author} {\bibfnamefont
  {C.}~\bibnamefont {Wang}}, \bibinfo {author} {\bibfnamefont {J.}~\bibnamefont
  {Blumoff}}, \bibinfo {author} {\bibfnamefont {K.}~\bibnamefont {Chou}},
  \bibinfo {author} {\bibfnamefont {M.~J.}\ \bibnamefont {Hatridge}}, \bibinfo
  {author} {\bibfnamefont {L.}~\bibnamefont {Frunzio}}, \bibinfo {author}
  {\bibfnamefont {M.~H.}\ \bibnamefont {Devoret}}, \bibinfo {author}
  {\bibfnamefont {L.}~\bibnamefont {Jiang}},\ and\ \bibinfo {author}
  {\bibfnamefont {R.~J.}\ \bibnamefont {Schoelkopf}},\ }\bibfield  {title}
  {\bibinfo {title} {Quantum memory with millisecond coherence in circuit
  {QED}},\ }\href {https://doi.org/10.1103/PhysRevB.94.014506} {\bibfield
  {journal} {\bibinfo  {journal} {Phys. Rev. B}\ }\textbf {\bibinfo {volume}
  {94}},\ \bibinfo {pages} {014506} (\bibinfo {year} {2016})}\BibitemShut
  {NoStop}%
\bibitem [{\citenamefont {Heidler}\ \emph {et~al.}(2021)\citenamefont
  {Heidler}, \citenamefont {Schneider}, \citenamefont {Kustura}, \citenamefont
  {Gonzalez-Ballestero}, \citenamefont {Romero-Isart},\ and\ \citenamefont
  {Kirchmair}}]{heidler2021}%
  \BibitemOpen
  \bibfield  {author} {\bibinfo {author} {\bibfnamefont {P.}~\bibnamefont
  {Heidler}}, \bibinfo {author} {\bibfnamefont {C.~M.~F.}\ \bibnamefont
  {Schneider}}, \bibinfo {author} {\bibfnamefont {K.}~\bibnamefont {Kustura}},
  \bibinfo {author} {\bibfnamefont {C.}~\bibnamefont {Gonzalez-Ballestero}},
  \bibinfo {author} {\bibfnamefont {O.}~\bibnamefont {Romero-Isart}},\ and\
  \bibinfo {author} {\bibfnamefont {G.}~\bibnamefont {Kirchmair}},\ }\bibfield
  {title} {\bibinfo {title} {Non-{Markovian} effects of two-level systems in a
  niobium coaxial resonator with a single-photon lifetime of 10 milliseconds},\
  }\href {https://doi.org/10.1103/PhysRevApplied.16.034024} {\bibfield
  {journal} {\bibinfo  {journal} {Phys. Rev. Applied}\ }\textbf {\bibinfo
  {volume} {16}},\ \bibinfo {pages} {034024} (\bibinfo {year}
  {2021})}\BibitemShut {NoStop}%
\bibitem [{\citenamefont {Paik}\ \emph {et~al.}(2011)\citenamefont {Paik},
  \citenamefont {Schuster}, \citenamefont {Bishop}, \citenamefont {Kirchmair},
  \citenamefont {Catelani}, \citenamefont {Sears}, \citenamefont {Johnson},
  \citenamefont {Reagor}, \citenamefont {Frunzio}, \citenamefont {Glazman},
  \citenamefont {Girvin}, \citenamefont {Devoret},\ and\ \citenamefont
  {Schoelkopf}}]{paik2011}%
  \BibitemOpen
  \bibfield  {author} {\bibinfo {author} {\bibfnamefont {H.}~\bibnamefont
  {Paik}}, \bibinfo {author} {\bibfnamefont {D.~I.}\ \bibnamefont {Schuster}},
  \bibinfo {author} {\bibfnamefont {L.~S.}\ \bibnamefont {Bishop}}, \bibinfo
  {author} {\bibfnamefont {G.}~\bibnamefont {Kirchmair}}, \bibinfo {author}
  {\bibfnamefont {G.}~\bibnamefont {Catelani}}, \bibinfo {author}
  {\bibfnamefont {A.~P.}\ \bibnamefont {Sears}}, \bibinfo {author}
  {\bibfnamefont {B.~R.}\ \bibnamefont {Johnson}}, \bibinfo {author}
  {\bibfnamefont {M.~J.}\ \bibnamefont {Reagor}}, \bibinfo {author}
  {\bibfnamefont {L.}~\bibnamefont {Frunzio}}, \bibinfo {author} {\bibfnamefont
  {L.~I.}\ \bibnamefont {Glazman}}, \bibinfo {author} {\bibfnamefont {S.~M.}\
  \bibnamefont {Girvin}}, \bibinfo {author} {\bibfnamefont {M.~H.}\
  \bibnamefont {Devoret}},\ and\ \bibinfo {author} {\bibfnamefont {R.~J.}\
  \bibnamefont {Schoelkopf}},\ }\bibfield  {title} {\bibinfo {title}
  {Observation of high coherence in {Josephson} junction qubits measured in a
  three-dimensional circuit {QED} architecture},\ }\href
  {https://doi.org/10.1103/PhysRevLett.107.240501} {\bibfield  {journal}
  {\bibinfo  {journal} {Phys. Rev. Lett.}\ }\textbf {\bibinfo {volume} {107}},\
  \bibinfo {pages} {240501} (\bibinfo {year} {2011})}\BibitemShut {NoStop}%
\bibitem [{\citenamefont {Reagor}\ \emph {et~al.}(2013)\citenamefont {Reagor},
  \citenamefont {Paik}, \citenamefont {Catelani}, \citenamefont {Sun},
  \citenamefont {Axline}, \citenamefont {Holland}, \citenamefont {Pop},
  \citenamefont {Masluk}, \citenamefont {Brecht}, \citenamefont {Frunzio},
  \citenamefont {Devoret}, \citenamefont {Glazman},\ and\ \citenamefont
  {Schoelkopf}}]{reagor2013}%
  \BibitemOpen
  \bibfield  {author} {\bibinfo {author} {\bibfnamefont {M.}~\bibnamefont
  {Reagor}}, \bibinfo {author} {\bibfnamefont {H.}~\bibnamefont {Paik}},
  \bibinfo {author} {\bibfnamefont {G.}~\bibnamefont {Catelani}}, \bibinfo
  {author} {\bibfnamefont {L.}~\bibnamefont {Sun}}, \bibinfo {author}
  {\bibfnamefont {C.}~\bibnamefont {Axline}}, \bibinfo {author} {\bibfnamefont
  {E.}~\bibnamefont {Holland}}, \bibinfo {author} {\bibfnamefont {I.~M.}\
  \bibnamefont {Pop}}, \bibinfo {author} {\bibfnamefont {N.~A.}\ \bibnamefont
  {Masluk}}, \bibinfo {author} {\bibfnamefont {T.}~\bibnamefont {Brecht}},
  \bibinfo {author} {\bibfnamefont {L.}~\bibnamefont {Frunzio}}, \bibinfo
  {author} {\bibfnamefont {M.~H.}\ \bibnamefont {Devoret}}, \bibinfo {author}
  {\bibfnamefont {L.}~\bibnamefont {Glazman}},\ and\ \bibinfo {author}
  {\bibfnamefont {R.~J.}\ \bibnamefont {Schoelkopf}},\ }\bibfield  {title}
  {\bibinfo {title} {Reaching 10 ms single photon lifetimes for superconducting
  aluminum cavities},\ }\href {https://doi.org/10.1063/1.4807015} {\bibfield
  {journal} {\bibinfo  {journal} {Applied Physics Letters}\ }\textbf {\bibinfo
  {volume} {102}},\ \bibinfo {pages} {192604} (\bibinfo {year}
  {2013})}\BibitemShut {NoStop}%
\bibitem [{\citenamefont {Romanenko}\ \emph {et~al.}(2020)\citenamefont
  {Romanenko}, \citenamefont {Pilipenko}, \citenamefont {Zorzetti},
  \citenamefont {Frolov}, \citenamefont {Awida}, \citenamefont {Belomestnykh},
  \citenamefont {Posen},\ and\ \citenamefont {Grassellino}}]{romanenko2020}%
  \BibitemOpen
  \bibfield  {author} {\bibinfo {author} {\bibfnamefont {A.}~\bibnamefont
  {Romanenko}}, \bibinfo {author} {\bibfnamefont {R.}~\bibnamefont
  {Pilipenko}}, \bibinfo {author} {\bibfnamefont {S.}~\bibnamefont {Zorzetti}},
  \bibinfo {author} {\bibfnamefont {D.}~\bibnamefont {Frolov}}, \bibinfo
  {author} {\bibfnamefont {M.}~\bibnamefont {Awida}}, \bibinfo {author}
  {\bibfnamefont {S.}~\bibnamefont {Belomestnykh}}, \bibinfo {author}
  {\bibfnamefont {S.}~\bibnamefont {Posen}},\ and\ \bibinfo {author}
  {\bibfnamefont {A.}~\bibnamefont {Grassellino}},\ }\bibfield  {title}
  {\bibinfo {title} {Three-dimensional superconducting resonators at ${T} < 20$
  mk with photon lifetimes up to $\tau=2$ s},\ }\href
  {https://doi.org/10.1103/PhysRevApplied.13.034032} {\bibfield  {journal}
  {\bibinfo  {journal} {Phys. Rev. Applied}\ }\textbf {\bibinfo {volume}
  {13}},\ \bibinfo {pages} {034032} (\bibinfo {year} {2020})}\BibitemShut
  {NoStop}%
\bibitem [{\citenamefont {Milul}\ \emph {et~al.}(2023)\citenamefont {Milul},
  \citenamefont {Guttel}, \citenamefont {Goldblatt}, \citenamefont {Hazanov},
  \citenamefont {Joshi}, \citenamefont {Chausovsky}, \citenamefont {Kahn},
  \citenamefont {\ifmmode~\mbox{\c{C}}\else \c{C}\fi{}ifty\"urek},
  \citenamefont {Lafont},\ and\ \citenamefont {Rosenblum}}]{milul2023}%
  \BibitemOpen
  \bibfield  {author} {\bibinfo {author} {\bibfnamefont {O.}~\bibnamefont
  {Milul}}, \bibinfo {author} {\bibfnamefont {B.}~\bibnamefont {Guttel}},
  \bibinfo {author} {\bibfnamefont {U.}~\bibnamefont {Goldblatt}}, \bibinfo
  {author} {\bibfnamefont {S.}~\bibnamefont {Hazanov}}, \bibinfo {author}
  {\bibfnamefont {L.~M.}\ \bibnamefont {Joshi}}, \bibinfo {author}
  {\bibfnamefont {D.}~\bibnamefont {Chausovsky}}, \bibinfo {author}
  {\bibfnamefont {N.}~\bibnamefont {Kahn}}, \bibinfo {author} {\bibfnamefont
  {E.}~\bibnamefont {\ifmmode~\mbox{\c{C}}\else \c{C}\fi{}ifty\"urek}},
  \bibinfo {author} {\bibfnamefont {F.}~\bibnamefont {Lafont}},\ and\ \bibinfo
  {author} {\bibfnamefont {S.}~\bibnamefont {Rosenblum}},\ }\bibfield  {title}
  {\bibinfo {title} {Superconducting cavity qubit with tens of milliseconds
  single-photon coherence time},\ }\href
  {https://doi.org/10.1103/PRXQuantum.4.030336} {\bibfield  {journal} {\bibinfo
   {journal} {PRX Quantum}\ }\textbf {\bibinfo {volume} {4}},\ \bibinfo {pages}
  {030336} (\bibinfo {year} {2023})}\BibitemShut {NoStop}%
\bibitem [{\citenamefont {Steffen}\ \emph {et~al.}(2006)\citenamefont
  {Steffen}, \citenamefont {Ansmann}, \citenamefont {McDermott}, \citenamefont
  {Katz}, \citenamefont {Bialczak}, \citenamefont {Lucero}, \citenamefont
  {Neeley}, \citenamefont {Weig}, \citenamefont {Cleland},\ and\ \citenamefont
  {Martinis}}]{steff2006}%
  \BibitemOpen
  \bibfield  {author} {\bibinfo {author} {\bibfnamefont {M.}~\bibnamefont
  {Steffen}}, \bibinfo {author} {\bibfnamefont {M.}~\bibnamefont {Ansmann}},
  \bibinfo {author} {\bibfnamefont {R.}~\bibnamefont {McDermott}}, \bibinfo
  {author} {\bibfnamefont {N.}~\bibnamefont {Katz}}, \bibinfo {author}
  {\bibfnamefont {R.~C.}\ \bibnamefont {Bialczak}}, \bibinfo {author}
  {\bibfnamefont {E.}~\bibnamefont {Lucero}}, \bibinfo {author} {\bibfnamefont
  {M.}~\bibnamefont {Neeley}}, \bibinfo {author} {\bibfnamefont {E.~M.}\
  \bibnamefont {Weig}}, \bibinfo {author} {\bibfnamefont {A.~N.}\ \bibnamefont
  {Cleland}},\ and\ \bibinfo {author} {\bibfnamefont {J.~M.}\ \bibnamefont
  {Martinis}},\ }\bibfield  {title} {\bibinfo {title} {State tomography of
  capacitively shunted phase qubits with high fidelity},\ }\href
  {https://doi.org/10.1103/PhysRevLett.97.050502} {\bibfield  {journal}
  {\bibinfo  {journal} {Phys. Rev. Lett.}\ }\textbf {\bibinfo {volume} {97}},\
  \bibinfo {pages} {050502} (\bibinfo {year} {2006})}\BibitemShut {NoStop}%
\bibitem [{\citenamefont {Cattaneo}\ and\ \citenamefont
  {Paraoanu}(2021)}]{cattaneo2021}%
  \BibitemOpen
  \bibfield  {author} {\bibinfo {author} {\bibfnamefont {M.}~\bibnamefont
  {Cattaneo}}\ and\ \bibinfo {author} {\bibfnamefont {G.~S.}\ \bibnamefont
  {Paraoanu}},\ }\bibfield  {title} {\bibinfo {title} {Engineering dissipation
  with resistive elements in circuit quantum electrodynamics},\ }\href
  {https://doi.org/https://doi.org/10.1002/qute.202100054} {\bibfield
  {journal} {\bibinfo  {journal} {Advanced Quantum Technologies}\ }\textbf
  {\bibinfo {volume} {4}},\ \bibinfo {pages} {2100054} (\bibinfo {year}
  {2021})}\BibitemShut {NoStop}%
\bibitem [{\citenamefont {Reed}\ \emph {et~al.}(2010)\citenamefont {Reed},
  \citenamefont {Johnson}, \citenamefont {Houck}, \citenamefont {DiCarlo},
  \citenamefont {Chow}, \citenamefont {Schuster}, \citenamefont {Frunzio},\
  and\ \citenamefont {Schoelkopf}}]{reed2010}%
  \BibitemOpen
  \bibfield  {author} {\bibinfo {author} {\bibfnamefont {M.~D.}\ \bibnamefont
  {Reed}}, \bibinfo {author} {\bibfnamefont {B.~R.}\ \bibnamefont {Johnson}},
  \bibinfo {author} {\bibfnamefont {A.~A.}\ \bibnamefont {Houck}}, \bibinfo
  {author} {\bibfnamefont {L.}~\bibnamefont {DiCarlo}}, \bibinfo {author}
  {\bibfnamefont {J.~M.}\ \bibnamefont {Chow}}, \bibinfo {author}
  {\bibfnamefont {D.~I.}\ \bibnamefont {Schuster}}, \bibinfo {author}
  {\bibfnamefont {L.}~\bibnamefont {Frunzio}},\ and\ \bibinfo {author}
  {\bibfnamefont {R.~J.}\ \bibnamefont {Schoelkopf}},\ }\bibfield  {title}
  {\bibinfo {title} {Fast reset and suppressing spontaneous emission of a
  superconducting qubit},\ }\href {https://doi.org/10.1063/1.3435463}
  {\bibfield  {journal} {\bibinfo  {journal} {Applied Physics Letters}\
  }\textbf {\bibinfo {volume} {96}},\ \bibinfo {pages} {203110} (\bibinfo
  {year} {2010})}\BibitemShut {NoStop}%
\bibitem [{\citenamefont {Moon}\ and\ \citenamefont {Girvin}(2005)}]{moon2005}%
  \BibitemOpen
  \bibfield  {author} {\bibinfo {author} {\bibfnamefont {K.}~\bibnamefont
  {Moon}}\ and\ \bibinfo {author} {\bibfnamefont {S.~M.}\ \bibnamefont
  {Girvin}},\ }\bibfield  {title} {\bibinfo {title} {Theory of microwave
  parametric down-conversion and squeezing using circuit {QED}},\ }\href
  {https://doi.org/10.1103/PhysRevLett.95.140504} {\bibfield  {journal}
  {\bibinfo  {journal} {Phys. Rev. Lett.}\ }\textbf {\bibinfo {volume} {95}},\
  \bibinfo {pages} {140504} (\bibinfo {year} {2005})}\BibitemShut {NoStop}%
\bibitem [{\citenamefont {Didier}\ \emph {et~al.}(2014)\citenamefont {Didier},
  \citenamefont {Qassemi},\ and\ \citenamefont {Blais}}]{didier2014}%
  \BibitemOpen
  \bibfield  {author} {\bibinfo {author} {\bibfnamefont {N.}~\bibnamefont
  {Didier}}, \bibinfo {author} {\bibfnamefont {F.}~\bibnamefont {Qassemi}},\
  and\ \bibinfo {author} {\bibfnamefont {A.}~\bibnamefont {Blais}},\ }\bibfield
   {title} {\bibinfo {title} {Perfect squeezing by damping modulation in
  circuit quantum electrodynamics},\ }\href
  {https://doi.org/10.1103/PhysRevA.89.013820} {\bibfield  {journal} {\bibinfo
  {journal} {Phys. Rev. A}\ }\textbf {\bibinfo {volume} {89}},\ \bibinfo
  {pages} {013820} (\bibinfo {year} {2014})}\BibitemShut {NoStop}%
\bibitem [{\citenamefont {Kono}\ \emph {et~al.}(2017)\citenamefont {Kono},
  \citenamefont {Masuyama}, \citenamefont {Ishikawa}, \citenamefont {Tabuchi},
  \citenamefont {Yamazaki}, \citenamefont {Usami}, \citenamefont {Koshino},\
  and\ \citenamefont {Nakamura}}]{kono2017}%
  \BibitemOpen
  \bibfield  {author} {\bibinfo {author} {\bibfnamefont {S.}~\bibnamefont
  {Kono}}, \bibinfo {author} {\bibfnamefont {Y.}~\bibnamefont {Masuyama}},
  \bibinfo {author} {\bibfnamefont {T.}~\bibnamefont {Ishikawa}}, \bibinfo
  {author} {\bibfnamefont {Y.}~\bibnamefont {Tabuchi}}, \bibinfo {author}
  {\bibfnamefont {R.}~\bibnamefont {Yamazaki}}, \bibinfo {author}
  {\bibfnamefont {K.}~\bibnamefont {Usami}}, \bibinfo {author} {\bibfnamefont
  {K.}~\bibnamefont {Koshino}},\ and\ \bibinfo {author} {\bibfnamefont
  {Y.}~\bibnamefont {Nakamura}},\ }\bibfield  {title} {\bibinfo {title}
  {Nonclassical photon number distribution in a superconducting cavity under a
  squeezed drive},\ }\href {https://doi.org/10.1103/PhysRevLett.119.023602}
  {\bibfield  {journal} {\bibinfo  {journal} {Phys. Rev. Lett.}\ }\textbf
  {\bibinfo {volume} {119}},\ \bibinfo {pages} {023602} (\bibinfo {year}
  {2017})}\BibitemShut {NoStop}%
\bibitem [{\citenamefont {Malnou}\ \emph {et~al.}(2018)\citenamefont {Malnou},
  \citenamefont {Palken}, \citenamefont {Vale}, \citenamefont {Hilton},\ and\
  \citenamefont {Lehnert}}]{malnou2018}%
  \BibitemOpen
  \bibfield  {author} {\bibinfo {author} {\bibfnamefont {M.}~\bibnamefont
  {Malnou}}, \bibinfo {author} {\bibfnamefont {D.~A.}\ \bibnamefont {Palken}},
  \bibinfo {author} {\bibfnamefont {L.~R.}\ \bibnamefont {Vale}}, \bibinfo
  {author} {\bibfnamefont {G.~C.}\ \bibnamefont {Hilton}},\ and\ \bibinfo
  {author} {\bibfnamefont {K.~W.}\ \bibnamefont {Lehnert}},\ }\bibfield
  {title} {\bibinfo {title} {Optimal operation of a {Josephson} parametric
  amplifier for vacuum squeezing},\ }\href
  {https://doi.org/10.1103/PhysRevApplied.9.044023} {\bibfield  {journal}
  {\bibinfo  {journal} {Phys. Rev. Appl.}\ }\textbf {\bibinfo {volume} {9}},\
  \bibinfo {pages} {044023} (\bibinfo {year} {2018})}\BibitemShut {NoStop}%
\bibitem [{\citenamefont {Dassonneville}\ \emph {et~al.}(2021)\citenamefont
  {Dassonneville}, \citenamefont {Assouly}, \citenamefont {Peronnin},
  \citenamefont {Clerk}, \citenamefont {Bienfait},\ and\ \citenamefont
  {Huard}}]{dass2021}%
  \BibitemOpen
  \bibfield  {author} {\bibinfo {author} {\bibfnamefont {R.}~\bibnamefont
  {Dassonneville}}, \bibinfo {author} {\bibfnamefont {R.}~\bibnamefont
  {Assouly}}, \bibinfo {author} {\bibfnamefont {T.}~\bibnamefont {Peronnin}},
  \bibinfo {author} {\bibfnamefont {A.~A.}\ \bibnamefont {Clerk}}, \bibinfo
  {author} {\bibfnamefont {A.}~\bibnamefont {Bienfait}},\ and\ \bibinfo
  {author} {\bibfnamefont {B.}~\bibnamefont {Huard}},\ }\bibfield  {title}
  {\bibinfo {title} {Dissipative stabilization of squeezing beyond 3 db in a
  microwave mode},\ }\href {https://doi.org/10.1103/PRXQuantum.2.020323}
  {\bibfield  {journal} {\bibinfo  {journal} {PRX Quantum}\ }\textbf {\bibinfo
  {volume} {2}},\ \bibinfo {pages} {020323} (\bibinfo {year}
  {2021})}\BibitemShut {NoStop}%
\bibitem [{\citenamefont {Ma}\ \emph {et~al.}(2019)\citenamefont {Ma},
  \citenamefont {Xie},\ and\ \citenamefont {Li}}]{ma2019}%
  \BibitemOpen
  \bibfield  {author} {\bibinfo {author} {\bibfnamefont {S.-l.}\ \bibnamefont
  {Ma}}, \bibinfo {author} {\bibfnamefont {J.-k.}\ \bibnamefont {Xie}},\ and\
  \bibinfo {author} {\bibfnamefont {F.-l.}\ \bibnamefont {Li}},\ }\bibfield
  {title} {\bibinfo {title} {Generation of superposition coherent states of
  microwave fields via dissipation of a superconducting qubit with broken
  inversion symmetry},\ }\href {https://doi.org/10.1103/PhysRevA.99.022302}
  {\bibfield  {journal} {\bibinfo  {journal} {Phys. Rev. A}\ }\textbf {\bibinfo
  {volume} {99}},\ \bibinfo {pages} {022302} (\bibinfo {year}
  {2019})}\BibitemShut {NoStop}%
\bibitem [{\citenamefont {Girvin}(2019)}]{girvin2019}%
  \BibitemOpen
  \bibfield  {author} {\bibinfo {author} {\bibfnamefont {S.~M.}\ \bibnamefont
  {Girvin}},\ }\bibfield  {title} {\bibinfo {title} {{Schr\"{o}dinger cat
  states in circuit QED}},\ }in\ \href
  {https://doi.org/10.1093/oso/9780198837190.003.0011} {\emph {\bibinfo
  {booktitle} {{Current Trends in Atomic Physics}}}}\ (\bibinfo  {publisher}
  {Oxford University Press},\ \bibinfo {year} {2019})\BibitemShut {NoStop}%
\bibitem [{\citenamefont {Howington}\ \emph {et~al.}(2019)\citenamefont
  {Howington}, \citenamefont {Opremcak}, \citenamefont {McDermott},
  \citenamefont {Kirichenko}, \citenamefont {Mukhanov},\ and\ \citenamefont
  {Plourde}}]{howington2019}%
  \BibitemOpen
  \bibfield  {author} {\bibinfo {author} {\bibfnamefont {C.}~\bibnamefont
  {Howington}}, \bibinfo {author} {\bibfnamefont {A.}~\bibnamefont {Opremcak}},
  \bibinfo {author} {\bibfnamefont {R.}~\bibnamefont {McDermott}}, \bibinfo
  {author} {\bibfnamefont {A.}~\bibnamefont {Kirichenko}}, \bibinfo {author}
  {\bibfnamefont {O.~A.}\ \bibnamefont {Mukhanov}},\ and\ \bibinfo {author}
  {\bibfnamefont {B.~L.~T.}\ \bibnamefont {Plourde}},\ }\bibfield  {title}
  {\bibinfo {title} {Interfacing superconducting qubits with cryogenic logic:
  Readout},\ }\href {https://doi.org/10.1109/TASC.2019.2908884} {\bibfield
  {journal} {\bibinfo  {journal} {IEEE Transactions on Applied
  Superconductivity}\ }\textbf {\bibinfo {volume} {29}},\ \bibinfo {pages} {1}
  (\bibinfo {year} {2019})}\BibitemShut {NoStop}%
\bibitem [{\citenamefont {Devoret}(1997)}]{devoret1997}%
  \BibitemOpen
  \bibfield  {author} {\bibinfo {author} {\bibfnamefont {M.~H.}\ \bibnamefont
  {Devoret}},\ }\bibfield  {title} {\bibinfo {title} {Quantum fluctuations in
  electrical circuits},\ }in\ \href@noop {} {\emph {\bibinfo {booktitle}
  {Proceedings of {L}es {H}ouches {S}ummer {S}chool, {S}ession {LXIII},
  1995}}},\ \bibinfo {editor} {edited by\ \bibinfo {editor} {\bibfnamefont
  {S.}~\bibnamefont {Reynard}}, \bibinfo {editor} {\bibfnamefont
  {E.}~\bibnamefont {Giacobino}},\ and\ \bibinfo {editor} {\bibfnamefont
  {J.}~\bibnamefont {Zinn-Justin}}}\ (\bibinfo  {publisher} {Elsevier},\
  \bibinfo {address} {Amsterdam},\ \bibinfo {year} {1997})\BibitemShut
  {NoStop}%
\bibitem [{\citenamefont {Vool}\ and\ \citenamefont
  {Devoret}(2017)}]{vool2017}%
  \BibitemOpen
  \bibfield  {author} {\bibinfo {author} {\bibfnamefont {U.}~\bibnamefont
  {Vool}}\ and\ \bibinfo {author} {\bibfnamefont {M.}~\bibnamefont {Devoret}},\
  }\bibfield  {title} {\bibinfo {title} {Introduction to quantum
  electromagnetic circuits},\ }\href {https://doi.org/10.1002/cta.2359}
  {\bibfield  {journal} {\bibinfo  {journal} {International Journal of Circuit
  Theory and Applications}\ }\textbf {\bibinfo {volume} {45}},\ \bibinfo
  {pages} {897} (\bibinfo {year} {2017})}\BibitemShut {NoStop}%
\bibitem [{\citenamefont {Rasmussen}\ \emph {et~al.}(2021)\citenamefont
  {Rasmussen}, \citenamefont {Christensen}, \citenamefont {Pedersen},
  \citenamefont {Kristensen}, \citenamefont {B\ae{}kkegaard}, \citenamefont
  {Loft},\ and\ \citenamefont {Zinner}}]{rasmussen2021}%
  \BibitemOpen
  \bibfield  {author} {\bibinfo {author} {\bibfnamefont {S.~E.}\ \bibnamefont
  {Rasmussen}}, \bibinfo {author} {\bibfnamefont {K.~S.}\ \bibnamefont
  {Christensen}}, \bibinfo {author} {\bibfnamefont {S.~P.}\ \bibnamefont
  {Pedersen}}, \bibinfo {author} {\bibfnamefont {L.~B.}\ \bibnamefont
  {Kristensen}}, \bibinfo {author} {\bibfnamefont {T.}~\bibnamefont
  {B\ae{}kkegaard}}, \bibinfo {author} {\bibfnamefont {N.~J.~S.}\ \bibnamefont
  {Loft}},\ and\ \bibinfo {author} {\bibfnamefont {N.~T.}\ \bibnamefont
  {Zinner}},\ }\bibfield  {title} {\bibinfo {title} {Superconducting circuit
  companion---an introduction with worked examples},\ }\href
  {https://doi.org/10.1103/PRXQuantum.2.040204} {\bibfield  {journal} {\bibinfo
   {journal} {PRX Quantum}\ }\textbf {\bibinfo {volume} {2}},\ \bibinfo {pages}
  {040204} (\bibinfo {year} {2021})}\BibitemShut {NoStop}%
\bibitem [{\citenamefont {Bravyi}\ \emph {et~al.}(2011)\citenamefont {Bravyi},
  \citenamefont {DiVincenzo},\ and\ \citenamefont {Loss}}]{bravyi2011}%
  \BibitemOpen
  \bibfield  {author} {\bibinfo {author} {\bibfnamefont {S.}~\bibnamefont
  {Bravyi}}, \bibinfo {author} {\bibfnamefont {D.~P.}\ \bibnamefont
  {DiVincenzo}},\ and\ \bibinfo {author} {\bibfnamefont {D.}~\bibnamefont
  {Loss}},\ }\bibfield  {title} {\bibinfo {title} {{S}chrieffer–{W}olff
  transformation for quantum many-body systems},\ }\href
  {https://doi.org/https://doi.org/10.1016/j.aop.2011.06.004} {\bibfield
  {journal} {\bibinfo  {journal} {Annals of Physics}\ }\textbf {\bibinfo
  {volume} {326}},\ \bibinfo {pages} {2793 } (\bibinfo {year}
  {2011})}\BibitemShut {NoStop}%
\bibitem [{\citenamefont {Klimov}\ and\ \citenamefont
  {Sanchez-Soto}(2000)}]{klimov2000}%
  \BibitemOpen
  \bibfield  {author} {\bibinfo {author} {\bibfnamefont {A.~B.}\ \bibnamefont
  {Klimov}}\ and\ \bibinfo {author} {\bibfnamefont {L.~L.}\ \bibnamefont
  {Sanchez-Soto}},\ }\bibfield  {title} {\bibinfo {title} {Method of small
  rotations and effective {H}amiltonians in nonlinear quantum optics},\ }\href
  {https://doi.org/10.1103/PhysRevA.61.063802} {\bibfield  {journal} {\bibinfo
  {journal} {Phys. Rev. A}\ }\textbf {\bibinfo {volume} {61}},\ \bibinfo
  {pages} {063802} (\bibinfo {year} {2000})}\BibitemShut {NoStop}%
\bibitem [{\citenamefont {{Jaynes}}\ and\ \citenamefont
  {{Cummings}}(1963)}]{jc1963}%
  \BibitemOpen
  \bibfield  {author} {\bibinfo {author} {\bibfnamefont {E.~T.}\ \bibnamefont
  {{Jaynes}}}\ and\ \bibinfo {author} {\bibfnamefont {F.~W.}\ \bibnamefont
  {{Cummings}}},\ }\bibfield  {title} {\bibinfo {title} {Comparison of quantum
  and semiclassical radiation theories with application to the beam maser},\
  }\href {https://doi.org/10.1109/PROC.1963.1664} {\bibfield  {journal}
  {\bibinfo  {journal} {Proc. IEEE}\ }\textbf {\bibinfo {volume} {51}},\
  \bibinfo {pages} {89} (\bibinfo {year} {1963})}\BibitemShut {NoStop}%
\bibitem [{\citenamefont {Shore}\ and\ \citenamefont
  {Knight}(1993)}]{shore1993}%
  \BibitemOpen
  \bibfield  {author} {\bibinfo {author} {\bibfnamefont {B.~W.}\ \bibnamefont
  {Shore}}\ and\ \bibinfo {author} {\bibfnamefont {P.~L.}\ \bibnamefont
  {Knight}},\ }\bibfield  {title} {\bibinfo {title} {The {J}aynes-{C}ummings
  model},\ }\href {https://doi.org/10.1080/09500349314551321} {\bibfield
  {journal} {\bibinfo  {journal} {J. Mod. Opt.}\ }\textbf {\bibinfo {volume}
  {40}},\ \bibinfo {pages} {1195} (\bibinfo {year} {1993})}\BibitemShut
  {NoStop}%
\bibitem [{\citenamefont {Breuer}\ and\ \citenamefont
  {Petruccione}(2002)}]{breuer2002book}%
  \BibitemOpen
  \bibfield  {author} {\bibinfo {author} {\bibfnamefont {H.-P.}\ \bibnamefont
  {Breuer}}\ and\ \bibinfo {author} {\bibfnamefont {F.}~\bibnamefont
  {Petruccione}},\ }\href@noop {} {\emph {\bibinfo {title} {The Theory of Open
  Quantum Systems}}}\ (\bibinfo  {publisher} {Oxford University Press},\
  \bibinfo {address} {New York},\ \bibinfo {year} {2002})\BibitemShut {NoStop}%
\bibitem [{\citenamefont {Manzano}(2020)}]{manzano2020}%
  \BibitemOpen
  \bibfield  {author} {\bibinfo {author} {\bibfnamefont {D.}~\bibnamefont
  {Manzano}},\ }\bibfield  {title} {\bibinfo {title} {{A short introduction to
  the Lindblad master equation}},\ }\href {https://doi.org/10.1063/1.5115323}
  {\bibfield  {journal} {\bibinfo  {journal} {AIP Advances}\ }\textbf {\bibinfo
  {volume} {10}},\ \bibinfo {pages} {025106} (\bibinfo {year}
  {2020})}\BibitemShut {NoStop}%
\bibitem [{\citenamefont {Sokolov}\ and\ \citenamefont
  {Wilhelm}(2020)}]{sokolov2020}%
  \BibitemOpen
  \bibfield  {author} {\bibinfo {author} {\bibfnamefont {A.~M.}\ \bibnamefont
  {Sokolov}}\ and\ \bibinfo {author} {\bibfnamefont {F.~K.}\ \bibnamefont
  {Wilhelm}},\ }\bibfield  {title} {\bibinfo {title} {Superconducting detector
  that counts microwave photons up to two},\ }\href
  {https://doi.org/10.1103/PhysRevApplied.14.064063} {\bibfield  {journal}
  {\bibinfo  {journal} {Phys. Rev. Applied}\ }\textbf {\bibinfo {volume}
  {14}},\ \bibinfo {pages} {064063} (\bibinfo {year} {2020})}\BibitemShut
  {NoStop}%
\bibitem [{\citenamefont {Strauch}(2004)}]{strauch2004}%
  \BibitemOpen
  \bibfield  {author} {\bibinfo {author} {\bibfnamefont {F.~W.}\ \bibnamefont
  {Strauch}},\ }\emph {\bibinfo {title} {Theory of Superconducting Phase
  Qubits}},\ \href@noop {} {Ph.D. thesis},\ \bibinfo  {school} {School
  University of Maryland} (\bibinfo {year} {2004})\BibitemShut {NoStop}%
\bibitem [{\citenamefont {Kr{\"a}mer}\ \emph {et~al.}(2018)\citenamefont
  {Kr{\"a}mer}, \citenamefont {Plankensteiner}, \citenamefont {Ostermann},\
  and\ \citenamefont {Ritsch}}]{kramer2018}%
  \BibitemOpen
  \bibfield  {author} {\bibinfo {author} {\bibfnamefont {S.}~\bibnamefont
  {Kr{\"a}mer}}, \bibinfo {author} {\bibfnamefont {D.}~\bibnamefont
  {Plankensteiner}}, \bibinfo {author} {\bibfnamefont {L.}~\bibnamefont
  {Ostermann}},\ and\ \bibinfo {author} {\bibfnamefont {H.}~\bibnamefont
  {Ritsch}},\ }\bibfield  {title} {\bibinfo {title} {{QuantumOptics.jl}: A
  {Julia} framework for simulating open quantum systems},\ }\href
  {https://doi.org/10.1016/j.cpc.2018.02.004} {\bibfield  {journal} {\bibinfo
  {journal} {Computer Physics Communications}\ }\textbf {\bibinfo {volume}
  {227}},\ \bibinfo {pages} {109} (\bibinfo {year} {2018})}\BibitemShut
  {NoStop}%
\bibitem [{\citenamefont {Born}(1926)}]{born1926}%
  \BibitemOpen
  \bibfield  {author} {\bibinfo {author} {\bibfnamefont {M.}~\bibnamefont
  {Born}},\ }\bibfield  {title} {\bibinfo {title} {{Z}ur {Q}uantenmechanik der
  {S}to{\ss}vorg\"ange.},\ }\href {https://doi.org/10.1007/BF01397477}
  {\bibfield  {journal} {\bibinfo  {journal} {Z. Physik}\ }\textbf {\bibinfo
  {volume} {37}},\ \bibinfo {pages} {863–867} (\bibinfo {year}
  {1926})}\BibitemShut {NoStop}%
\bibitem [{\citenamefont {Plenio}\ and\ \citenamefont
  {Huelga}(2008)}]{plenio2008}%
  \BibitemOpen
  \bibfield  {author} {\bibinfo {author} {\bibfnamefont {M.~B.}\ \bibnamefont
  {Plenio}}\ and\ \bibinfo {author} {\bibfnamefont {S.~F.}\ \bibnamefont
  {Huelga}},\ }\bibfield  {title} {\bibinfo {title} {Dephasing-assisted
  transport: quantum networks and biomolecules},\ }\href
  {https://doi.org/10.1088/1367-2630/10/11/113019} {\bibfield  {journal}
  {\bibinfo  {journal} {New Journal of Physics}\ }\textbf {\bibinfo {volume}
  {10}},\ \bibinfo {pages} {113019} (\bibinfo {year} {2008})}\BibitemShut
  {NoStop}%
\bibitem [{\citenamefont {Rebentrost}\ \emph {et~al.}(2009)\citenamefont
  {Rebentrost}, \citenamefont {Mohseni}, \citenamefont {Kassal}, \citenamefont
  {Lloyd},\ and\ \citenamefont {Aspuru-Guzik}}]{rebentrost2009}%
  \BibitemOpen
  \bibfield  {author} {\bibinfo {author} {\bibfnamefont {P.}~\bibnamefont
  {Rebentrost}}, \bibinfo {author} {\bibfnamefont {M.}~\bibnamefont {Mohseni}},
  \bibinfo {author} {\bibfnamefont {I.}~\bibnamefont {Kassal}}, \bibinfo
  {author} {\bibfnamefont {S.}~\bibnamefont {Lloyd}},\ and\ \bibinfo {author}
  {\bibfnamefont {A.}~\bibnamefont {Aspuru-Guzik}},\ }\bibfield  {title}
  {\bibinfo {title} {Environment-assisted quantum transport},\ }\href
  {https://doi.org/10.1088/1367-2630/11/3/033003} {\bibfield  {journal}
  {\bibinfo  {journal} {New Journal of Physics}\ }\textbf {\bibinfo {volume}
  {11}},\ \bibinfo {pages} {033003} (\bibinfo {year} {2009})}\BibitemShut
  {NoStop}%
\bibitem [{\citenamefont {Shi}\ \emph {et~al.}(2022)\citenamefont {Shi},
  \citenamefont {Guo}, \citenamefont {Su}, \citenamefont {Chi}, \citenamefont
  {Sheng}, \citenamefont {Jiang}, \citenamefont {Cao}, \citenamefont {Wu},
  \citenamefont {Tu}, \citenamefont {Sun}, \citenamefont {Chen},\ and\
  \citenamefont {Wu}}]{shi2022}%
  \BibitemOpen
  \bibfield  {author} {\bibinfo {author} {\bibfnamefont {L.}~\bibnamefont
  {Shi}}, \bibinfo {author} {\bibfnamefont {T.}~\bibnamefont {Guo}}, \bibinfo
  {author} {\bibfnamefont {R.}~\bibnamefont {Su}}, \bibinfo {author}
  {\bibfnamefont {T.}~\bibnamefont {Chi}}, \bibinfo {author} {\bibfnamefont
  {Y.}~\bibnamefont {Sheng}}, \bibinfo {author} {\bibfnamefont
  {J.}~\bibnamefont {Jiang}}, \bibinfo {author} {\bibfnamefont
  {C.}~\bibnamefont {Cao}}, \bibinfo {author} {\bibfnamefont {J.}~\bibnamefont
  {Wu}}, \bibinfo {author} {\bibfnamefont {X.}~\bibnamefont {Tu}}, \bibinfo
  {author} {\bibfnamefont {G.}~\bibnamefont {Sun}}, \bibinfo {author}
  {\bibfnamefont {J.}~\bibnamefont {Chen}},\ and\ \bibinfo {author}
  {\bibfnamefont {P.}~\bibnamefont {Wu}},\ }\bibfield  {title} {\bibinfo
  {title} {Tantalum microwave resonators with ultra-high intrinsic quality
  factors},\ }\href {https://doi.org/10.1063/5.0124821} {\bibfield  {journal}
  {\bibinfo  {journal} {Applied Physics Letters}\ }\textbf {\bibinfo {volume}
  {121}},\ \bibinfo {pages} {242601} (\bibinfo {year} {2022})}\BibitemShut
  {NoStop}%
\bibitem [{\citenamefont {Mandel}(1979)}]{mandel1979}%
  \BibitemOpen
  \bibfield  {author} {\bibinfo {author} {\bibfnamefont {L.}~\bibnamefont
  {Mandel}},\ }\bibfield  {title} {\bibinfo {title} {Sub-poissonian photon
  statistics in resonance fluorescence},\ }\href
  {https://doi.org/10.1364/OL.4.000205} {\bibfield  {journal} {\bibinfo
  {journal} {Opt. Lett.}\ }\textbf {\bibinfo {volume} {4}},\ \bibinfo {pages}
  {205} (\bibinfo {year} {1979})}\BibitemShut {NoStop}%
\bibitem [{\citenamefont {Agarwal}\ and\ \citenamefont
  {Tara}(1992)}]{agarwal92}%
  \BibitemOpen
  \bibfield  {author} {\bibinfo {author} {\bibfnamefont {G.~S.}\ \bibnamefont
  {Agarwal}}\ and\ \bibinfo {author} {\bibfnamefont {K.}~\bibnamefont {Tara}},\
  }\bibfield  {title} {\bibinfo {title} {Nonclassical character of states
  exhibiting no squeezing or sub-{P}oissonian statistics},\ }\href
  {https://doi.org/10.1103/PhysRevA.46.485} {\bibfield  {journal} {\bibinfo
  {journal} {Phys. Rev. A}\ }\textbf {\bibinfo {volume} {46}},\ \bibinfo
  {pages} {485} (\bibinfo {year} {1992})}\BibitemShut {NoStop}%
\bibitem [{\citenamefont {Klyshko}(1996)}]{klyshko1996}%
  \BibitemOpen
  \bibfield  {author} {\bibinfo {author} {\bibfnamefont {D.~N.}\ \bibnamefont
  {Klyshko}},\ }\bibfield  {title} {\bibinfo {title} {Observable signs of
  nonclassical light},\ }\href
  {https://doi.org/https://doi.org/10.1016/0375-9601(96)00091-6} {\bibfield
  {journal} {\bibinfo  {journal} {Phys. Lett. A}\ }\textbf {\bibinfo {volume}
  {213}},\ \bibinfo {pages} {7} (\bibinfo {year} {1996})}\BibitemShut {NoStop}%
\bibitem [{\citenamefont {Innocenti}\ \emph {et~al.}(2022)\citenamefont
  {Innocenti}, \citenamefont {Lachman},\ and\ \citenamefont
  {Filip}}]{innocenti2022}%
  \BibitemOpen
  \bibfield  {author} {\bibinfo {author} {\bibfnamefont {L.}~\bibnamefont
  {Innocenti}}, \bibinfo {author} {\bibfnamefont {L.}~\bibnamefont {Lachman}},\
  and\ \bibinfo {author} {\bibfnamefont {R.}~\bibnamefont {Filip}},\ }\bibfield
   {title} {\bibinfo {title} {Nonclassicality detection from few {F}ock-state
  probabilities},\ }\href {https://doi.org/10.1038/s41534-022-00538-y}
  {\bibfield  {journal} {\bibinfo  {journal} {npj Quantum Information}\
  }\textbf {\bibinfo {volume} {8}},\ \bibinfo {pages} {30} (\bibinfo {year}
  {2022})}\BibitemShut {NoStop}%
\bibitem [{\citenamefont {Sperling}\ \emph
  {et~al.}(2012{\natexlab{b}})\citenamefont {Sperling}, \citenamefont {Vogel},\
  and\ \citenamefont {Agarwal}}]{sperling12c}%
  \BibitemOpen
  \bibfield  {author} {\bibinfo {author} {\bibfnamefont {J.}~\bibnamefont
  {Sperling}}, \bibinfo {author} {\bibfnamefont {W.}~\bibnamefont {Vogel}},\
  and\ \bibinfo {author} {\bibfnamefont {G.~S.}\ \bibnamefont {Agarwal}},\
  }\bibfield  {title} {\bibinfo {title} {Sub-binomial light},\ }\href
  {https://doi.org/10.1103/PhysRevLett.109.093601} {\bibfield  {journal}
  {\bibinfo  {journal} {Phys. Rev. Lett.}\ }\textbf {\bibinfo {volume} {109}},\
  \bibinfo {pages} {093601} (\bibinfo {year} {2012}{\natexlab{b}})}\BibitemShut
  {NoStop}%
\bibitem [{\citenamefont {Bartley}\ \emph {et~al.}(2013)\citenamefont
  {Bartley}, \citenamefont {Donati}, \citenamefont {Jin}, \citenamefont
  {Datta}, \citenamefont {Barbieri},\ and\ \citenamefont
  {Walmsley}}]{bartley13}%
  \BibitemOpen
  \bibfield  {author} {\bibinfo {author} {\bibfnamefont {T.~J.}\ \bibnamefont
  {Bartley}}, \bibinfo {author} {\bibfnamefont {G.}~\bibnamefont {Donati}},
  \bibinfo {author} {\bibfnamefont {X.-M.}\ \bibnamefont {Jin}}, \bibinfo
  {author} {\bibfnamefont {A.}~\bibnamefont {Datta}}, \bibinfo {author}
  {\bibfnamefont {M.}~\bibnamefont {Barbieri}},\ and\ \bibinfo {author}
  {\bibfnamefont {I.~A.}\ \bibnamefont {Walmsley}},\ }\bibfield  {title}
  {\bibinfo {title} {Direct observation of sub-binomial light},\ }\href
  {https://doi.org/10.1103/PhysRevLett.110.173602} {\bibfield  {journal}
  {\bibinfo  {journal} {Phys. Rev. Lett.}\ }\textbf {\bibinfo {volume} {110}},\
  \bibinfo {pages} {173602} (\bibinfo {year} {2013})}\BibitemShut {NoStop}%
\bibitem [{\citenamefont {Sperling}\ \emph {et~al.}(2013)\citenamefont
  {Sperling}, \citenamefont {Vogel},\ and\ \citenamefont
  {Agarwal}}]{sperling13b}%
  \BibitemOpen
  \bibfield  {author} {\bibinfo {author} {\bibfnamefont {J.}~\bibnamefont
  {Sperling}}, \bibinfo {author} {\bibfnamefont {W.}~\bibnamefont {Vogel}},\
  and\ \bibinfo {author} {\bibfnamefont {G.~S.}\ \bibnamefont {Agarwal}},\
  }\bibfield  {title} {\bibinfo {title} {Correlation measurements with on-off
  detectors},\ }\href {https://doi.org/10.1103/PhysRevA.88.043821} {\bibfield
  {journal} {\bibinfo  {journal} {Phys. Rev. A}\ }\textbf {\bibinfo {volume}
  {88}},\ \bibinfo {pages} {043821} (\bibinfo {year} {2013})}\BibitemShut
  {NoStop}%
\bibitem [{\citenamefont {Rivas}\ and\ \citenamefont {Luis}(2009)}]{rivas2009}%
  \BibitemOpen
  \bibfield  {author} {\bibinfo {author} {\bibfnamefont {A.}~\bibnamefont
  {Rivas}}\ and\ \bibinfo {author} {\bibfnamefont {A.}~\bibnamefont {Luis}},\
  }\bibfield  {title} {\bibinfo {title} {Nonclassicality of states and
  measurements by breaking classical bounds on statistics},\ }\href
  {https://doi.org/10.1103/PhysRevA.79.042105} {\bibfield  {journal} {\bibinfo
  {journal} {Phys. Rev. A}\ }\textbf {\bibinfo {volume} {79}},\ \bibinfo
  {pages} {042105} (\bibinfo {year} {2009})}\BibitemShut {NoStop}%
\bibitem [{\citenamefont {Zhang}\ and\ \citenamefont
  {Sanderson}(2009)}]{Zhang2009}%
  \BibitemOpen
  \bibfield  {author} {\bibinfo {author} {\bibfnamefont {J.}~\bibnamefont
  {Zhang}}\ and\ \bibinfo {author} {\bibfnamefont {A.~C.}\ \bibnamefont
  {Sanderson}},\ }\href@noop {} {\emph {\bibinfo {title} {Adaptive Differential
  Evolution}}}\ (\bibinfo  {publisher} {Springer},\ \bibinfo {address}
  {Berlin/Heidelberg, Germany},\ \bibinfo {year} {2009})\BibitemShut {NoStop}%
\bibitem [{\citenamefont {Goldstein}(1980)}]{goldstein1980}%
  \BibitemOpen
  \bibfield  {author} {\bibinfo {author} {\bibfnamefont {H.}~\bibnamefont
  {Goldstein}},\ }\href@noop {} {\emph {\bibinfo {title} {Classical
  Mechanics}}},\ \bibinfo {edition} {2nd}\ ed.\ (\bibinfo  {publisher}
  {Addison-Wesley Publishing Company},\ \bibinfo {year} {1980})\BibitemShut
  {NoStop}%
\end{thebibliography}%
\end{document}